**Charge control of semiconductor surfaces as elucidated by *ab initio* calculations – a review**

by Stanisław Krukowski*, Pawel Kempisty and Pawel Strak

Institute of High Pressure Physics, Polish Academy of Sciences, Sokołowska 29/37, 01-142 Warsaw, Poland

*Corresponding author: Stanislaw Krukowski, stach@unipress.waw.pl

Abstract

Recent progress in the investigation of the role of charge on semiconductor surfaces has been reviewed. The review begins with contributions to the calculations and data analysis. This new procedure includes the application of the Laplace correction method in *ab initio* calculations. The new analysis incorporates the projected density of states (PDOS) and Crystal Orbital Hamilton Population (COHP) and averaging of the electric potential to derive its smoothed long-range variation in space, parallel to plots of real-space band profiles. These methods include the discovery of spurious Coulomb interactions between the separated subsystems, which have different Fermi levels. The *ab initio* use of a single Fermi level may cause incorrect electron redistribution, an artificial charge of the separated subsystems, and spurious interactions. The quantum nature of the charge influence on semiconductor surfaces stems from the delocalization of electrons, which leads to the emergence of an external surface dipole, which is important for determining the workfunction and plays a role in the proposed thermalization of the adsorbate via electron tunnelling. The kinetic energy loss of the adsorbate (i.e., its thermalization) occurs via the tunnelling of electrons into the solid interior owing to the strong external dipole electric field. The other charge-related quantum effect is related to the known subsurface dipole charge layer. New simulations of the variation of the electric potential within a slab model show the band bending at the semiconductor surfaces that induces the Surface States Stark Effect (SSSE) and misrepresents the surface band diagrams. This underlines the role of pinning the Fermi level and its connection to subsurface dipoles. The charge balance determines the occupation of the surface states and the symmetry and periodicity of surface reconstructions. The occupation of the surface states may be changed by adsorption, both by the new donated electrons and by the emergence of new quantum states. As their numbers could be different, this leads to a jump in the Fermi level pinning and adsorption energy at selected critical coverages. Thus, the Fermi level becomes free, subsurface dipoles disappear, and bands become flat. The adsorption energy jump may reach several electronvolts, which may change the adsorbate equilibrium vapor pressure by several orders of magnitude. Such a flat-band state is likely to occur during the growth of crystals that fall within such pressure intervals. Additionally, quantum effects may include resonant bonding involving several states, which leads to fractional occupation. The existence of resonant states resolves the existing inconsistencies between bonding and lattice symmetry in nitrides. This effect is observed not only at the stability points but also at the activated complex position in the diffusion jumps. Additionally, the effects include the quantum state energy increase during the jump, shifting them to the vicinity of the Fermi level or even above. According to quantum statistics, the state occupation is reduced, which affects the energy barrier for diffusion. Thus, these results demonstrate the role of quantum effects in the charge control of semiconductor surfaces.

**Contents**

## I. Introduction

Crystalline semiconductors are an important class of solid-state systems that have been the focus of intensive research over the past 80 years [1,2]. They are indispensable for the development of modern civilization. This stems from their unique properties, which were first investigated in fundamental research and subsequently applied in technology. These applications are always related to the investigation, control, and manipulation of an electric charge.

Therefore, the charge is the most essential factor in semiconductor physics and chemistry. Charge manipulation is indispensable in computer logic, control of the electric current flow, detection of changes in the environment, absorption and emission of electromagnetic radiation, generation and measurements of vibration including sound, measurements of time, frequency, etc. These functions are possible because of the different functionalities of specially designed semiconductor structures, built at the depth of the semiconductors and close to and on the surface. Therefore, the surface charge state is an important factor affecting the functioning of semiconductor devices [3].

The emergence of locally unbalanced charges at semiconductor surfaces creates electric fields in their surrounding [3-6]. This induces changes in the semiconductor bulk by the emergence of a screening charge that limits field penetration into the semiconductor bulk. This long-range effect of the charge balance is related to the creation of a surface electric dipole on the internal side of the surface. The properties of the dipole, such as the charge separation and dipole density, depend on the mobile charge in the semiconductor interior. In general, the dipole may have different orientations depending on the accumulated charge and location of the surface donors or acceptors. In addition to this effect, an additional second dipole was present on the other side of the surface. The latter charge effect is related exclusively to the quantum mechanical properties of solid state bonding. Accordingly, this is typical for all solids, with the most dramatic consequences for semiconductors. The wavefunctions of the valence band states extend behind the outermost layers of atoms or, more precisely, behind the outermost layer of atomic nuclei, thus creating an external layer of negative charge. The positive charge of these atomic nuclei creates an electric dipole layer, preventing electrons from escaping, that is, constituting the electron work function [3,7-10]. The external dipole layer was recognized early by Lang and Kohn [9,10]. This phenomenon is different from the previous one, as the latter generates a high-magnitude electric dipole field narrow layer. The field was strongly localized in the direction perpendicular to the surface. Simultaneously, it is slowly varying, long-range, in a direction parallel to the surface [9, 10]. Unlike the subsurface dipole, the external dipole always points in the same direction, such that the electron energy is lower in the solid interior.

These two effects are purely physical in nature, that is, they are identical for all semiconductor surfaces. They could have a chemical counterpart in a relative secondary sense, that is, their amplitude can be modified. In addition, semiconductor surfaces have properties that are different in nature. They are related to the bonding between atoms, that is, they are chemical in nature. These are the bonds or quantum states associated with the bonding between atoms, that is, the valence band (VB) states. In standard semiconductors, these bonds are interpreted in terms of the overlap between $sp^3$ hybridized orbitals of metals and nonmetals, such as Ga and As. This picture is not universal, as shown by soft X-ray emission spectroscopy of nitrides: AlN [11] and GaN[12]. From ab initio simulations, it was found that the valence band consists of two separate subbands: the upper band created by metal $sp^3$ hybridized orbitals and nitrogen *p* orbitals [11-13]. The lower part consists of the nitrogen *s* and *d* orbitals of gallium [12,14]. In the case of AlN, the latter contribution is missing. Therefore, the bonding of nitrogen in tetrahedral coordination is created via four resonant states [14]. This has drastic consequences both in the bonding of the bulk and at the surfaces.

Surface atoms have broken bonds, that is, with no overlap, because their neighbors are missing [3]. Therefore, these states have energies higher than those of the VB states and are usually located in the bandgap. These states may be occupied, fully or partially occupied, or empty. They can be modified by the relocation of atoms, which leads to the emergence of their overlap with other atoms, changing their energies and occupations. This changes the surface symmetry, that is, new reconstructions emerge. [3]. In addition, the adsorption of other species could have similar effects. These adsorption processes may also involve charge transfer between the different surface states. Therefore, the interplay between the charge distribution, reconstruction, and possible adsorption of species is extremely complex. These phenomena are predominantly quantum. In summary, the elucidation of these phenomena by *ab initio* calculations and their relation to other approaches and the final verification by the experiment is the subject of the present review.

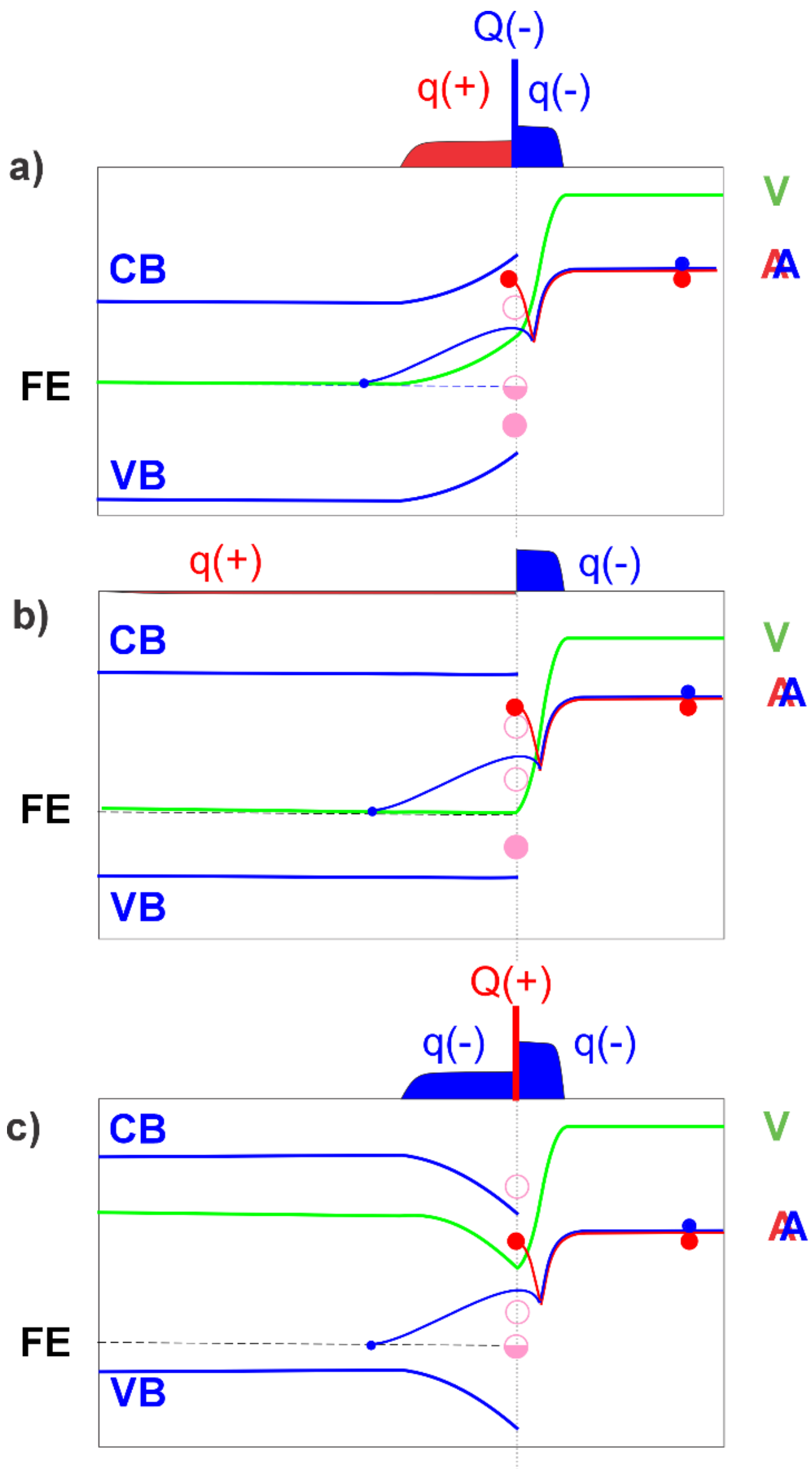

Fig. 1. Schematic presentation of basic effects affecting the properties of semiconductor surfaces underlining the role of the charge. Blue lines represent the conduction band (CB) and valence band(VB), the green line presents the electric potential (V) expressed as electron energy. Broken black line represents the Fermi energy (FE) in three cases: (a) FE pinned at the surface acceptor state, (b) FE free, (c) FE pinned at the surface donor state. Above the diagrams the excess charge is plotted using blue and red color to denote negative (q(-) and Q(-)) and positive (q(+) and Q(+)) charge, respectively. The extended charge (q) and localized (Q) are presented. The line marked by double red/blue colors (AA) shows the energy of electrically neutral adsorbate which is represented by two connected filled circles: large red for positive nucleus and small blue for electrons, respectively. In the high field region the adsorbate is dissociated into positive nucleus which is decelerated and negative electron - accelerated, respectively. The electron tunnels into the solid interior and is ultimately thermalized to the Fermi energy by the emission of phonons. The rosy circles at the surface

represent occupied and empty surface states. During attachment of the adsorbate both the electrons and the states are added. In the case of surplus of the electrons the state of the surface is shifted in the following way :$(c) \rightarrow (b) \rightarrow (a)$, in the opposite case (i.e. electron shortage) the shift is reversed $(a) \rightarrow (b) \rightarrow (c)$. When the number of new electrons and states below the Fermi level is equal, there is no shift.

The basic principles governing these phenomena are presented in a schematic diagram of the semiconductor surface profiles, as illustrated in Fig. 1. The semiconductor surface properties are affected by the presence of charged surface states and extended charge density on both the internal and external sides of the surface. The variants presented in Fig 1 (a-c) correspond to the different charges present on the surface states and on the internal side of the surface, that is, the subsurface dipole. The external dipole, which is purely of quantum origin, is always directed in the same direction. This dipole is responsible for the work function, that is, the barrier that maintains electrons inside the crystal. These two effects are related to the charge density (point and extended) and the corresponding band bending and potential change at the surface resulting from Fermi level pinning.

As shown in the equilibrium properties, charge effects also affect the molecular processes at the surface. As shown, the adsorption of a new species adds a specified number of electrons and quantum states. The relative balance between these two factors is crucial.

- In the case of a surplus of electrons, the partially occupied state is completely filled, and the ensuing motion of the Fermi level upward from the pinning state leads to a free position and ultimately to the final pinning by the higher energy state. In the case of the Fermi level free, that is, fully occupied states, the upward motion of the Fermi level is reduced to the final part, that is, it leads to the partial filling of the higher energy surface state and pinning of the Fermi level there.
- In the case of a surplus of newly added states, adsorption leads to the evacuation of the partially filled states, and the Fermi level motion goes downward from a pinned to a free position, and ultimately to partial evacuation of the lower energy state and pinning of the Fermi level there. In the case of the Fermi level free state, the sequence of events reduces the partial evacuation of the lower energy state and pinning the Fermi level there.

These scenarios describe the final results of the adsorption process and can be treated using electron counting rule (ECR) formalism.

The external dipole plays a critical role in the adsorption process. As shown in Fig. 1, the potential energy has a minimum value that is necessary to stabilize the adsorbate at the surface. Therefore, the approaching adsorbate attains the kinetic energy. This must be dissipated to thermalize the adsorbate at the surface and avoid reflection. The process is attained by two factors related to the external dipole: acceleration and tunneling of the electron into the interior, and deceleration of the positively charged adsorbate.

In addition, the periodicity of surface reconstruction is controlled by the charge balance, which may be affected by the adsorbate. This can be understood in the framework of resonating bond formalism, which must be applied to the nitrogen component of nitrides. This equilibrium picture is supplemented by an analysis of diffusion, which describes the influence of the Fermi level position on the initial and saddle point states, including the peculiarities of nitrogen bonding.

These ideas are presented in detail below: To create a coherent picture, they will be preceded by a detailed presentation of the calculation models based on *ab initio* methods addressed to surfaces and the developed analytical tools. The methodological part will be followed by a presentation of recent developments in the field along with the above short description. Finally, the results are summarized.

## II. The *ab initio* calculation models of semiconductor surfaces

The calculation models used for the simulation of the surface properties stem from the difficult compromise between the limited computing resources available for calculations and the long-distance coupling existing in these systems. In recent years, *ab initio* calculations have relied predominantly on density functional theory (DFT) formalism, which solves a set of coupled Kohn-Sham nonlinear equations for one-electron eigenfunctions and the Poisson equation for electric potential [15-19]. The Kohn–Sham equation consists of a linear part resulting from direct expression of the many-body Schrodinger equation wavefunction in terms of electron density, and the additional exchange-correlation (XC) functional that approximates the electron exchange and correlation terms, which cannot be directly converted into an electron density term. The solution of the Kohn-Sham equation is expressed as a set electron eigenfunctions, which are used to calculate electron density. These eigenfunctions were expressed as a series of functional basis sets. In practice, any set of normalized orthogonal functions may be used; nevertheless, the requirement of numerical efficiency limits the possible choices to several of the most widely used ones. The first choice is a plane wave set that is widely used, including the Vienna Ab initio Simulation Package (VASP) [20-23], Quantum

Espresso [24,25], and CASTEP [26-28]. These planar wavefunctions are marked by momentum vector values $\vec{k}$. The maximal value of the momentum vector is determined by the energy cutoff value, which is set arbitrarily using the maximal kinetic energy cutoff value $E_{cut} = \frac{\hbar^2 k^2}{2m}$. The density of the $\vec{k}$ points is determined by the size of the system $(L_i, i = x, y, z)$ by the period boundary conditions (PBC) $\left(k_i = \frac{2\pi}{L_i}\right)$. The other popular choice stems from quantum chemistry employing molecular orbital sets, such as in the Spanish Initiative for Electronic Simulations with Thousands of Atoms (SIESTA) [29], or Gaussian atom-centered functions in CRYSTAL [30]. SIESTA uses numerical finite radius atomic orbitals, which are multiplied by spherical harmonics to simulate angular dependence [31]. In solid and surface simulations, the Poisson equation is solved using the Fast Fourier Transform (FFT), which is essentially identical to the plane-wave functional basis set, as it is significantly faster than any other method. This choice entails severe limitations for the shape of the simulation system, which is rectangular with periodic boundary conditions (PBC) in three directions. Some of these consequences are discussed below.

For surface simulations, where realistic models often involve slabs containing hundreds of atoms, computational cost becomes a limiting factor. Therefore, full-electron simulations are not routinely performed. The size is reduced by limiting the basis to a fraction of all sets, which are denoted as valence electrons. The interaction with the core electrons is expressed using a pseudopotential, which is considerably different from the Coulomb term in the full-electron Hamiltonian. Several different types are used: norm-conserving pseudopotentials (NCPP), ultrasoft pseudopotentials (USPP), or the combined Projector Augmented-Wave (PAW) method. The NCPP is numerically more demanding that preserves the norm of the wavefunction of the all-electron calculations outside the cutoff radius.  The USPP has relaxed norm-conservation, which leads to improved efficiency, whereas PAW is the most accurate. Several PAWs are implemented in the VASP code, which translates into high accuracy [32,33]. In Quantum Espresso, all three pseudopotential types can be used [34,35]. In CASTEP, both the NCPP and USPP are available [26]. In parallel, SIESTA uses NCPP in the separable Kleinman-Bylander form [36] in the Troulliers-Martin design [37,38]. A convenient solution exists in form of application of the SIESTA distributed code *ATOM* for the generation of NCPP by all-electron calculations developed originally by Froyen and subsequently modified by the authors of the SIESTA code [39]. In addition, there is a routine conversion of Quantum Espresso NCPP into a Siesta application. Finally, CRYSTAL is based on Gaussian-type effective core potentials (ECP) [30].

The other equally important problem in *ab initio* DFT calculations is related to the approximation of the correlation-exchange term in the Kohn-Sham Hamiltonian. These terms cannot be directly converted into a density-dependent form. Therefore, approximate expressions must be used. Over the years, continuous efforts have been directed towards this problem, resulting in the use of numerous approximate expressions. They can be divided into the Local Density Approximation (LDA), Generalized Gradient Approximation (GGA), Meta-GGA and Hybrid functionals. In surface investigations, it is important to use large slabs, that is, a large number of atoms, in the model. For this reason, the results presented in this review are primarily based on semi-local GGA functionals, such as PBE and PBESol [40-42], or on the GGA-1/2 method [43-46], which improves the bandgap while remaining computationally tractable. Despite known limitations in gap prediction, these functionals provide reliable descriptions of surface-related charge effects, including Fermi level pinning, surface dipole formation, and electrostatic potential variation, which are essential for analyzing adsorption and reconstruction phenomena. Meta-GGA and hybrid functionals, such as Strongly Constrained and Appropriately Normed (SCAN) or Heyd-Scuseria-Ernzernhof (HSE) [47,48], although more accurate in predicting absolute band gaps and localized state positions, are generally not feasible for such large systems.

In addition, the requirement for numerical efficiency enforces the application of mathematical methods based on a Fourier transform, which imposes severe limitations on the construction of possible models. This has several consequences as it requires integration over the momentum space. An efficient method is to use 19$^{th}$ century Gauss quadrature [49]. An efficient method for implementing this approximation is to sum over the Monkhorst-Pack grid [50]. Naturally, this enforces periodic boundary conditions in all three directions. In the plane parallel to the surface, this is a natural choice and is subject to the system size only. Essentially, this choice reduces the effective simulations to the flat surface only. Therefore, these systems are also denoted as supercells, as they contain several single unit cells of the crystalline lattice. In the direction perpendicular to the surface, the simulations have to overcome long-range coupling, which acts against the periodicity of the potential and wavefunctions. The solution is to use a slab of atoms, that is, several layers of atoms in the lattice of the solid surrounded by an empty space, that is, having two surfaces. The first is natural, that is, the real one that is simulated, and the second is artificial, as it replaces the crystal bulk. In this unnatural manner, the basic periodicity in the dimension perpendicular to the surface is recovered. The boundaries of the simulated volume in the perpendicular direction can be selected in any way; the preferable choices are inside the atomic layers or in the vacuum space. This entails an infinite periodic chain of slabs in the direction perpendicular to the surface.

The slab as a whole must be electrically neutral, which is a necessary condition for the stability of an infinite chain of slabs. Nevertheless, three different factors can contribute to the charge separation in the slab, creating an electric dipole and subsequently the potential difference across the slab: spontaneous polarization, polarity of the surfaces, and localization of the charge on the surface states. All of these factors contribute to the potential difference across the slab. As a result, this potential difference must be compensated for by an artificially designed opposite term. The first remedy was proposed by Neugebauer and Scheffler, who introduced a compensating dipole layer in vacuum space [51]. The next solution was proposed by Bengtsson, who also proposed a dipole correction [52]. Krukowski et al. proposed Laplace correction method that compensates the field in a vacuum space [53]. In the latter case, in the absence of any special measures, the potential difference is compensated for by the additional uniform electric field in the vacuum space [53].

In fact, the Neugebauer and Scheffler solution compensates the slab dipole by an additional compensating dipole, which is inserted in the vacuum space at a distance close to the midpoint of the vacuum spacer between slab copies. This compensation could be affected by errors related to the imprecise representation of the charge density by a finite grid and miscalculation of the slab dipole [54]. In addition, the presence of a dipole affects the wavefunctions of the slab, despite the fact that they are at some distance. Thus, this solution could potentially deteriorate the convergence in the self-consistent field (SCF) solution loop of nonlinear equations [53]. The Bengtsson method is based on the periodic solution of the Poisson equation, with a potential jump added to the Hartree potential. Therefore, the Bengtsson method is essentially the introduction of a direct potential jump at the boundary of the simulated volume obtained from the dipole calculated for this volume [52]. This again suffers from an imprecise representation of the charge density field [54]. The finite-size dipole is replaced by a potential jump. Finally, in the paper by Krukowski et al. the Laplace correction was used as the field compensation method [53]. The method adds a solution to the Laplace equation, that is, a uniform electric field opposite to the field in the vacuum space. Therefore, the field is compensated within the precision of the determination of the field in the empty space. Naturally, this is with finite precision only, which creates a possible source of error in the application of this method. In summary, in any situation, the electric potential is different between both sides of the slab, which is exactly the case of the surface finite layer of atoms at real semiconductor surfaces. Therefore, no additional energy term must be added to the Laplace correction method [53].

As previously mentioned, the surface dipole in the slab calculations may be related to several factors. First, the spontaneous polarization is possible in the crystals having symmetry, permitting the emergence of a vectorial quantity. In wurtzite semiconductors, spontaneous polarization is possible, while in cubic zinc blende lattices, it is not [55,56]. The spontaneous polarization of wurtzite nitrides has been simulated theoretically and experimentally investigated by several authors [53,57-65].

*Table 1. Fundamental properties of group III nitride wurtzite semiconductors: experiment and theory (a and c – lattice constants, S – area for single surface site, P3 – spontaneous polarization, third component, q – polarization charge for single lattice site.)*

| $MeN_{exp/th}$ | $a$ (Å) | $c$ (Å) | $S_{wz}$ (Å$^2$) | $P_3$ $(C/m^2)$ | q ($10^{-3}$e) | $E_g$(eV) |
|---|---|---|---|---|---|---|
| $BN_{exp}$ | 2.550 | 4.227 | 5.631 | | | 6.8 |
| $BN_{th}$ | 2.5417 | 4.2019 | 5.595 | 0.061 | 3.81 | 6.77 |
| $AlN_{exp}$ | 3.111 | 4.981 | 8.382 | | | 6.09 |
| $AlN_{th}$ | 3.1126 | 4.9815 | 8.390 | 0.059 | 3.68 | 6.19 |
| $GaN_{exp}$ | 3.1890 | 5.1864 | 8.807 | | | 3.47 |
| $GaN_{th}$ | 3.1955 | 5.2040 | 8.843 | 0.011 | 0.687 | 3.41 |
| $InN_{exp}$ | 3.5705 | 5.703 | 11.040 | | | 0.90 |
| $InN_{th}$ | 3.5705 | 5.7418 | 11.040 | 0.014 | 0.874 | 0.65 |

*The "exp" and "th" mark the experiment and theory, respectively. The x-ray data were collected from: BN [66], AlN [67], GaN [68], InN [60]. The area for single site for MeN(0001) surface was calculated from $S_{wz} = a^2\sqrt{3}/2$. The spontaneous polarization are not accessible in experiment, the theoretical values were collected from Ref 62 [62]. The other theoretical values are given in Refs 58-64 [58-64]. The polarization charge for the single site was calculated from the formula: $q_{sp-pol} = P_3/S_{wz}$. The experimental bandgaps were collected from: BN [70,71], AlN [72], GaN [73,74], InN [75-77]. The theoretical bandgaps depend on the potential used, the quoted data are collected from the authors publications.*

From these data, it follows that relatively good agreement was obtained for the lattice parameters. In the case of polarization values, the theoretical data obtained from Ref 61-63 are close, but the values obtained in Ref. 64 are much larger. The obtained polarization values were relatively low, especially for GaN and InN, indicating that the electric fields associated with spontaneous polarization were relatively small. This was confirmed by the work of Kempisty et al. for a GaN(0001) surface, where the compensation of the leftover polarization field was achieved using small, fractional charge hydrogen pseudoatoms of the charge $\Delta Z \approx 0.01$ [78].

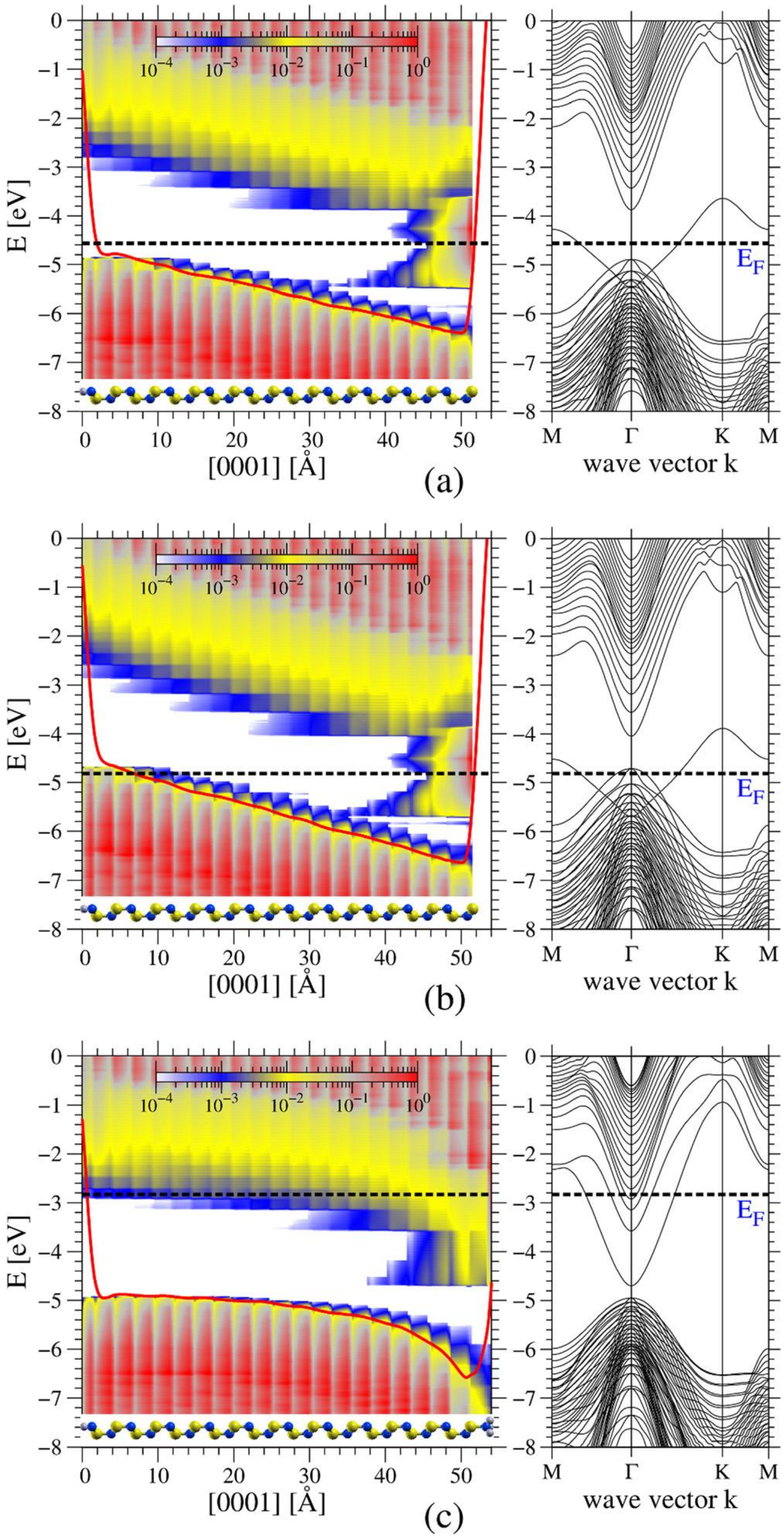

Fig. 2. Left diagram—the electric potential distribution, averaged in the plane perpendicular to c-axis, shown as electron energy (red line), and the density of states projected on the atom quantum states (P-DOS), showing the spatial variation of the valence and conduction bands in the slabs having 20 DALs used for a simulation of a surface donor. The shades are (in arbitrary units): red—above 0.5, grey—0.1 yellow—0.01, blue— 0.001, white below 0.0001. Right diagram—the band diagram of the slab. The diagrams represent: (a) clean GaN(0001) surface with termination by Z = 0.715 hydrogen pseudoatoms ; (b) clean GaN(0001) surface with termination by Z = 0.635 hydrogen pseudoatoms; (c) GaN(0001) surface covered by 1 ML of ammonia with Z = 0.735 hydrogen termination pseudoatoms. The schemes of the GaN slabs are presented beneath the spatial band variation diagrams. Reproduced Fig. 9 from Ref. 65 [65].

The use of hydrogen pseudoatoms at the opposite termination surface was proposed to reduce the effect of broken bond states in both quantum mechanical and electric potential aspects [79,80]. This is related to the presence of the charge located on the surface states and the ensuing emergence of an electric field at the surface [4, 5, 65]. The simulations recover such fields in a finite thickness layer; therefore, the obtained energies of the states within the slab are affected by their position. This effect is illustrated in Fig. 1, copied from Ref. 68. where this effect is discussed.

Similar results were recently obtained by Yoo et al. in simulations of ZnO slab supercell calculations [81]. The natural consequence of the field presence is a skewed band diagram in which the projection of the states, necessary to create the typical band diagram in momentum space, leads to an apparent reduction of the bandgap, as shown in Fig. 2. This effect was not considered in previous publications on semiconductor surfaces. The volume-surface band structure of ideal and relaxed nonpolar GaN surfaces calculated using the local density approximation (LDA) was first presented by Northrup and Neugebauer [82]. In their diagram, the band states are presented as shaded areas, whereas the surface states are presented as a line in the energy- momentum (parallel) coordinate system. This approach, which is necessary to avoid overshadowing of the surface states, was employed in a number of later publications [82-87]. These diagrams are not incorrect; nevertheless, they do not capture the complexity of the semiconductor surfaces.

In fact, the surface states are part of the total band structure of the surface slab, which is composed of bulk and surface states. Examples of such diagrams are presented in Fig. 3, where the band diagrams of a slab presenting clean polar AlN $(0001)$ and AlN$(000\bar{1})$ surfaces are presented [88]. The calculated slabs contained 24 double atomic layers (DAL) of the Al-N slabs, without any coverage on either side. The image presents band diagrams in momentum and real space as well as partial density of states (PDOS).

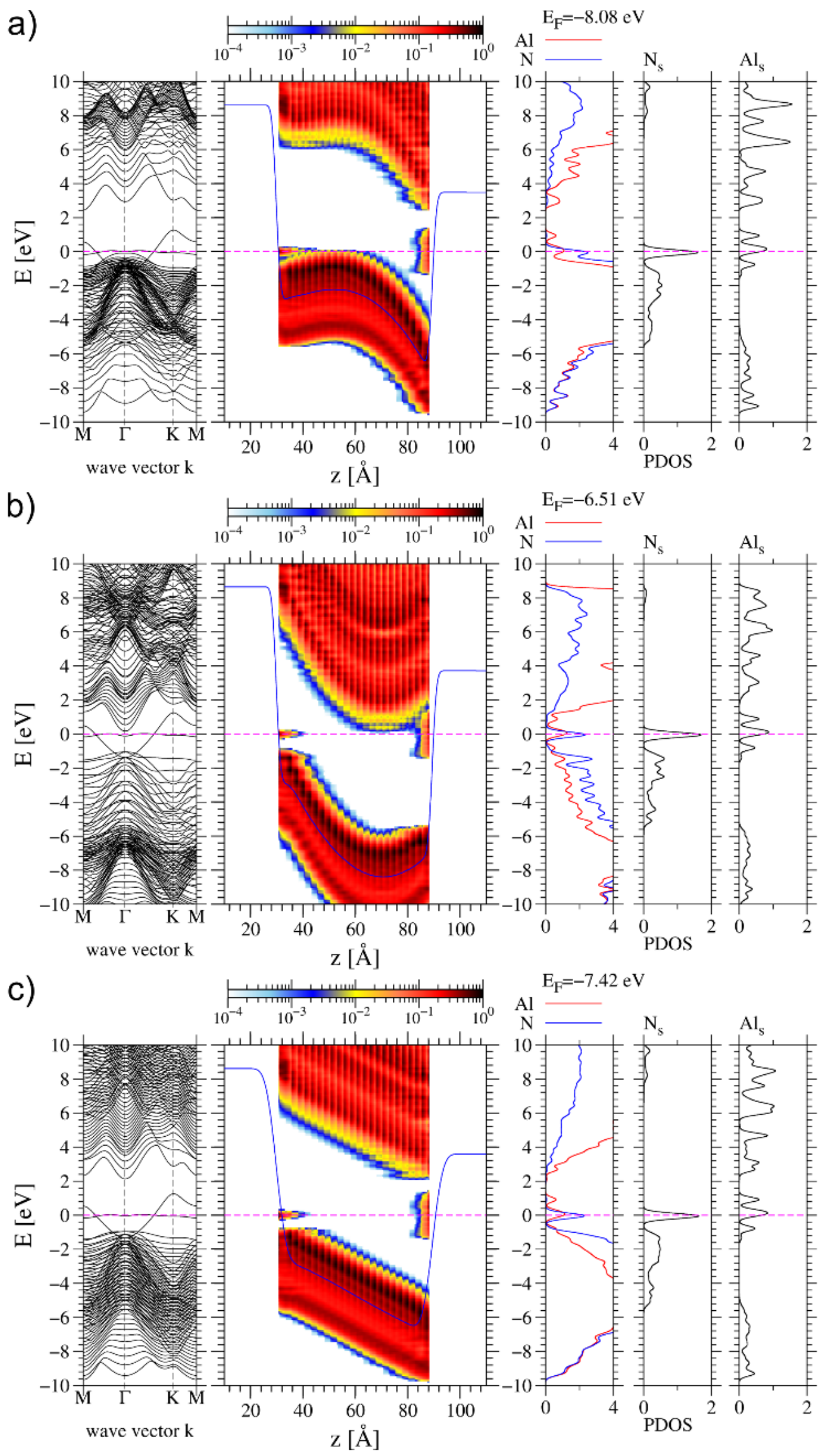

Fig. 3. Band diagram in momentum (left) and position space (middle) of the WZ 24 DALs AlN slabs and the density of states (DOS) of surface Al atoms (right): (a) p-type, (b) n-type, (c) SI. The blue line superimposed on the position space diagram is electric potential, derived as above, plotted in units of electron energy. The colors in the central panel represent electron density according to the scale at the top. Reproduced Fig. 4 from Ref. 88 [88].

As shown, surface states exist on both surfaces. The states associated with the Al-terminated surface, located close to the conduction band minimum (CBM), have a large dispersion of approximately 2 eV. This is related to the overlap of the Al states between the neighboring top Al atoms. The states associated with the N-terminated surface have reduced dispersion, of order of 0.3 eV. They are located close to the valence band maximum (VBM). Their dispersion is smaller, related to the small extension of the nitrogen wavefunctions and consequently much smaller overlap between neighbor functions. These three diagrams represent p-type, n-type, and semi-insulating (SI) AlN bulk, which is reflected by the position of the Fermi level inside the slab, close to the conduction band, to the valence band, and to the midgap, respectively. The latter diagram is specific, associated with the absence of bulk charge, leading to a linear dependence of the electric potential. The other two diagrams have strongly nonlinear potential profiles, confirming the presence of charge in the bulk [88]. In addition, the field is present at both edges of the slab, leading to the energy shift of the surface and band states, which was identified as the surface states Stark effect (SSSE) [78,89,90]. This identification is correct because the surface state energies are affected by the electric field present in the slab.

The presence of a field inside the slab is closely related to the partial occupation of the surface states. In fact, these surface states are broken bond states, that is, states without overlap with the states of the neighboring atoms as they are missing. In essence, they originate from the valence band, and their higher energy is due to the absence of an overlap. Thus, their energies are typically within the bandgap. Examples of such states are the surface states of the AlN(0001) polar slabs, as presented in Fig. 3. These states are partially filled, so the Fermi energy has to be there; hence, they pin the Fermi level.

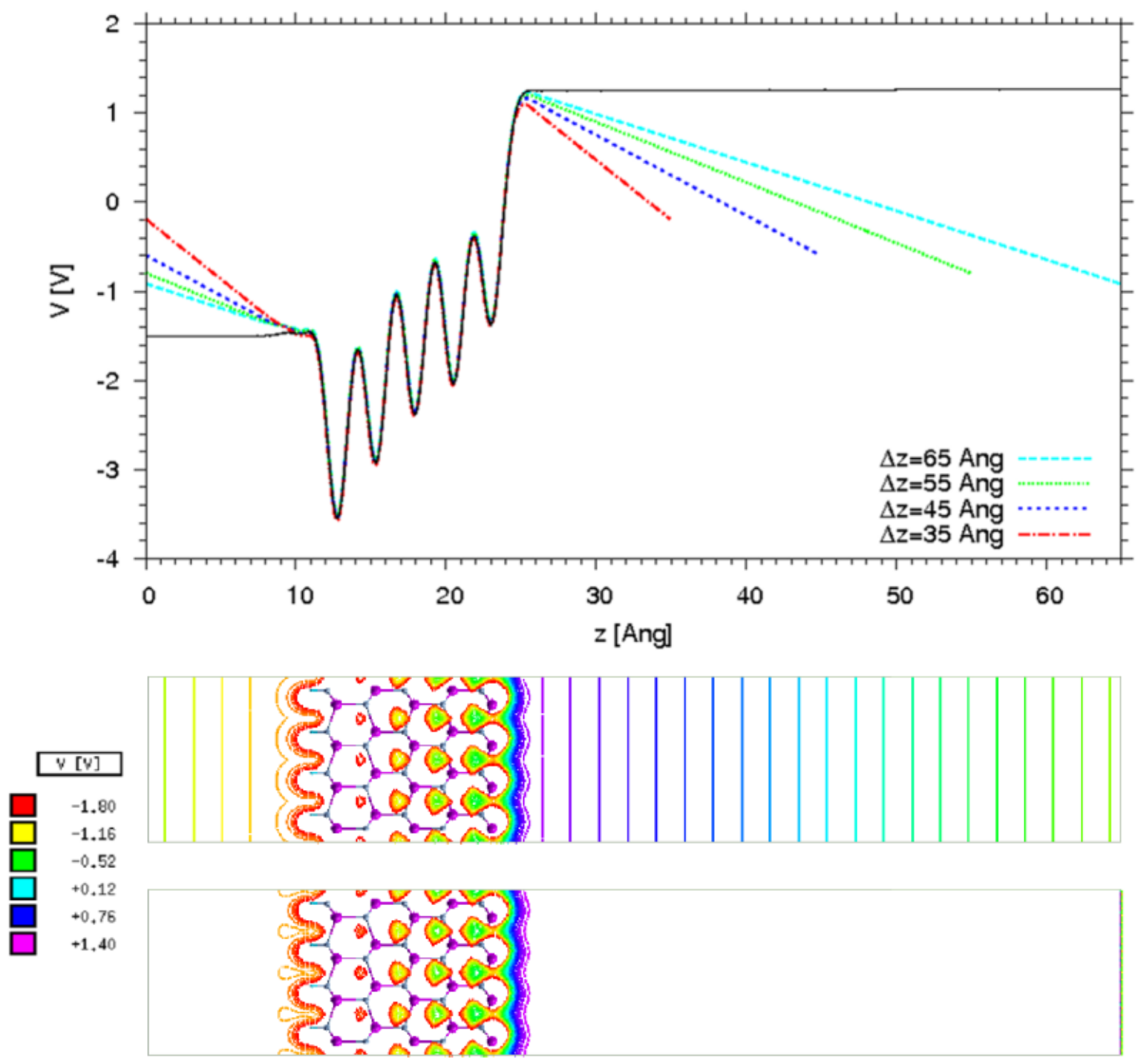

Fig. 4. Electric potential distribution (SIESTA) for different thickness of the spacer, obtained with and without Laplace correction. Top diagram presents the potential profiles along the channeling path. The black line correspond to the solution with, the color lines - without Laplace correction. Lower two diagrams show the potential sections across the slab with (upper) and without (lower) gradient correction, drawn for the largest width of the spacer. Reproduced Fig. 4 from Ref. 34 [53].

In general, the presence of the Fermi level pinned at both sides of the slab stabilizes the potential difference with respect to the applied external field. This result, plotted in Fig. 4, was obtained by Krukowski et al. [53]. The potential difference may be manipulated by a change in the electron charge in the termination hydrogen pseudoatoms or by the distance between these atoms and the bottommost slab atoms. This led to a difference in the field within the slab, as shown in Fig. 5.

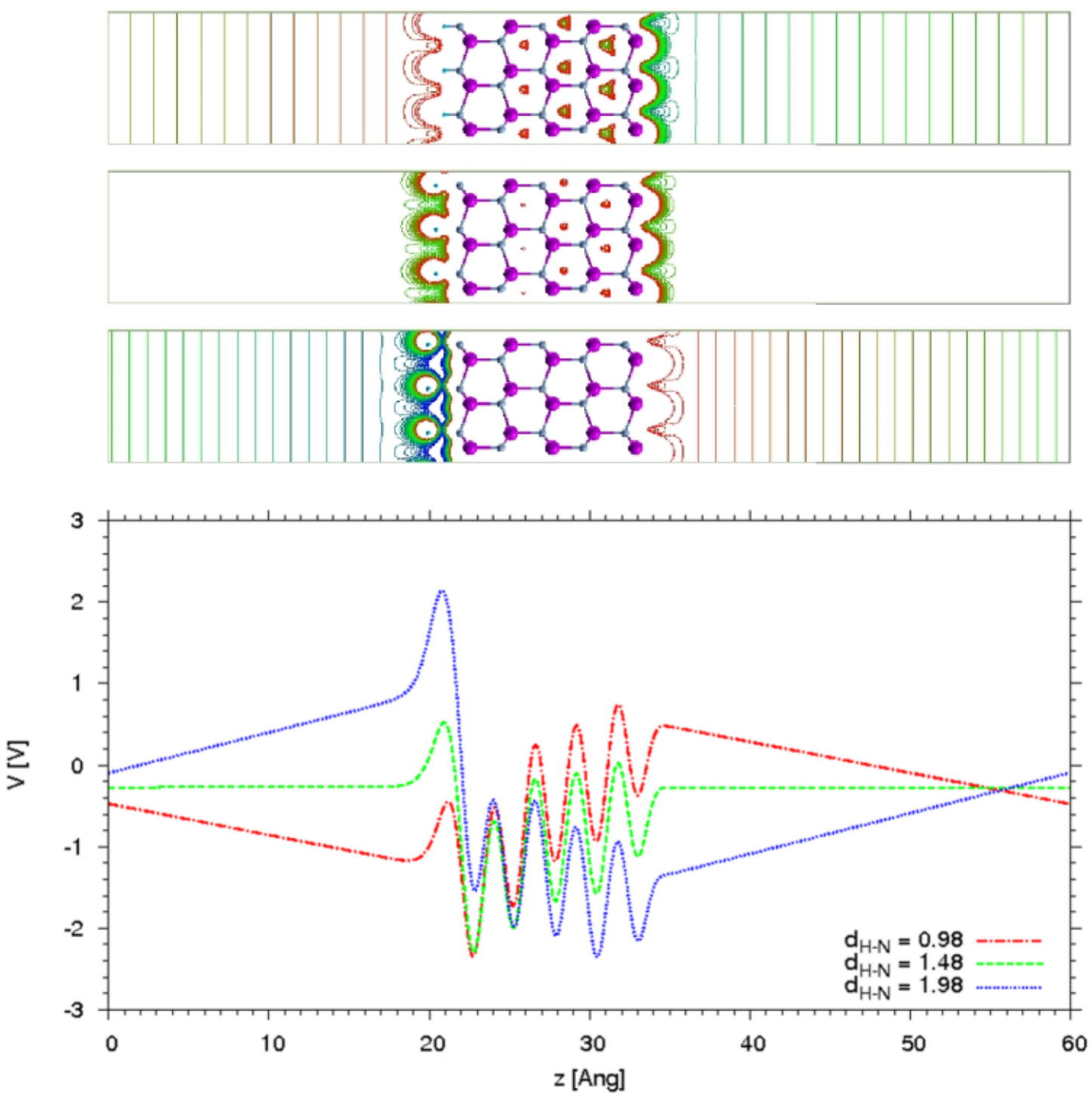

Fig. 5. Electric potential distribution in 3x3 Ga-N slab, fulfilling PBC, obtained for three different distances between N bottom atoms and H saturation atoms in SIESTA: a) 0.98Å, b) 1.48 Å and c) 1.98 Å. Top- the sections across the slab, bottom - the potential profiles along the channeling path. Reproduced Fig. 2 from Ref. 53 [53].

The potential profiles presented in Figs. 4 and 5 are affected by the presence of atomic cores, resulting in high amplitude fluctuations. These fluctuations were removed by the averaging method described in Ref. 91 [91]. This method allows one to obtain a long-range variation of the electric potential, which is presented as the electron energy in Fig. 3 (i.e. multiplied by electron charge). This also proves that the band variation in real space strictly follows the change in the electric potential, confirming the identification of the SSSE.

At both slab edges, the diagram shows a steep change in the electric potential of the order of several volts. This is evidence that the external dipole of the surface contributes to the work function of the semiconductor, which was first identified by Lang and Kohn [9,10]. This dipole may be affected by the adsorption of species at the surface, which is sometimes related to charge transfer and the creation of additional dipoles. Both the dipoles contribute to the overall dipole of the slab, as shown in Fig. 3.

It has to be added, that in the investigations of the adsorption processes, additional spurious effects may arise which can affect adversely the precision of the obtained results. The separation of the quantum system into two isolated subsystems may lead to different positions of the Fermi levels in both parts. This is a pure quantum effect; in fact, it is a combination of the quantum state energy and the Fermi-Dirac statistics. Formally, overlap always exists because the wavefunction asymptotic behavior at far distance is always exponential. Thus, the two subsystems have a coherent wavefunction at any separation. Nevertheless, such weak mathematical cohesion is destroyed by thermal fluctuations, because the systems are at some nonzero temperature. Therefore, in reality, remote subsystems are essentially independent, possibly having different Fermi energies. Standard DFT procedures use a single common Fermi level for the entire system. This may entail an artificial redistribution of the charge between the subsystems. An example could be the slab and separate adsorbate. This charge effect is not due to the tunneling of electrons; in fact, it is mere reshuffling of the electrons during occupation of the states by numerical procedure. Nevertheless, this may create a significant physical effect, such as an electric interaction between the charged species and slab. This is illustrated in the case of Ga atoms near the GaN slab, as shown in Fig. 6.

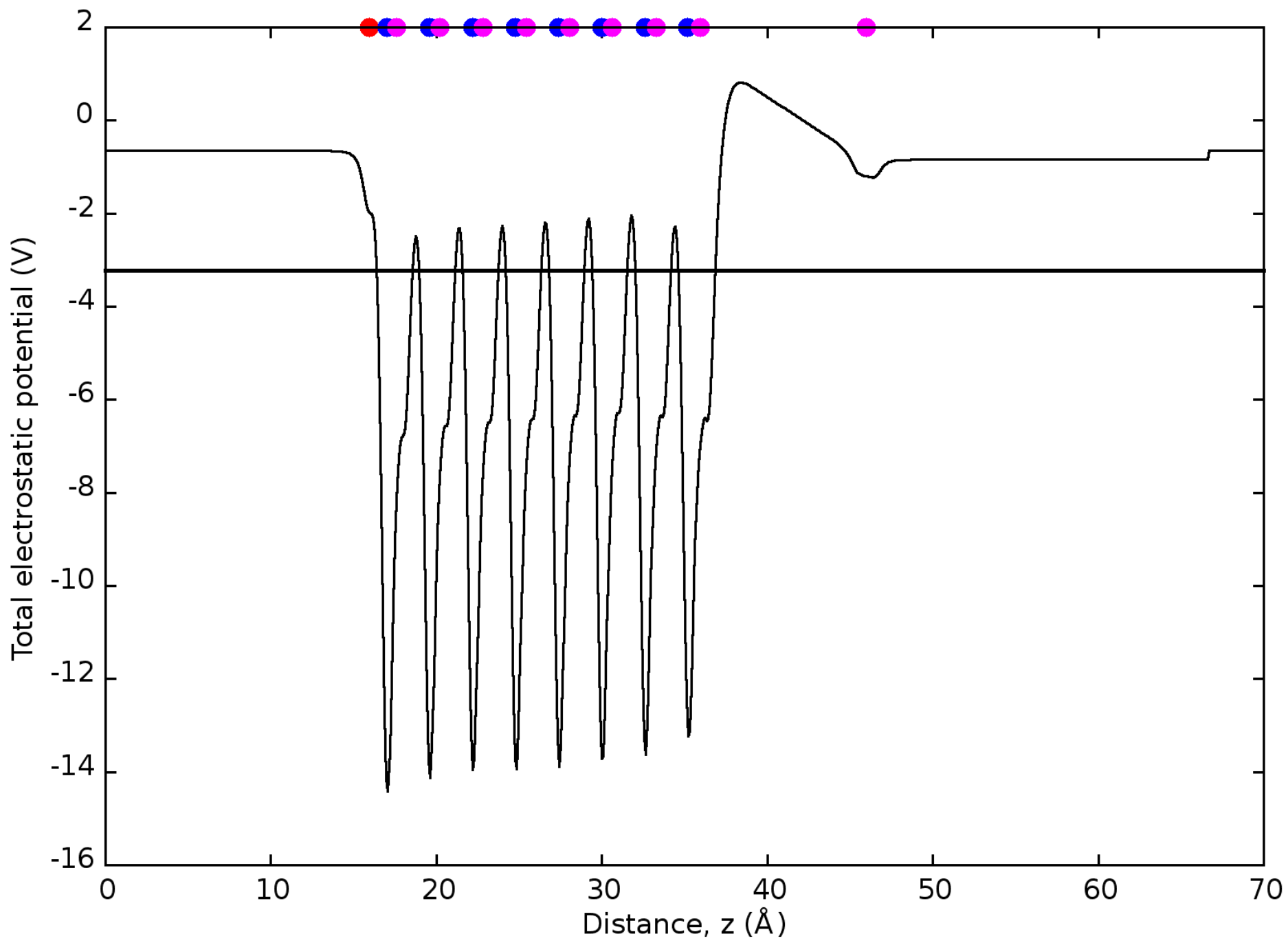

Fig. 6. Electric potential profile averaged in a plane perpendicular to the c-axis, plotted across the slab representing a bare GaN(0001) surface and d Ga atom at the distance of 10 Å from the surface. Positions of atoms are denoted by circles (blue – Ga, magenta – N, red – H) on top of the diagram. Reproduced Fig. 4. from Ref. 92 [92].

This effect may be beneficial for simulations of the interaction of plasma-charged species: atoms and molecules. On the contrary, the charge redistribution may affect the calculations of the adsorption energy using the standard procedure of translation of the species away from the surface and determination of the total system energy as a function of the distance. It is assumed that above the critical distance, at which the overlap of the wavefunctions of the slab and adsorbate is negligible, the energy attains a constant asymptotic value. In the case of the charged species, this is not observed because the Coulomb interaction is long ranged. This is demonstrated in Fig. 7, where the energy of the GaN(0001) surface – Ga system is plotted [92]. The total energy varies considerably at large distances; therefore, precise determination of the adsorption using this procedure is not possible. The other procedure, based on separate calculations of the species and slab, is evidently superior.

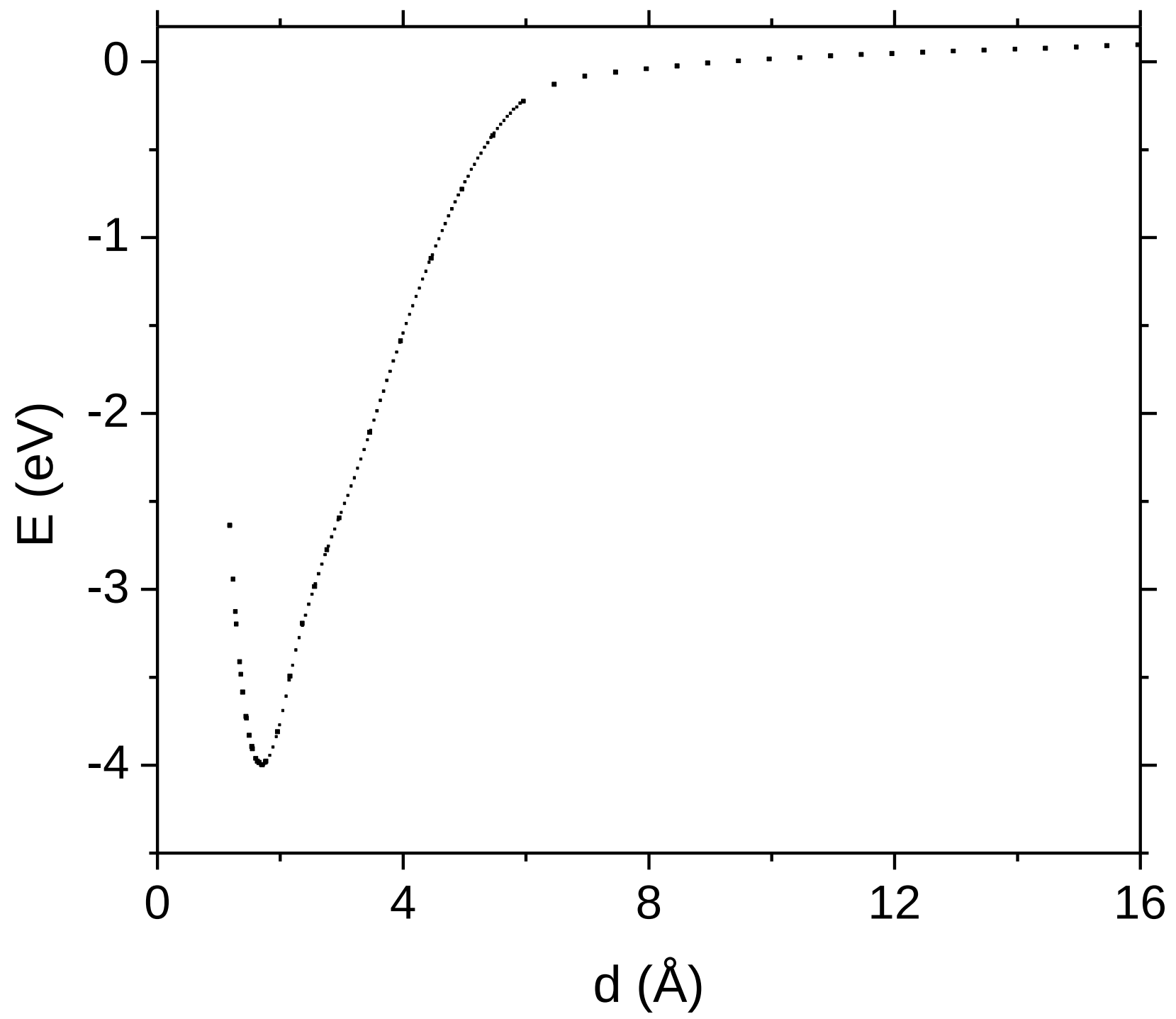

Fig. 7. Energy E of the system as a function of the distance d between Ga adatom and the bare GaN(0001) surface. Zero energy level is set for a large distance from the surface. The distance is measured as a difference between the position of the Ga adatom and a position of the topmost layer of Ga surface atoms. Reproduced Fig. 3 from Ref. 92 [92].

In the critical discussion of the slab representation of semiconductor surfaces, it is necessary to reiterate that the separation of the two quantum subsystems cannot be complete. Therefore, a residual overlap is always present. This is visible in the overlap of the wavefunctions of the real and terminated (opposite) surface atoms of the slab, as shown in Fig. 8. It follows from these data that the far distance decay of the wavefunction is exponential and always universal, as it is the solution of the linear equation. Therefore, the overlap between the two surfaces does not vanish but it is reduced in an exponential asymptotic manner. The statement claiming that the interaction between these two surfaces is not observed are not precise. The overlap influence is exponentially reduced, therefore no critical slab width can be identified. This indicates that the requirement for the number of atomic layers in the slab simulations was severe and required a large number of atomic layers. It increases the considered system to several hundreds of atoms, which is particularly unfortunate, as high precision calculations employing Heyd-Scuseria-Ernzerhof (HSE) are not possible for such a large number of atoms [48]. Other approximations, such as PBE or PBESol, are much less precise in the determination of the energy of quantum states, leading to a bandgap error of the order of 30% [40,42]. However, they recover mechanical properties well, so they are useful in the determination of surface structures, for example, surface reconstruction. The other possible choice is to use the band correction scheme of Ferreira et al., known as the GGA-1/2 approximation, which gives the proper band gap energies, effective masses, and band structures [43-46]. This approach provides incorrect mechanical properties; therefore, these calculations may be used only in the final determination of the energy of quantum states, without any relaxation of the atomic positions.

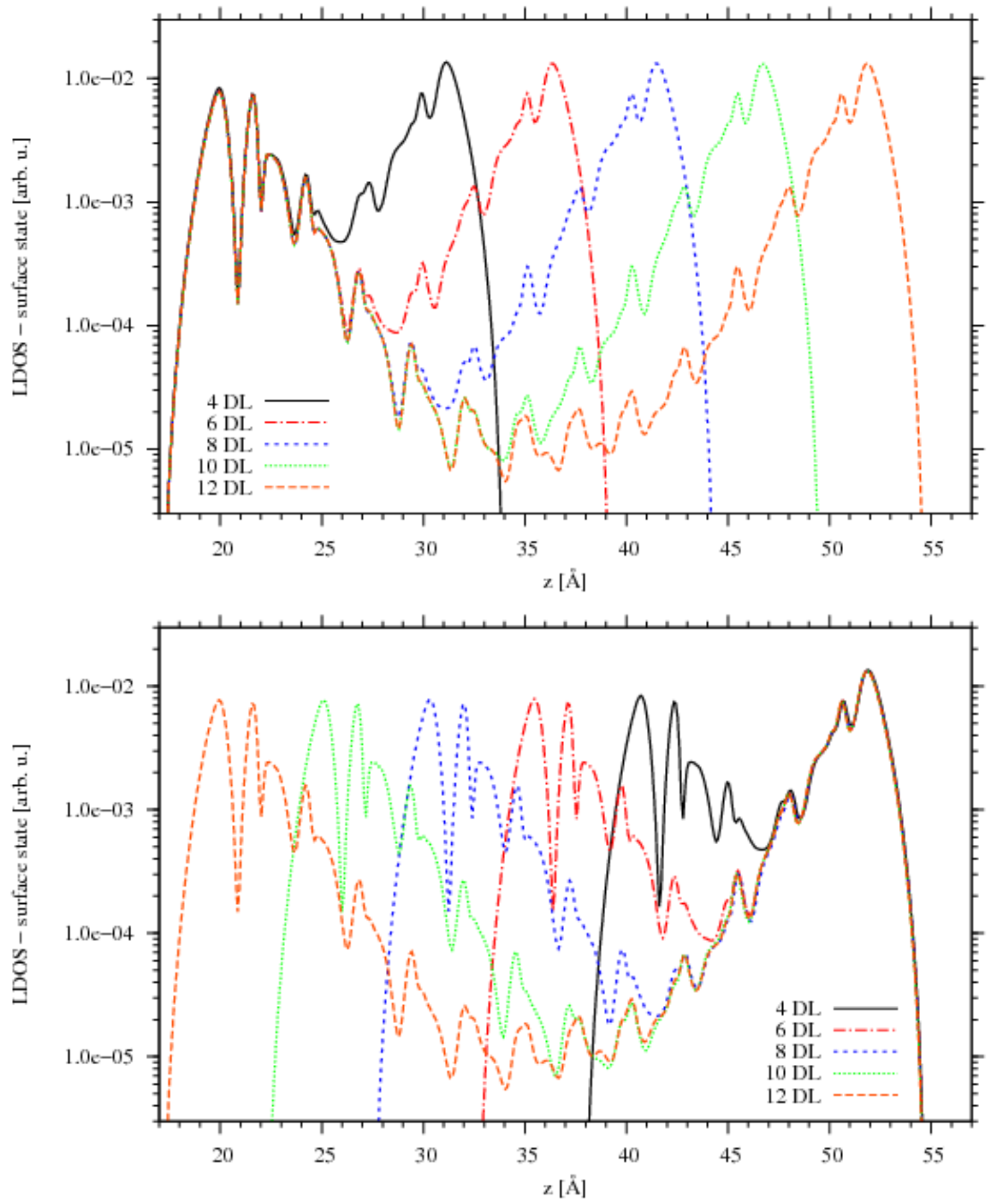

Fig. 8. Electron density profiles of the sum of the surface states related to real and artificial surfaces. The colors denote (1 x 1) GaN (0001) slabs, having four (black), six (red), eight (blue), ten (green) and twelve (yellow) double atomic layers, all terminated by H atoms. The plots correspond to left and right-adjusted diagrams. Reproduced Fig. 4 from Ref. 89 [89].

From the above plotted diagram it follows that the wavefunction overlap of the real and termination surfaces is considerable for slab thicknesses of 4 and 6 DALs. In the cases of 8 and 10 DALs, this was much lower. The extension from 8 to 10 DALs brings a relatively smaller reduction, while the shift to 12 DALs brings only a minor improvement. Thus, the slab for high-precision calculation should have at least eight DALs, whereas 10 or 12 DALs can be used for verification purposes.

These considerations prove that *ab initio* simulations of the properties of semiconductor surfaces still face considerable obstacles to the route towards full precise determination of their properties. The goal is to simulate a relatively thick slab of size in a plane sufficiently large to capture the basic symmetry properties of the surfaces. As it will be shown, the size of the simulated slab may severely distort the symmetry of obtained results. The same condition applies to the slab thickness. Here, the quantum independence should be correlated with proper electric potential and charge modelling, which again increases the size of the simulated systems. The precise requirements depend on the properties of the simulated system, as they differ considerably from one case to another. Nevertheless, the standard requirement for simulations of the semiconductor surface properties is several hundreds of atoms. This effectively precludes the application of highly precise calculations such as the HSE approximation. Thus, less precise versions of ab initio calculations, such as PBE and related approximations, are the appropriate choice. However, insight into the properties and processes on semiconductor surfaces is evidently incomplete. Therefore, the optimal approach is to direct research towards understanding the basic processes and phenomena, even at the cost of high precision. In the following sections, we describe the state of the knowledge in several aspects, showing that the basic understanding of the physics and chemistry of semiconductor surfaces requires basic research and formulation of the basic ideas in the first instance.

## III. Bonding in semiconductors – charge in the bulk and at the surface

It is widely accepted that the vast majority of semiconductor crystals, including elemental crystals such as diamond, silicon, or germanium, as well as compound crystals such as GaAs, InSb, AlP, CdSe, and HgTe, are characterized by the cubic symmetry of the lattice (i.e., zinc blende or diamond) and tetrahedral coordination. This is closely related to the hybridized $sp^3$ bonding of both the elemental and compound [94, 94]. In the latter case, covalent bonding is supplemented by a partial ionic contribution. Owing to the symmetry of zinc blende lattice, the ionic contribution is neutralized in the bulk, such that the charge separation effect does not manifest itself in the form of spontaneous polarization [55,56]. The list of elemental and compound semiconductors and their lattice symmetries related to spontaneous polarization are presented in Table 2.

*Table 2. Lattice symmetry of elemental and compound, most frequently applied semiconductors: ZB/D – Zinc blende (compound) or Diamond cubic lattice, W – wurtzite hexagonal lattice, H – hexagonal lattice, RH – rhombohedral lattice [66-69,95-101].*

| Type | Compound | Lattice symmetry | Lattice parameters (Å) | Bandgap(eV) |
|---|---|---|---|---|
| IV | C (diamond) | D | $a = 3.567$ | $E_g = 5.47$ (I) |
| | Si | D | $a = 5.431$ | $E_g = 1.12$ (I) |
| | Ge | D | $a = 5.658$ | $E_g = 0.66$ (I) |
| | Sn (gray tin) | D | $a = 6.489$ | $E_g \approx 0.00$ |
| IV-IV | SiC (3C) | ZB | $a = 4.3596$ | $E_g = 2.36$ (I) |
| | SiC (2H) | W | $a = 3.08$<br>$c = 5.05$ | $E_g = 3.3$ (I) |
| | SiC (4H) | H | $a = 3.073$<br>$c = 10.053$ | $E_g = 3.26$ (I) |
| | SiC (6H) | H | $a = 3.081$<br>$c = 15.12$ | $E_g = 3.02$ (I) |
| | SiC | RH | $a = 3.08$<br>$c = 37.7$ | $E_g = 2.4$ (I) |
| III-V | BN | W | $a = 2.55$<br>$c = 4.227$ | $E_g = 6.4$ (D) |
| | AlN | W | $a = 3.111$<br>$c = 4.981$ | $E_g = 6.0$ (D) |
| | GaN | W | $a = 3.1890$<br>$c = 5.1864$ | $E_g = 3.47$ (D) |
| | InN | W | $a = 3.5705$<br>$c = 5.703$ | $E_g = 0.7$ (D) |
| | BP | ZB | $a = 4.538$ | $E_g = 2.0$ (I) |
| | AlP | ZB | $a = 5.451$ | $E_g = 2.45$ (I) |
| | GaP | ZB | $a = 5.4505$ | $E_g = 2.26$ (I) |
| | InP | ZB | $a = 5.8687$ | $E_g = 1.34$ (D) |
| | BAs | ZB | $a = 4.78$ | $E_g = 1.46$ (I) |
| | AlAs | ZB | $a = 5.6611$ | $E_g = 2.16$ (I) |
| | GaAs | ZB | $a = 5.6533$ | $E_g = 1.515$ (D) |
| | InAs | ZB | $a = 6.0583$ | $E_g = 0.36$ (D) |
| | BSb | ZB | $a = 5.22$ | $E_g = 1.24$ (I) |
| | AlSb | ZB | $a = 6.1355$ | $E_g = 1.62$ (I) |
| | GaSb | ZB | $a = 6.0959$ | $E_g = 0.726$ (D) |
| | InSb | ZB | $a = 6.4794$ | $E_g = 0.17$ (D) |
| II-VI | ZnO | W | $a = 3.25$<br>$c = 5.20$ | $E_g = 3.37$ (D) |
| | CdS | W | $a = 4.14$<br>$c = 6.72$ | $E_g = 2.42$ (D) |
| | CdSe | W | $a = 4.30$<br>$c = 7.01$ | $E_g = 1.74$ (D) |
| | CdSe | ZB | $a = 6.05$ | $E_g = 1.74$ (D) |
| | CdTe | ZB | $a = 6.48$ | $E_g = 1.50$ (D) |
| | HgS | ZB | $a = 5.87$ | $E_g = 0.2$ (I) |
| | HgSe | ZB | $a = 6.4794$ | $E_g \approx 0.00$ |
| | HgTe | ZB | $a = 6.4794$ | $E_g = -0.3$ |

*Crystals with cubic lattice symmetry (ZB/D) are spontaneous polarization – free, those of hexagonal (W, H) and rhombohedral (RH) lattice have nonzero spontaneous polarization. D and I denote direct and indirect bandgap, respectively.*

These data can be used as a reference for *ab initio* simulations during the parameterization of energy functionals.

In fact, the ionic contribution is directly proportional to the energy difference of the $sp^3$hybridized states [93]. It is absent in the case of elemental semiconductors and relatively small in the majority of II-VI and III-V compounds. This was directly related to cubic lattice structure of the crystals. In the case of the group III metal nitrides, such as BN, AlN, GaN, and InN, the ionic part is considerably larger. As a result, the lattice structure is changed to hexagonal, tetrahedrally coordinated wurtzite. In this structure, spontaneous polarization is permitted in both pure compounds and mixed crystals [55,56]. These effects are accompanied by different bonds in these semiconductors. Despite the tetrahedral structure, GaN bonding includes hybridized $sp^3$ orbitals of gallium and separated *s* and *p* orbitals of nitrogen. In addition, the *d* orbitals of Ga were involved. Therefore, the bonding overlap includes separate overlaps of gallium *d* and nitrogen *s* orbitals. The second bonding involves hybridized $sp^3$ orbitals of gallium and *p* orbitals of nitrogen. This creates two separate sub-bands of the valence band (VB), which were measured experimentally using soft X-ray spectroscopic measurements by Magnuson et al. [12]. By *ab initio* calculations, the existence of two sub-bands was confirmed, and the nature of the bonding was identified [12,13]. In the case of AlN, sub-band separation was also identified, with the lower sub-band due to nitrogen s orbitals only [11].

Sub-band identification does not solve problems; in fact, it creates more problems. The nitrogen atom is in a tetrahedral surrounding and is connected to four Ga atoms. However, the bonding in the upper sub-band was created from three nitrogen *p* states. The lower *d*-*s* sub-band is not compatible with tetrahedral coordination. Therefore, four bonds are necessary. This is possible because of the existence of resonant bonds proposed long ago by German chemist F. A. Kekulé [102]. The concept of delocalization of electrons within $sp^2$ bonds was formulated in terms of the probability of occupation of the nonorthogonal states, which is fully compatible with quantum field theory. In addition, π bonds are incorporated in the direction perpendicular to the ring [103]. A slightly different formulation was proposed by P. W. Anderson, which involved different features of quantum mechanics, namely wave theory. This describes the quantum states as resonant, that is, the wave states in full mutual resonance, thus having the same frequency, that is, energy [104]. This formulation was subsequently applied to describe the high-temperature superconductivity of copper oxide structures [105]. Subsequently, this concept was extended to other systems [105-110].

The GaN case was recently considered during the study of N adatoms at the GaN(0001) surface [14]. Accordingly, the quantum states of the N atom in the bulk GaN were obtained and are presented in Fig. 9. As shown in the PDOS diagram for N atoms in the bulk, two basic states are present at approximately $E = -5\ eV$eV and at $E = -9\ eV$. As proved by the COHP data, these subbands are created first by hybridized $sp^3$ orbitals of gallium and second by $p$ orbitals of nitrogen and gallium $d$ and nitrogen $s$ orbitals, respectively.

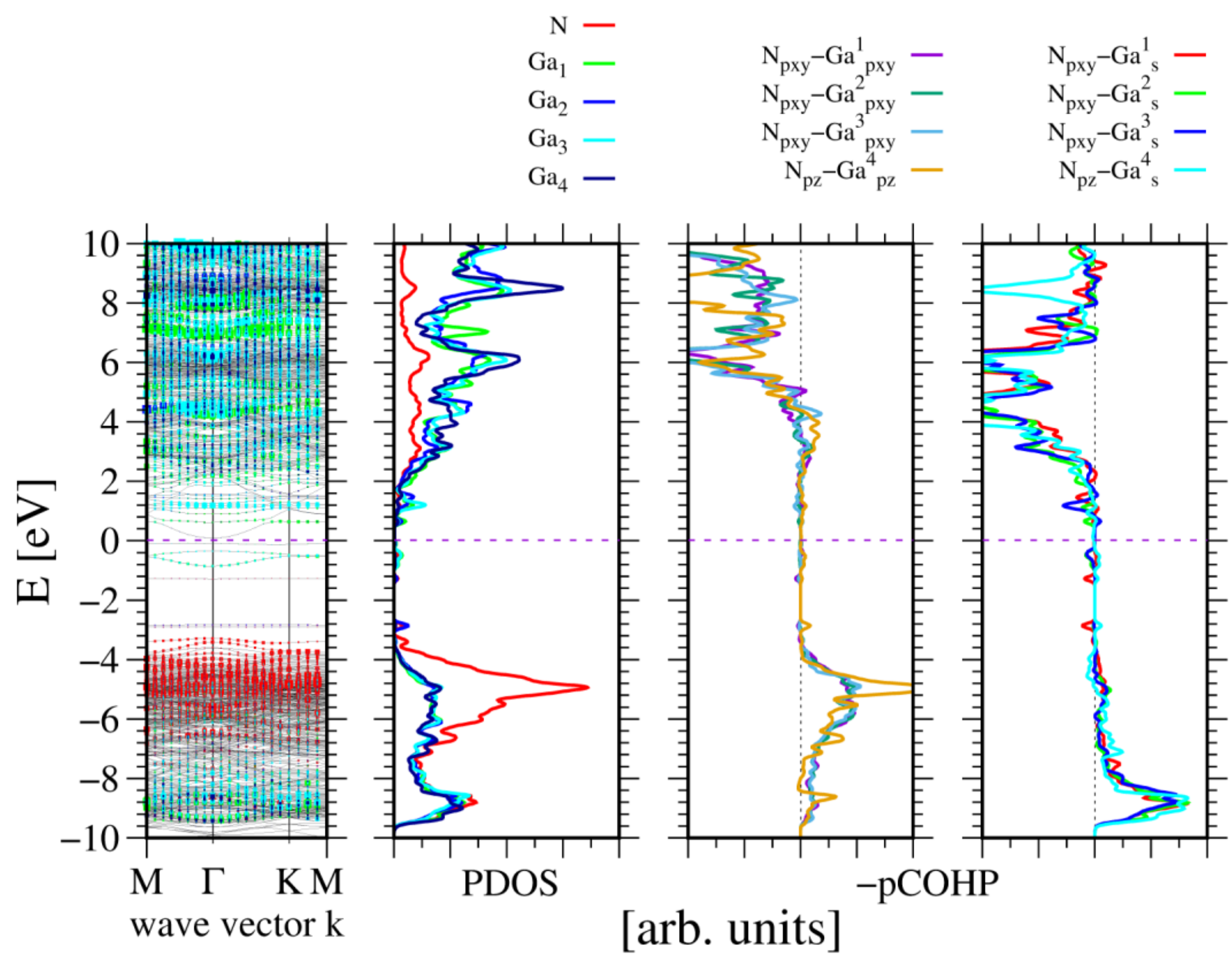

Fig 9. Energies of the quantum states of nitrogen and gallium states located in the GaN bulk of $(2\sqrt{3}\ \times 2\sqrt{3})$ slab representing clean GaN(0001) surface with single N adatom located in H3 site. The panels represents, from the left: energy of the quantum states in the momentum space, projected density of states (PDOS) of the N and Ga atoms, and the right two panel - Crystal Orbital Hamilton Population (COHP) [111,112]. The COHP data correspond to N adatom and the closest topmost Ga atoms. The Fermi energy is set to zero. COHP positive values correspond to the bonding overlap.

The COHP data indicated that the upper state was due to the four states created by the overlap of $|N_{2p}\rangle$ states and $|Ga_{4sp^3}\rangle$ of the four atoms. Thus, this number is four, whereas they are created from three $|N_{2p}\rangle$ states. Thus, the number is extended to four because these states are resonant, as in the case of benzene [102,103]. Naturally, these states must be nonorthogonal; they are occupied with a probability $P = 3/4$. Because the number is four, it is sufficient to create a tetrahedral bonding pattern in the wurtzite lattice symmetry. This was confirmed by the angles created by these bonds, which were in the plane $\varphi = 110.14°$ or $\varphi = 109.73°$, that is, close to the ideal tetrahedron. The c-axis angles are slightly different: $\varphi = 106.92°$ and $\varphi = 109.32°$. It is worth to underline that these states are located deep below Fermi energy, their occupation probability stems from electron distribution among large number of states. The number of electrons in these resonant states is six; thus, the occupation probability $P = 3/4$. These states have important consequences for charge analysis of surface states.

## IV. Work function – external dipole layer

Wide applications of electron vacuum devices require the creation of cold cathodes, which are effective electron sources. For some advanced applications, it is necessary to efficiently emit electrons with approximately equal kinetic energy. Thus, cold emission is required, as it is the optimal method. Therefore, the electron affinity (EA), that is, the energy difference between the conduction band minimum and the vacuum, should be possibly low, or even negative [3-5,7,8]. The two other EA-related quantities, the work function (WF) and the ionization energy (IE), are defined using the energy difference between the Fermi level and the valence band maximum, respectively [3-5,7-8]. In the case of metals, these values are identical, or more precisely, the two could not be defined properly, so the common value defined by the work function was adopted [7,8]. In fact, EA and IE are independent of band bending caused by the dipole subsurface layer discussed above. This is because the dipole shifts the energy of the quantum states with respect to the energy of the surface states. The latter remains unchanged with respect to the vacuum level, that is, the far-distance energy value. However, these values are derived from the measurements of the work function, which is affected by the Fermi level pinning by the surface states, adsorbate, etc. [113,114]. This causes problems both experimentally and theoretically; thus, the ionization energy of AlN was disputed [113-118]. The absolute values of these three quantities (EA, IE, and WF) were derived from emission experiments, which may be affected by the number of factors [113]. Their absolute values, were finally determined to be $EA[AlN(0001)] = 1.9\ eV$ [85] and $3.2\ eV \leq EA[GaN(0001)] \leq 3.4\ eV$ [88]. The *ab initio* calculations gave for GaN $EA^{DFT}[GaN(0001)] \cong 3.73\ eV$ [49]. More extensive ab initio calculation give smaller value for different surface coverage $EA^{DFT}[GaN(0001)] \cong 2.93\ eV$ [88].

In fact, ab initio calculations can provide a more precise assessment, which is limited by the calculation precision. This is illustrated by the calculation test for pure Al, which is presented in Fig. 10. The diagram may seem superfluous, but it is nevertheless useful for presenting the calculation procedure. The difference in the potential profile at infinity and at the interface is obtained, that is, the electrostatic contribution to work-function. In the case of pure Al, the potential profile is symmetric, with the work-function value $WF[Al] = 3.94\ eV$ .

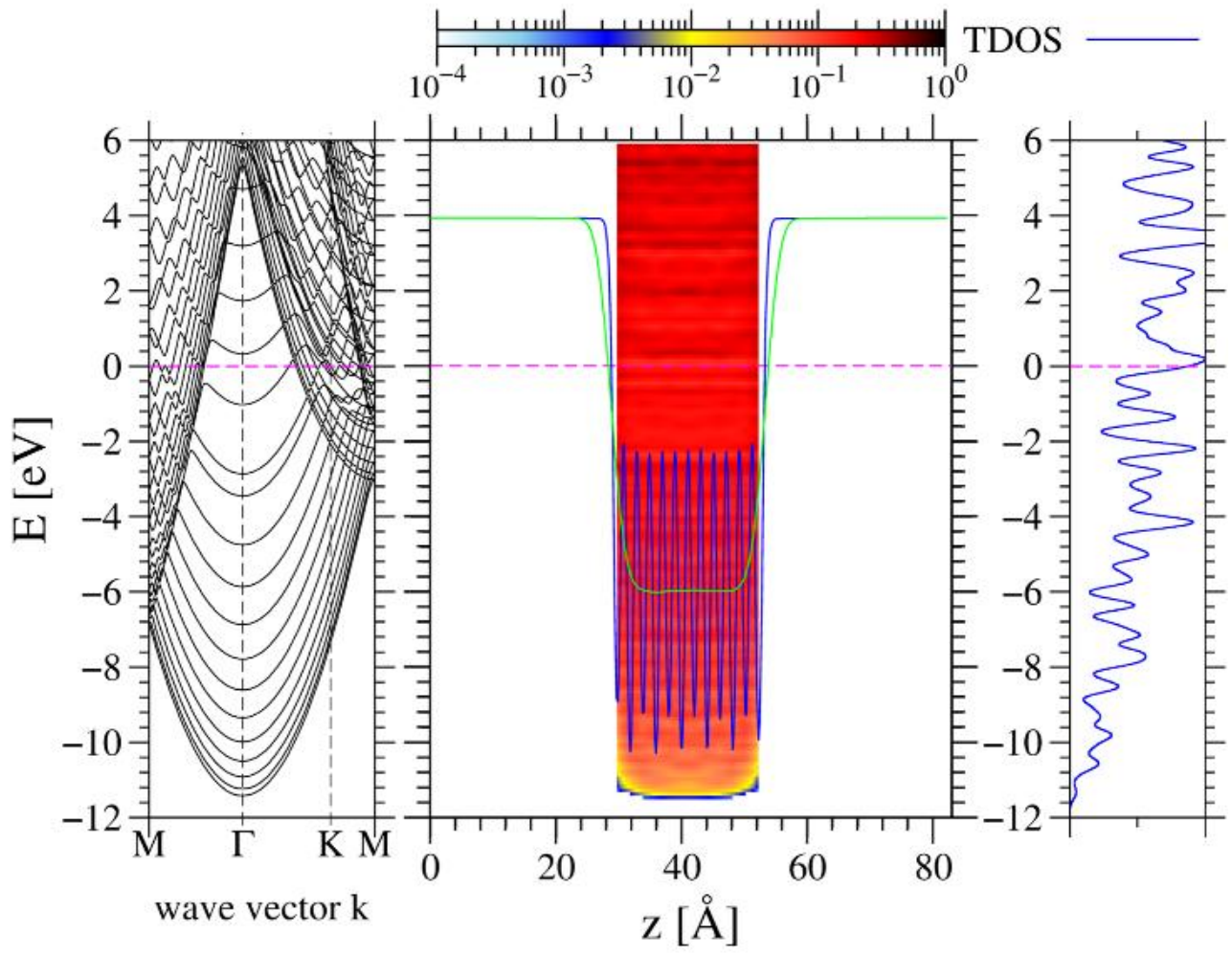

FIG. 10. Band diagram in momentum (right) and position space (middle) of face centered cubic (fcc) Al slab and the total density of states (total DOS) of the slab (right). The blue line superimposed on the position space diagram is electric potential, averaged in the plane parallel to surface of the slab, plotted in the units of electron energy whilst the green line is curve additionally smoothed along c-axis. The colors in the central panel represent electron density according to the scale at the top. Reproduced Fig. 2 from Ref 88 [88]

An important issue is related to the work-function determination procedure. Formally, it is defined as the potential energy of motion of the unit probe charge $q = 1$ from the cluster (or finite size solid) surface to infinity. In fact, this is burdened by the error caused by the Coulomb attraction contribution, because the cluster attains the opposite charge. For conductors, this can be calculated using the mirror charge ($q_m = -1$ ) with the mirror charge located such position that the equipotential surface coincides with the conductor surface. This is equivalent to the external potential completely screened. Similarity between the real conductor surface and its mirror charge equivalent ends. In real conductor interior, there is no field inside and the charges are only at the surface. Thus the Coulomb interaction contribution is larger for smaller distances between the probe and mirror charges. Therefore, the location of the mirror surface is critical. Microscopically, the position of the mirror surface is determined by screening inside the conductor, that is, the density of the mobile charge. For higher mobile charge density the Coulomb error is larger.

Therefore, the standard definition is burdened by the Coulomb interaction error of the probe unit charge. This can be remedied by redefining the procedure. The unit charge must be replaced by an infinitely small charge in the potential energy determination, that is, $q \to \varepsilon \to 0$. Because the Coulomb interaction energy is proportional to both charges, $E_{Coul} \sim q\, q_m \sim \varepsilon^2$ therefore the interaction contribution scaled by the charge, is $V_{Coul} = V_{Coul}/q \;\sim \varepsilon \;\to 0$. Therefore, the Coulomb error in such a defined procedure disappears at this limit, and the measurement is precise.

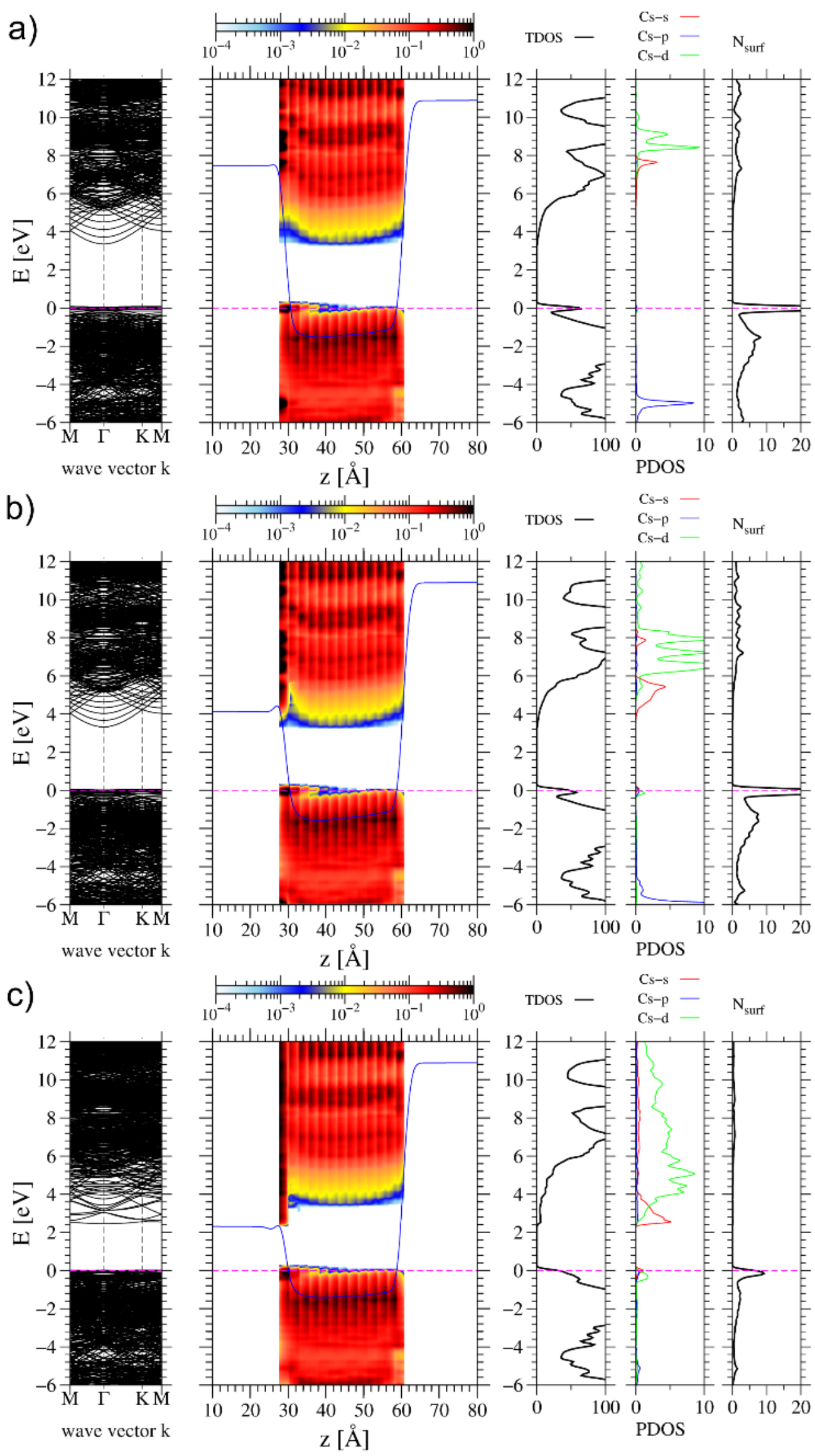

Fig. 11. Band diagram in momentum (left panel) and position space (middle panel) of the following number of Cs atoms: (a) single, (b) four, (c) eight atoms, attached to 4 x 4 slab representing wurtzite N-side GaN(000$\underline{1}$) surface (i.e. 0.0625, 0.25 and 0.5 ML coverage). The panels on the right present density of states (DOS): total, Cs adatom and N surface atoms. The blue line superimposed on the position space diagram is electric potential, derived as above, plotted in units of electron energy. The colors in the central panel represent electron density according to the scale at the top. Reproduced Fig. 8 from Ref. 88 [88].

It should be noted that the potential profiles plotted in Fig. 3 cannot be used directly to determine these three surface characteristics (EA, WF, and IE). These potentials were calculated in the case of a periodic potential along the c-axis; therefore, a far distance value is absent. The potential is corrected using an additional compensation gradient in the vacuum space [52,53] or Laplace correction field [54]. Therefore, the potential level is different on both sides of the slab, as shown in Fig. 3.

The potential profiles obtained are still very useful in different aspects. The adsorption of additional species may affect the surface characteristics (EA, WF, and IE) by additional surface dipoles when the adsorbed species are charged. This was studied for nitride surfaces covered by cesium adsorbed layers [88]. It is well known that the Cs *6s* state has a low binding energy, which is high on the absolute energy scale. In fact, the Cs atom attached to $GaN(0001)$ surface is stripped of the electron and attached by electrostatic forces only [88]. This creates an additional electric dipole opposite to the fundamental dipole of the solid. Therefore, the electric potential barrier is reduced from $9V\ to\ 5\ V$ i.e. by approximately $4\ V$ (Fig. 11). Nevertheless, the charge is compensated by the opposite charge at the surface, thus not changing the potential profile within the slab, as presented. The identified effect, which is in fact the creation of the opposite dipole by the positive charge of electron-stripped Cs atoms, explains the drastic decrease of the work function by Cs coverage of the AlN surface.

The external surface dipole layer and the associated potential jump may be important for adsorption processes on solid surfaces. Solid surfaces in contact with vapor undergo constant bombardment by vapor species. As a result, a fraction of these are attached, whereas the others are repelled. Attachment at the surface requires energy loss during scattering, that is, in a relatively short time. Then, the adsorbate could be immobilized in the potential well at the surface, attaining vibrational energy in accordance with the temperature of the solid, that is, the adsorbate is thermalized [119]. The kinetic energy of the adsorbate at the temperature $T = 1000\ K$ is of the order of $kT = 86.17\ meV$. The benchmark energy of the adsorbate attachment is $\Delta E = 1\ eV$, which is at least one order of magnitude higher. Thus, the parallel motion of the adsorbate is relatively slow, that is, the adsorbate moves almost perpendicularly to the surface. This external dipole contribution is specific to the adsorption of any species on a solid surface.

It is worth noting that the interaction between the adsorbates and solid essentially belongs to the realm of chemical kinetics. This is an enormous subject and includes all types of interactions connected with the complex dynamics of electrons and atoms; therefore, these complex issues will not be discussed here. Shortly, these subjects include atom tunneling across potential barriers and purely quantum effect [120-122]. The classical effects include hot electrons role in chemical reactions [123] Finally, hot-atoms exchange in HD molecules reaction during desorption [124]

Therefore, it is natural to use the scenario of the creation of the "hot" atoms during adsorption to thermalize the adsorbate [125–129]. This induces adsorbate sliding along the surface over a large distance to dissipate kinetic energy. The scenario can work in part, but it does not capture the essence of kinetic energy loss. The duration of the scattering event may be estimated assuming that the extension of the potential well is $d = 0.5\ nm$ and the velocity $v_{th} = \sqrt{\frac{\Delta E}{2M}} \approx 450\ m/s$, thus the collision time is $\tau_{coll} \approx 10^{-12}\ s$ [119]. Hence, the duration of the entire collision process is no longer than the typical frequency of Brillouin zone boundary phonons in GaN, which is of the order of 15–20 THz [130,130]. Thus, the collision time is on the order of the single-phonon frequency. The emission of phonon bunches has also been proposed as a thermalization channel [131]. Thermalization requires the emission of a large number of phonons in this period; thus, nonlinear effects would result in the creation of surface defects.

Although the bombardment s intense, numerous surface defects are not observed. Therefore, a faster and more effective mechanism should be responsible for this effect. This channel should involve electronic degrees of freedom because only these processes are fast enough to be effective in a short time scale. Therefore, an electron tunneling mechanism has been proposed [119]. The existence of a strong electric field induced by the dipole layer causes an electron to jump from the approaching atoms into the solid interior. The estimated time of the jump is $\tau_{jump} \leq 10^{-14}\ s$ therefore, it is sufficiently fast to be completed within the collision time. Subsequently, the high electron remains in the solid interior, which can dissipate the excess kinetic energy. This process may last longer but is essential for the course of the collision process because of the retardation of the positively charged adsorbate. This is due to the interaction with the dipole layer electric field, which acts repulsively owing to the opposite positive charge of the adsorbate. A scheme of the electric potential is shown in Fig. 12.

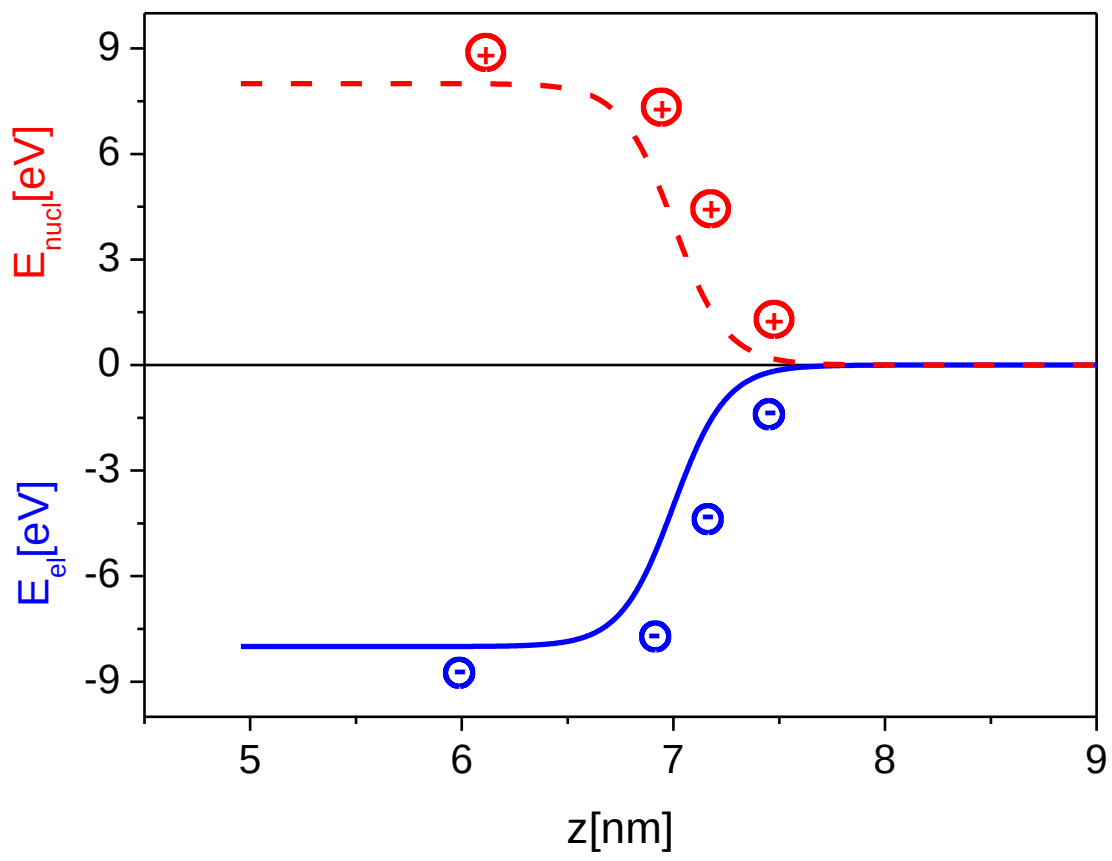

Fig. 12. The potential energy of the electron (blue solid line) and the positive ion (red dashed line) derived from AlN slab by *ab initio* method modeling. Reproduced Fig. 5 from Ref 191 [119].

It should be noted that these are preliminary investigations. The proposed scenario assumes that the Born-Oppenheimer approximation, which is customarily used to describe adsorption processes, should be abandoned. In this approximation, the kinetic energy of the atomic motion is neglected, and the energy is calculated using immobile atoms. This is equivalent to the incorporation of an additional, imaginary external force that stops the adsorbate at each point along the adsorption path. In reality, this is not the case because the adsorbate attains kinetic energy. The dissipation by the proposed electrostatic repulsion involves the potential collectively, such that the force is exerted on a large number of surface atoms directly and also via collective interaction with the electron charge. This new scenario needs to be further investigated as it could be helpful in understanding the number of properties. This inevitably involves a surface dipole layer and the associated electric potential.

## V. Subsurface charged dipole layer formation

In striking contrast to the external dipole layer influence discussed above, subsurface dipole layer effects have been well described for a long time [3–5]. The layer emerges as a consequence of pinning of the Fermi level by the fractionally occupied surface states. The band bending necessary to attain uniformity of the Fermi level over the entire conductive solid is related to the accumulation of the mobile charge, that is, screening. This led to the creation of a subsurface dipole. In contrast to the external dipole previously described by Lang and Kohn, the subsurface dipole is a partially classical effect. This effect is described by the screening length determined from the Lindhard theory of charged gas screening [132]. In the long-distance limit, this result is compatible with the Thomas-Fermi screening length $\lambda_{TF}$ for degenerate electron gas [3]:

$$\lambda_{TF} = \sqrt{\frac{2\varepsilon\varepsilon_o E_F}{3ne^2}} \tag{1a}$$

In the case of a nondegenerate electron gas, the Debye-Hückel screening length $\lambda_{DH}$ [3]:

$$\lambda_{DH} = \sqrt{\frac{\varepsilon\varepsilon_o kT}{ne^2}} \tag{1b}$$

where $\varepsilon$ is the dielectric constant which for the case of GaN is: $\varepsilon = 10.28$. Other quantities are k – Boltzmann constant, $\varepsilon_o$ is the permittivity of vacuum, $e$ is the elementary charge, $n$ – charge density, and $T$ – temperature in Kelvins. The screening length depends on the temperature and charge density. This value may be evaluated for typical conditions: for $T = 300\,K$ and typical electron density for n-type GaN, i.e. $n = 10^{18} cm^{-3}$ and $n = 10^{19} cm^{-3}$ these lengths are: $\lambda_{DH} = 3.83\,nm$ and $\lambda_{DH} = 1.21\,nm$, respectively. Thus, these conditions can be recovered using *ab initio* simulation. Such simulations were performed for a GaN(0001) surface by Kempisty and Krukowski [133]. In these simulations, two different cases of GaN(0001) surfaces were considered: bare and under full H – coverage. Both p- and n-type doping were considered. Doping was performed by adding a uniform charge background over the part of the volume occupied by the cores of the functional basis set in SIESTA; thus, the total charge added is equivalent to a single electron. For a positive charge background, an additional negative charge must be added to the occupation numbers, which is equivalent to n-type doping. In the case of a negative charge, the number of electrons occupying quantum states is smaller by one, which corresponds to p–type doping. Therefore, the doping level may be estimated assuming single charge for 20 elementary cells, where the cell volume for GaN is $V_{GaN} = 46.0$ Å$^3$. The calculations were performed for the 20 GaN DAL slab; therefore, the doping level was $n = p \cong 10^{21}\,cm^{-3}$. In order to achieve the convergence, the electron temperature was set to $T = 1000\,K$. From the estimate it was that $\lambda_{DH}(10^{21}\,cm^{-3}, 1000\,K) =$ 2.2 Å$^3$. The simulation results for a clean GaN(0001) surface are shown in Fig. 13.

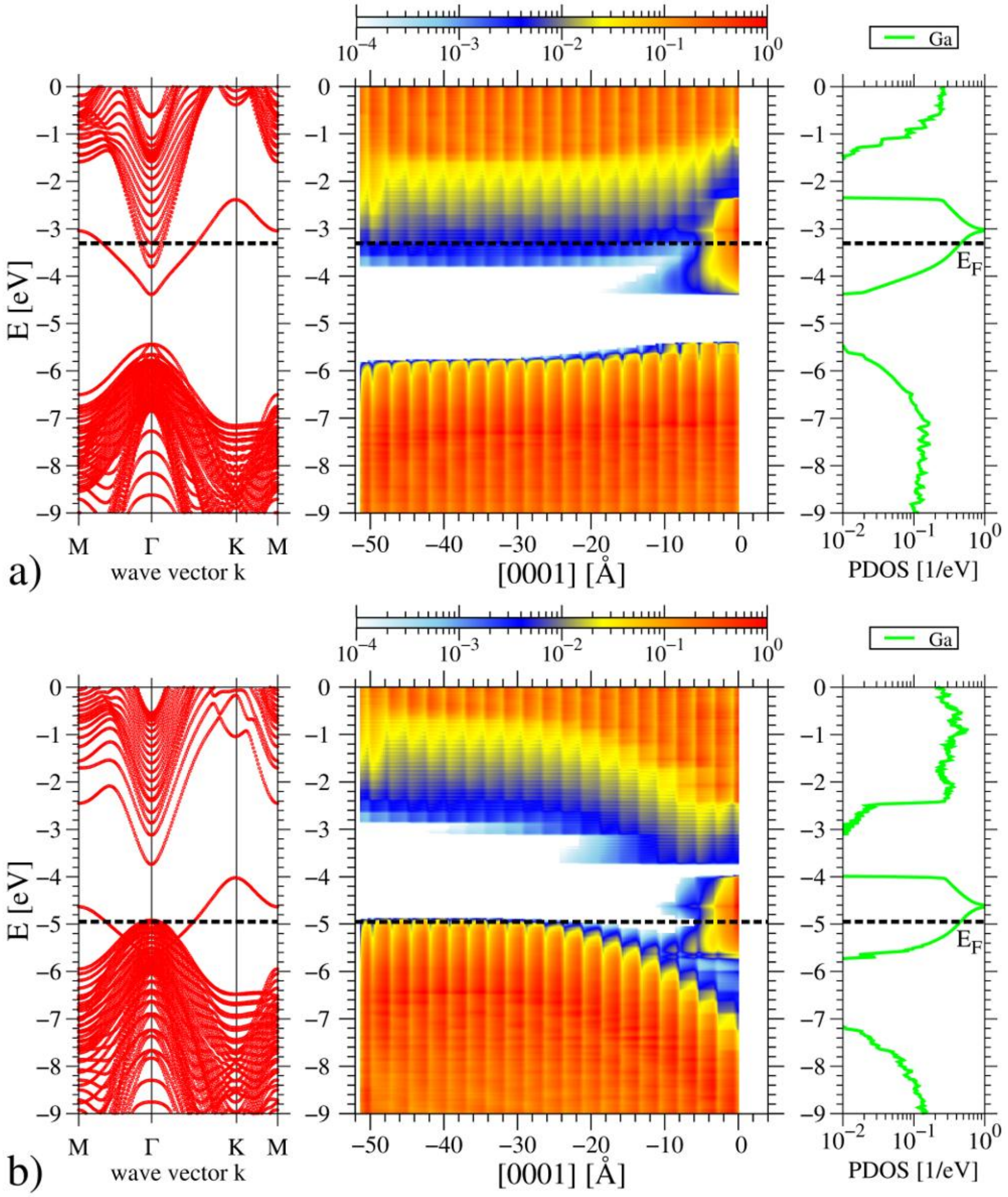

Fig. 13. Dispersion relations (left), space alignment of the bands derived from atom projected density of states (P-DOS) (middle) and DOS projected on surface gallium atoms (right) for clean GaN(0001) surface: a) n-type; b) p-type, obtained from 1 x 1 slab. The electronic temperature was 1000K. Reproduced Fig. 1 from Ref. 133 [133].

It is well known that a clean GaN(0001) surface terminated by Ga-broken bond atoms has a surface with energy located close to the CBM. This state has a wide dispersion of approximately 2 eV and is partially filled. In the case of n-type doping, the Fermi level is pinned close to the CBM such that there is no band bending or the nonlinear potential profile as the opposite side is not pinned. A drastic difference s observed for the p-type doping. Here again, the Fermi level is pinned by the Ga-broken bond state close to the CBM. The amplitude of the band bending is close to the bandgap energy, that is, $E_{g} \sim 3.47\ eV$. This is a large difference; therefore, the linear theory is not precise. The estimated screening length is $\lambda_{DH}(10^{21}\ cm^{-3}, 1000\ K) = 2.2\ \text{Å}^3$. In the simulation, this length s close to $\lambda_{DH} \sim 10\ \text{Å}^3$. Naturally, only the doping level was estimated, and this could also contribute to the observed results.

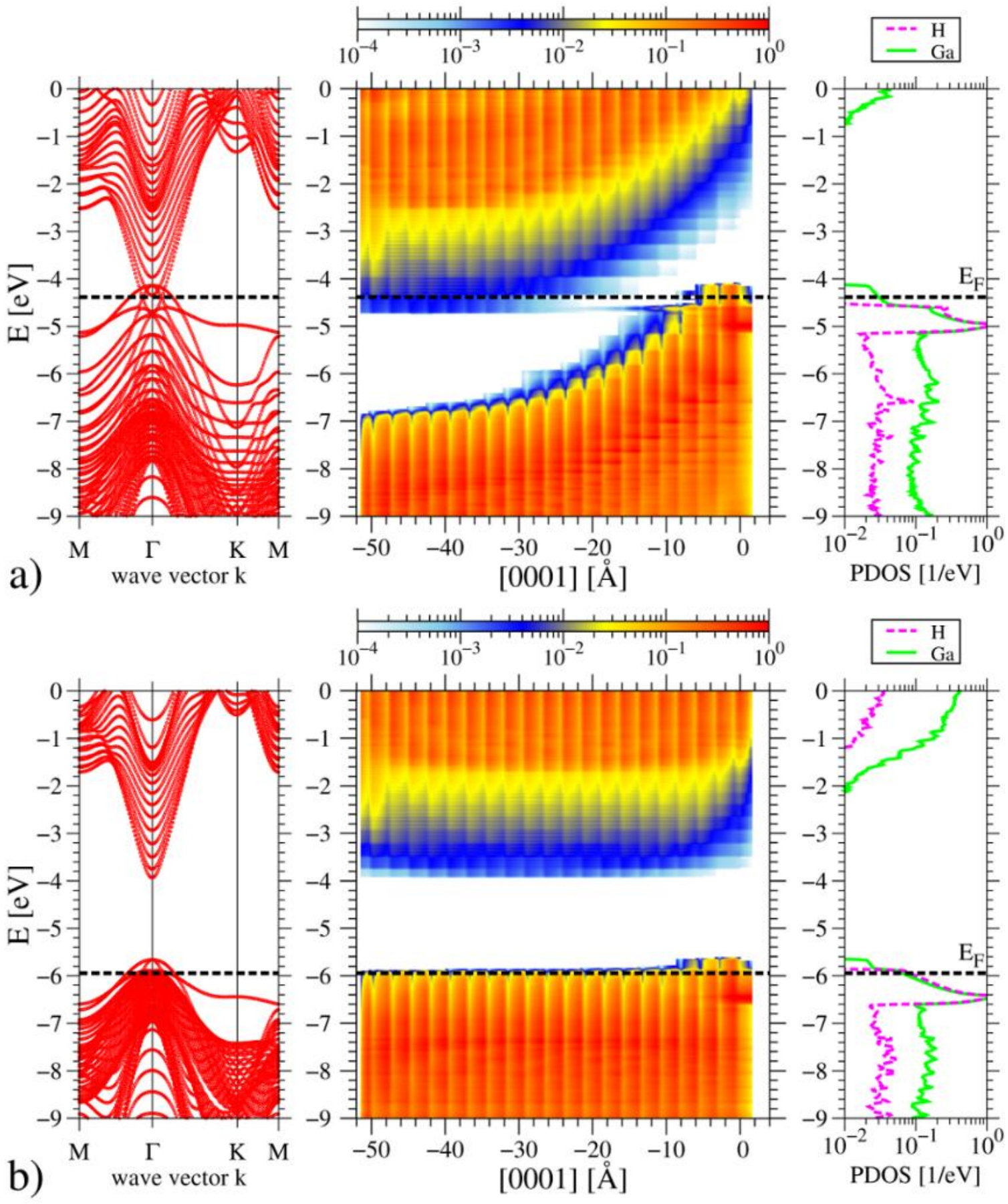

Fig. 14. Dispersion relations (left), space alignment of the bands derived from atom projected density of states (P-DOS) (middle) and DOS projected on surface gallium and hydrogen atoms (right), for fully hydrogen-covered GaN(0001) surface: a) n-type; b) p-type, both obtained from $(1 \times 1)$ slab. The electronic temperature was 1000K. Reproduced Fig. 2 from Ref. 133 [133].

Similar data were obtained for the Fermi level close to the VBM, as shown in Fig. 14. The results were obtained for a GaN(0001) surface fully covered by H atoms at the top positions [133]. This creates a Ga-H bonding state of the energy close, but below the VBM. The state has $n_e = 1\frac{3}{4}$ electrons, one from H and $3/4$ from the Ga broken bond, that is, it is not fully occupied. Therefore, this state pins the Fermi level at the surface. In the case of p-type materials, the pinning is close to the Fermi level in the bulk, so that no band bending occurs. In the case of n-type band bending, is again close to the bandgap energy, that is, $E_g \sim 3.47\ eV$. Again, the band bending extends much wider than estimated. This may be attributed to several factors, such as a charge carrier density lower than the estimated value, or nonlinear effects.

In general, the majority of simulations use undoped slabs; thus, screening is much less effective. Therefore, in most cases, the potential profiles are linear, which is related to the absence of bulk charge and the difference in the energy of the states pinning the Fermi level at both surfaces [53]. Such a situation is typical, and is shown in Fig. 2 (a) and (b), as well as in Fig 3 (c). In other cases, such as Fig. 2 (c) and Fig 3 (a) and (b), the profile is nonlinear, which is related to the penetration of the Fermi level in the bands and the existence of the band charge. In the case of Fig. 3 (c), the Fermi level is in the conduction band for more than half of the slab thickness, which leads to downward band bending close to the Ga(0001) surface. In the case of Fig. 3 (a), the Fermi level penetrates the valence band, so that the holes are present close to $GaN(000\bar{1})$ surface. Band bending indicates the presence of positive charge in the bulk material. In Fig. 3 (b), this is the opposite, and a negative charge is present in the bulk. Generally, these cases correspond to the presence of mobile charges and screening of charged surface states.

An interesting application of these methods was recently presented by the publication of ab initio investigations of two-dimensional electron gas (2DEG) at AlN/GaN heterointerface within more complex structures, which is the basis of high electron mobility transistors (HEMTs) [134]. GaN-based HEMTs are important devices for future high-frequency applications [135]. Owing to spontaneous and piezoelectric polarization, 2DEG emerges at the GaN/AlN interfaces without additional doping, allowing these devices to operate at high frequencies. The critical parameter is the high electron mobility, which can be attained with a relatively high gas density. This was achieved at the GaN/AlN interface close to the surface. The developed method has recently been used in ab initio investigations. The results of the ab initio calculations for such a system are presented in Fig. 15 (a).

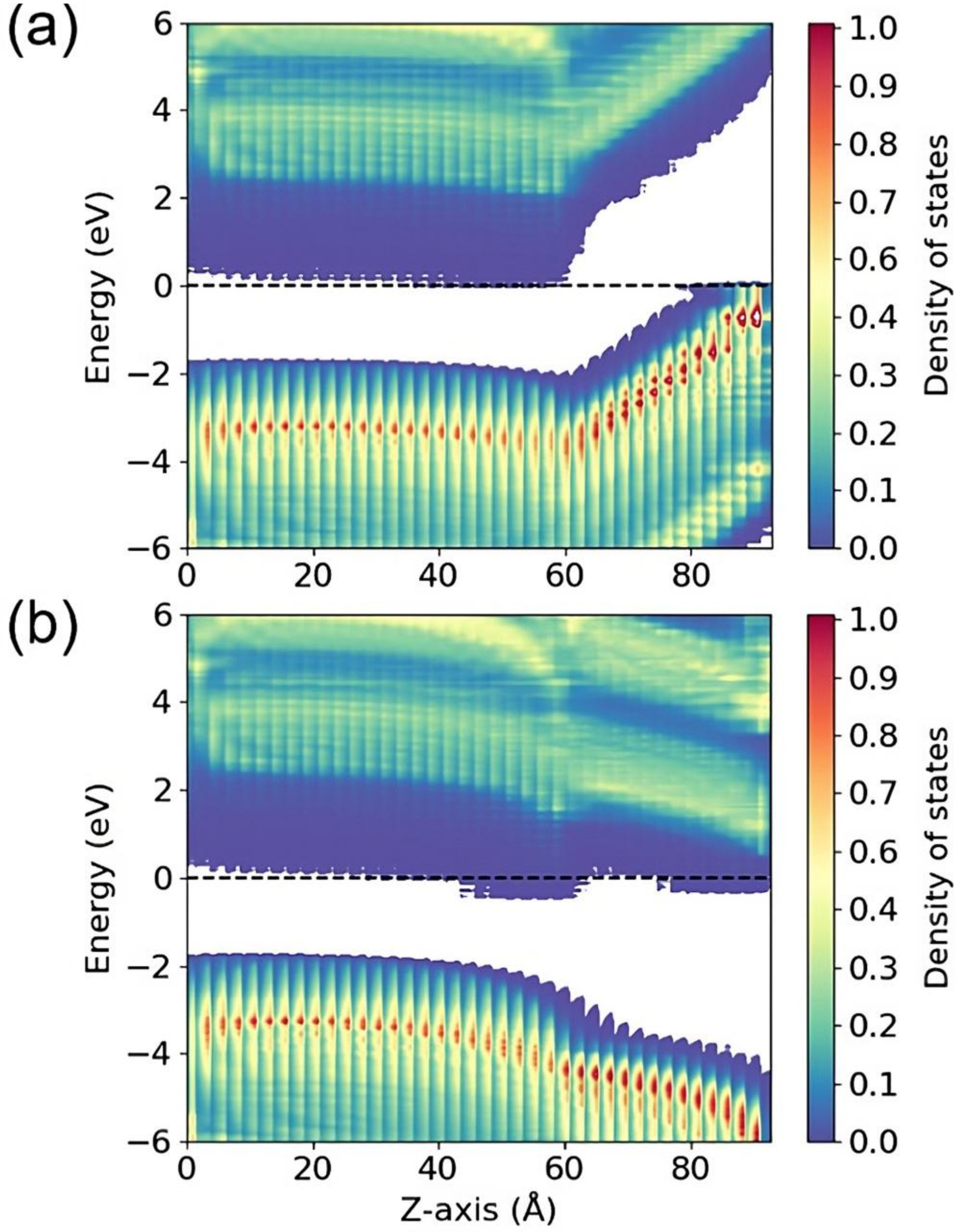

Fig. 15 . Spatially projected DOS profiles for the different pseudohydrogen atoms used to passivate the both nitrogen-side of GaN and aluminum-side of AlN surfaces of GaN/AlN slab: (a) $Z_N = 0.733$ and $Z_{Al} = 1.260$, (b) ) $Z_N = 0.733$ and $Z_{Al} = 1.380$. Reproduced from Fig 3 of Ref. 134 [134].

The obtained slab profile indicates the intersection of the conduction band minimum and Fermi level on the GaN side of the AlN/GaN interface, indicating the possible presence of 2DEG at the interface. This is shown by the charge profiles presented in Fig. 16.

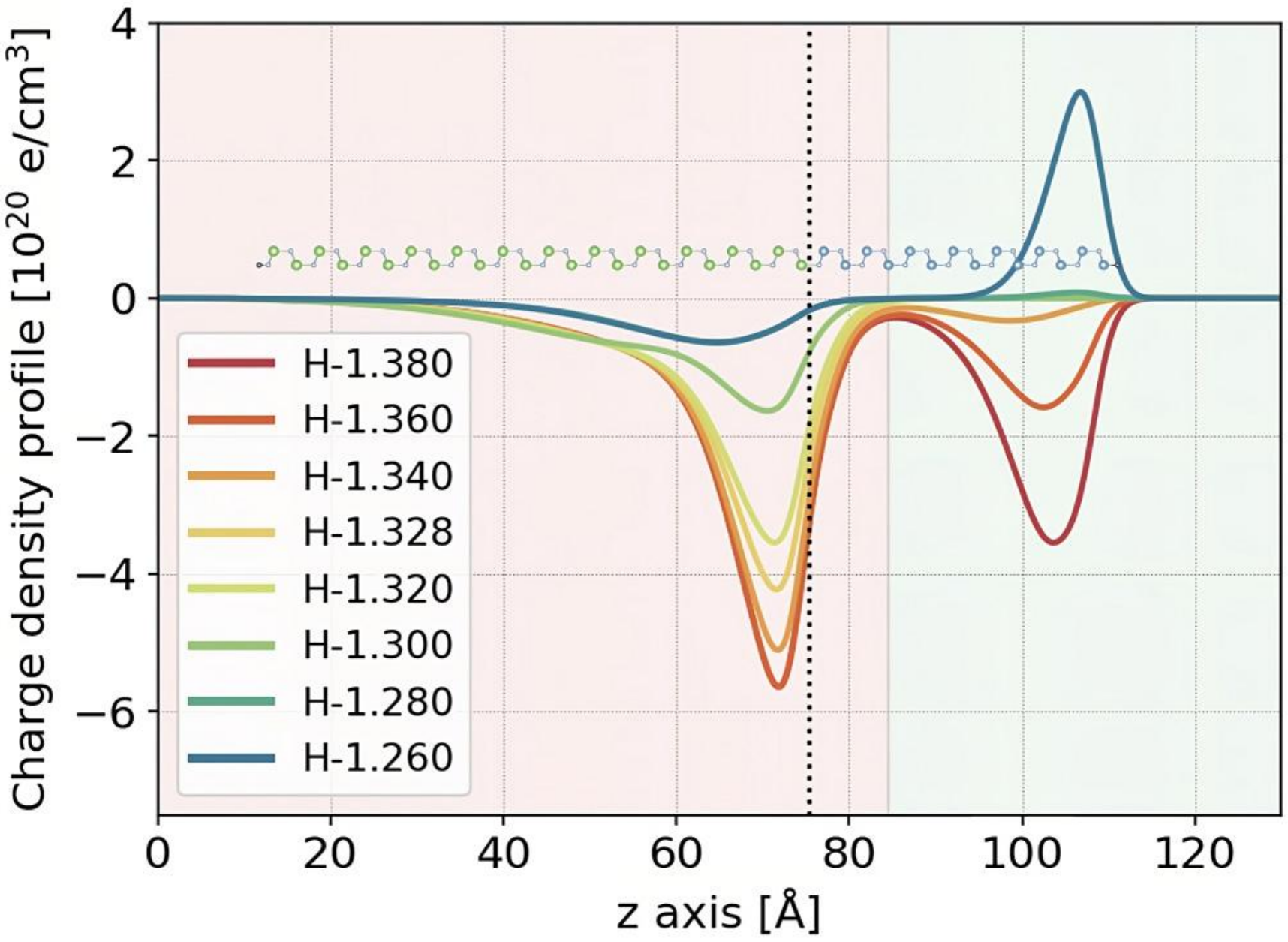

Fig. 16. Charge density profiles of 2DEG obtained along z-axis of the AlN/GaN slab obtained for different values of hydrogen pseudoatoms at AlN surface. The first peak is located at GaN/AlN interface, the second is at AlN surface. Reproduced for Fig. 4 of Ref. 134 [134].

In real devices, an additional oxide passivation layer is deposited above the AlN layer. Typically this layer is obtained by the deposition of $Al_2O_3$ as the most natural choice. This triple GaN/AlN/$Al_2O_3$ structure was also investigated in Ref 91. The results are shown in Fig 17.

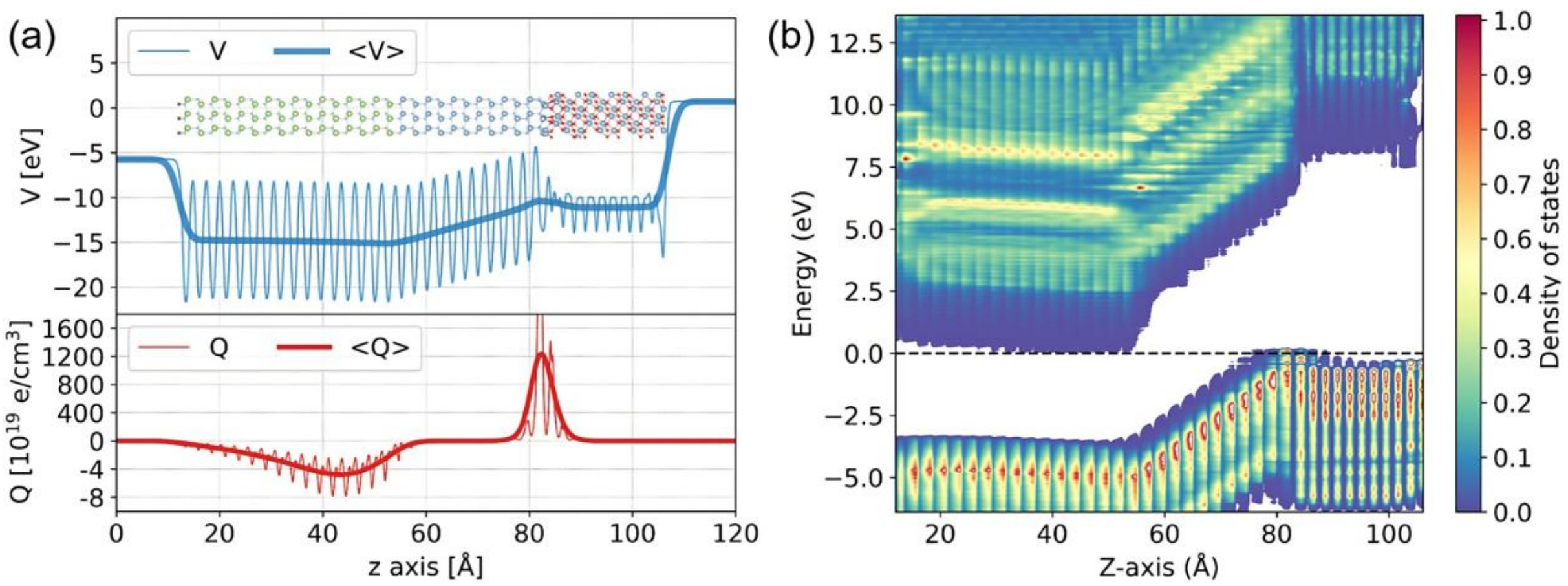

Fig. 17. Electric properties of $GaN/AlN/Al_2O_3$ structure: (a) Electric potential profiles plotted in the units of electron energy (b) band diagram. Reproduced Fig 3 from Ref. 134 [134].

In summary, it is worth stressing that *ab initio* calculations are capable of capturing the characteristic features of doping in the bulk of solids and its consequences to the subsurface dipole layer. It was possible to simulate the mobile charge at the surface and to screen the surface charge in the interior. This is due to the presence of the mobile charge, such that the Fermi level penetrates the bands close to the surface. A precise account of screening is possible in the case of very heavy doping, of the order of $n = p \cong 10^{20} \div 10^{21}\ cm^{-3}$. For lower doping the effective mass approaches can be used [4,5].

## VI. Charge role in surface reconstruction

The reconstruction of a semiconductor surface, which differs from the surfaces of other substances, is one of the most prominent features of semiconductor surfaces. In the case of the surfaces of other types of substances, the phenomenon of reduction of translational symmetry is diminished and relatively rarely observed. Nevertheless, in such cases, a small displacement is sufficient to remove symmetry. Therefore, small-amplitude reconstructions are observed for the surfaces of other crystals. Reconstructions that could drastically change the properties of the surfaces, and are large in magnitude, are mainly observed in the case of semiconductor surfaces. Therefore, the main efforts of the investigation of the reconstructions were directed towards the investigation of the reconstructions of semiconductor surfaces. They have been intensively investigated for many years and a large number of patterns have been identified [3, 136, 137].

Several principal rules governing reconstruction selection were identified in these investigations. In general, these rules indicate the transformation to a semiconducting (insulator) surface by

(i) atomic relaxation, which leads to a lower total energy of the system owing to the separation of the empty and occupied states,

(ii) creation of new bonds,

(iii) relaxation according to the electron counting rule (ECR).

Generally, these arguments are well based, as the transformations (i-iii) lead to the modification of the overlap of the quantum states by atomic motion, so that the energies of the occupied states decrease and the empty states increase. As the occupied states are the only ones that contribute to the total energy of the system, the total energy value decreases, that is, a more stable structure is attained.

ECR was first proposed by Chadi for the tight-binding analysis of As-rich reconstruction of GaAs surfaces [138], and then applied by Pashley in the interpretation of transmission electron microscopy (TEM) images of GaAs(0001) – (2 × 4) surfaces [139]. The rule is essentially a simple calculation of the electron occupation of the surface states, assuming the fractional charge contribution from the broken bond state derived from the chemical valence concept [138-140]. Thus, the fractional charge contribution from the broken Ga bond is $q_{Ga} = 3/4$. Similarly, the fractional charge contributions from the broken N and Si bonds were $q_N = 5/4$ and $q_{Si} = 1$, respectively. The contributions of the other elements from the same row in the Periodic Table are identical. The ECR calculation proceeds as follows. Assume that the total number of valence electrons is $Q = \sum_i q_i$ and the number of quantum states below the supposed Fermi energy is $N = \sum_j n_j$. Then, the occupation of all the states is $f = Q/N$. The following results could be obtained:

i) $f = 2$. The Fermi level is located above the higher energy of all included states and below the energy of the first higher-energy quantum state not included. Naturally, each state is double-spin-degenerated. If the spin degeneration is removed, the *f* value should be changed to 1.

ii) $f > 2$. The number of states was insufficient. Additional states should be added until the condition $f \leq 2$ is attained. If the condition is attained, it is assumed that the lower energy states have occupation $f = 2$, and the charge and the value are subtracted. The ratio of the residue of the charge to the number of states gives the fractional charge occupation of the highest energy state.

iii) $f < 2$. The highest energy states or states are fractionally occupied. Then, the highest states are removed, and the calculation is repeated to obtain the fractional occupation according to (ii).

The ECR procedure is equivalent to the  well-known treatment of HOMO and LUMO orbitals in quantum chemistry [141]. ECR calculation is a very simple version of the determination of the occupation of one-electron states in the DFT procedure. In the latter, it is assumed that the occupation of the one-electron states is determined by Fermi-Dirac (F-D) statistics, that is, the occupation fraction is a real number obtained from the F-D distribution at the selected temperature. ECR mimics this calculation in a simplified manner. In fact the surface states are taken into account only, those degenerated with valence band (VB) and conduction band (CB), are assumed to be occupied or empty, respectively. The important addendum is the assumption that those in the bandgap (i.e., broken bond states) are fractionally occupied. The latter is correct when these surface states are located far from the band states in the energy scale. Thus, ECR can be applied in the case of bandgap surface states located deep in the bandgap. Nevertheless, in contrast to the Fermi-Dirac occupation numbers, ECR has the advantage of being informative and intuitively appealing when addressed to a small number of states. Therefore, it has been applied in a large number of *ab initio* simulations [82-90]. Subsequently, it was extended (EECR) for use in adsorption on semiconductor surfaces [133].

ECR provides correct guidance for the treatment of the structures of semiconductor surfaces. The minimum energy criterion is generally applicable to any system in equilibrium. This deviation stems from the fact that such a simple formulation does not consider all contributions to energy [142]. This formulation is not sufficiently elaborate to describe phenomena such as those existing on semiconductor surfaces [3]. Even those limited to pure elemental semiconductors are not within the scope of ECR applications.

The stoichiometric gallium nitride Ga-terminated GaN(0001) surface was originally identified as reconstruction-free [85]. This finding was confirmed by several other reports [86-89]. In addition, the Fermi level was determined to be pinned at the surface Ga broken bond, approximately 0.5 eV below the conduction band minimum (CBM) [86-90]. As a rule, these calculations employ small $(2 \times 2)$ slabs. Later investigations discovered $(2 \times 1)$ row structure [143,144]. It was discovered that the difference between these two structures was very small, and that the unreconstructed surface was due to persistent metastability.

Recently, more detailed investigations of the stoichiometric GaN(0001) surface have been undertaken using larger slabs, including $(4 \times 4)$, $(6 \times 4)$ and $(8 \times 4)$ [145]. It was shown that the reconstruction of GaN(0001) led to the division of the top Ga atom surface into $sp^3$ and $sp^2$ coordinated Ga atoms. This leads to the division of the Ga-broken bond band into two separate sub-bands, as shown in Fig. 18.

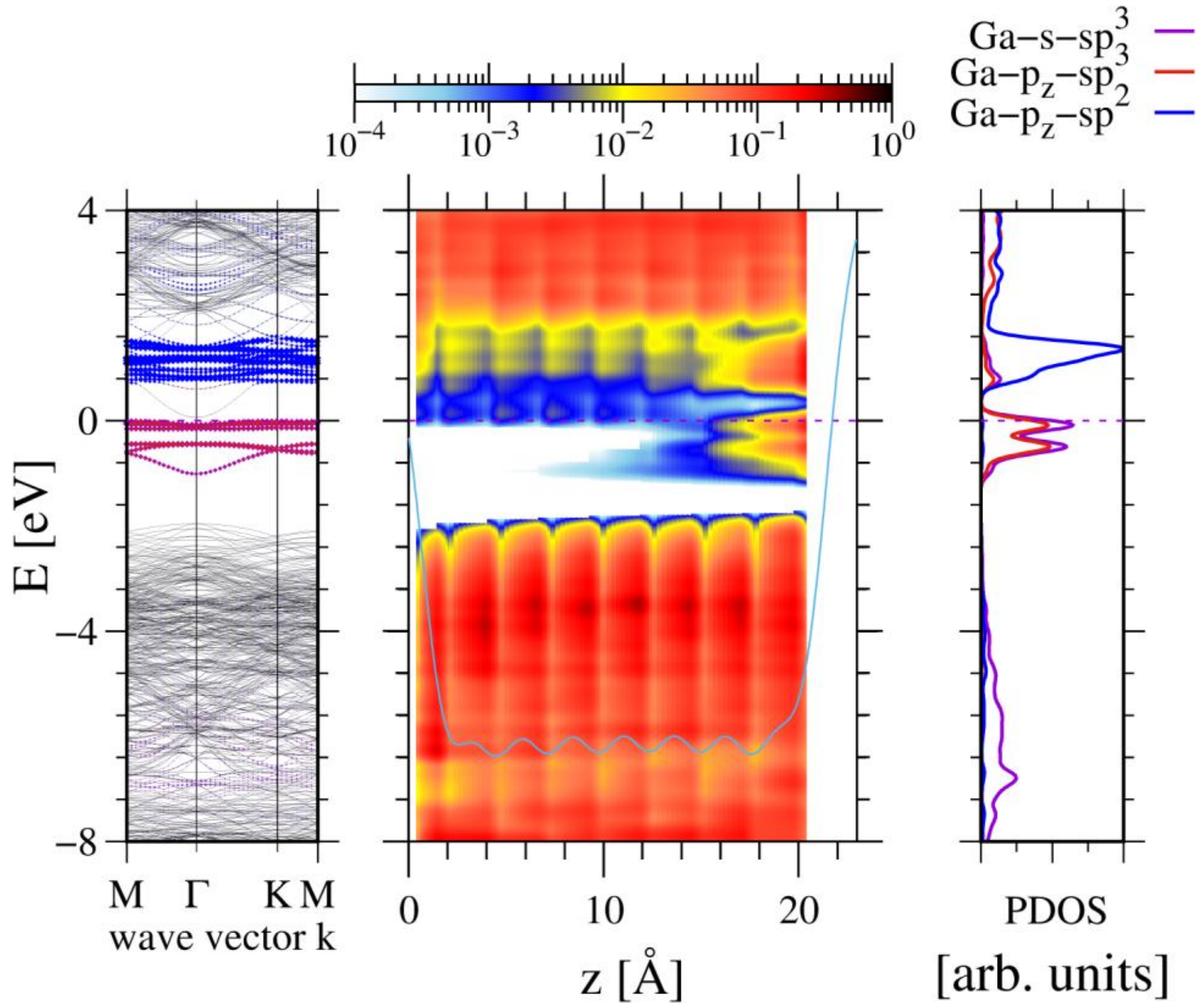

Fig. 18. Electronic properties of $(4 \times 4)$ 8 double Ga-N atomic layers (DALs) thick slab, representing stoichiometric Ga-terminated GaN(0001) surface: left – band diagram in momentum space, center - energy bands in real space, plotted along c-axis (blue line – averaged electric potential profile plotted in the units of electron energy), right - partial density of states (PDOS) plotted for the top layer Ga atoms. The blue and magenta color represent surface layer Ga atoms: $Ga4sp^3$ hybridized and $Ga4p_z$ orbital states -in conjunction to $Ga4sp^2$ hybridized states, respectively. Reproduced from Fig. 1 Ref. 145 [145].

The first is created by the gallium $sp^3$ hybridized orbitals, and the second by the $p_s$ states of $sp^2$ hybridized topmost Ga atoms. The existence of such a separation is confirmed by the plots of the GaN slabs in Fig. 19.

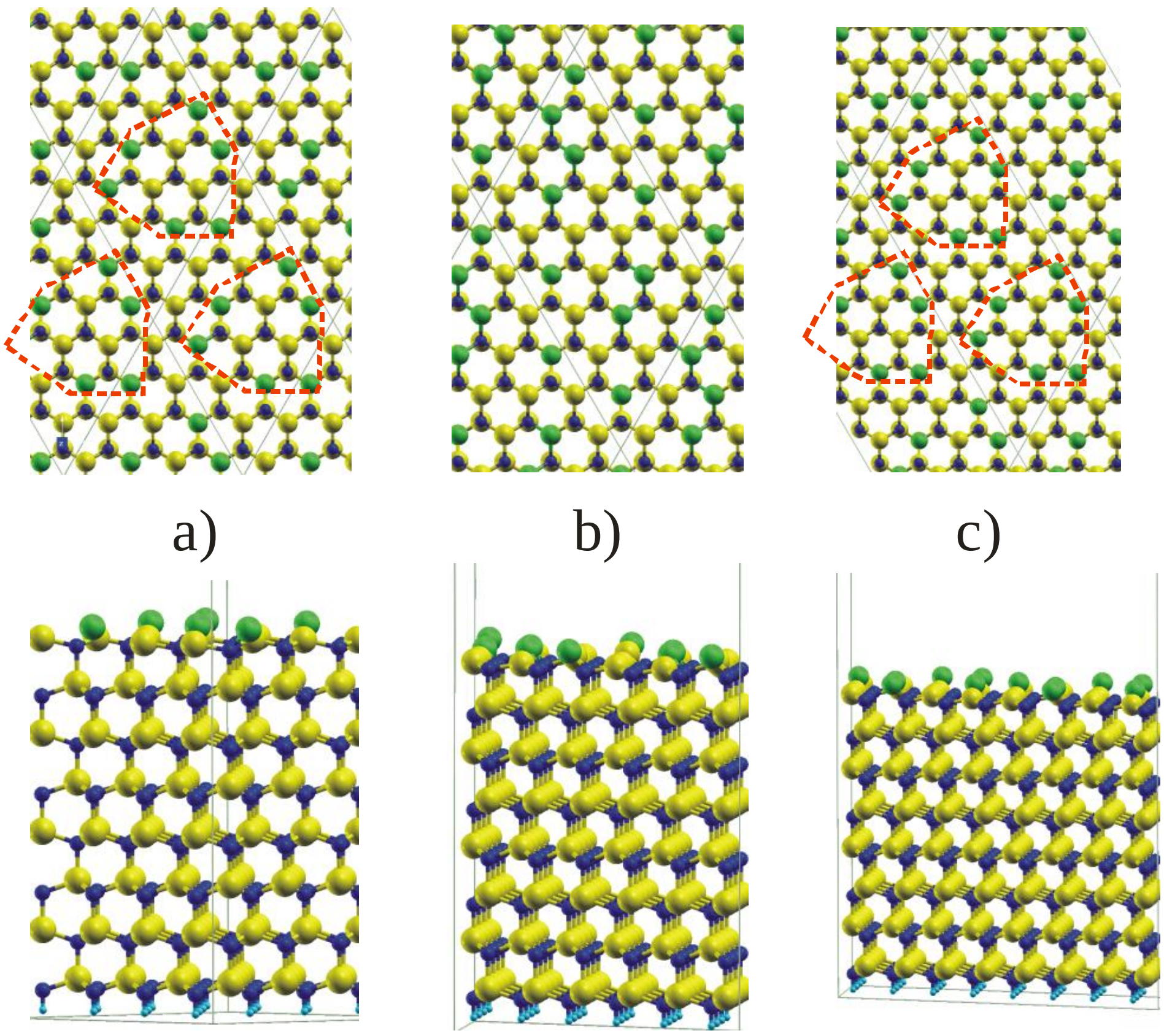

Fig. 19. Gallium nitride 8 Ga-N double atomic layer (DALs) thick slabs representing Ga-terminated GaN(0001) surface: a) $(4\ \times 4)$ slab, b) $(6\ \times 4)$ slab, c) $(8\ \times 4)$ slab. The upper and lower rows present the top and side view, respectively. The balls represent the following atoms: blue – nitrogen, yellow – gallium, green – surface layer gallium $(sp^3)$ hybridized , cyan – hydrogen termination pseudoatoms. The red color broken lines mark the basic units of $(4\ \times 4)$ reconstruction pattern. Reproduced from Fig. 2 of Ref. 145 [145].

As shown, a fraction of the topmost atoms are located in the plane of the nitrogen layer atoms, clearly indicating $sp^2$ reconstruction. In these figures, a $(4\ \times 4)$ reconstruction unit is identified. In fact, this size was determined by the ECR charge balance as follows: All Ga-N bonds are occupied; therefore, the excess charge for the surface Ga top layer atom broken bond (i.e., vertical) is $(3/4)$ electrons (elementary charge) for a single Ga topmost atom. Thus, the ECR-based electron charge balance assumes occupation by two electrons in the vertical bond state in $sp^3$ configuration, that is, a fraction of *(1-x)* sites, and 0 electrons in $p_z$ orbital in $sp^2$ hybridized atom, that is, a fraction of *x* sites:

$$(1 - x) \times 2 + 0 \times x \ = \frac{3}{4} \tag{2}$$

From this relationship, it follows that the fraction of $sp^2$ hybridized Ga atoms is $x = \frac{5}{8}$. The remaining fraction $1 - x = \frac{3}{8}$ Ga atoms is $sp^3$ hybridized. In fact, in $(4 \times 4)$ slab, 10 Ga atoms should be located in the N atom plane, whereas the remaining six Ga atoms are predicted to be located in the higher, lattice-compatible positions. The calculated distribution of Ga atoms in the topmost plane was in perfect agreement with this prediction. Similarly, in the case of $(6 \times 4)$ slab, 15 and 9 Ga atoms are in the predicted position, in full agreement with the ECR argument. Finally, in the case of $(8 \times 4)$ slab, 20 and 12 Ga atoms are in $sp^2$ and $sp^3$ hybridized, respectively. In summary, the *ab initio* results fully confirm the prediction based on ECR analysis; therefore, the atoms are hybridized owing to electron redistribution. In the case of $(4 \times 4)$ and $(8 \times 4)$ slabs, the emerging $(4 \times 4)$ reconstruction was compatible with the slab periodicity. In the $(6 \times 4)$ case, this reconstruction was not observed. Thus, successful simulations of surface reconstruction require proper choice of the simulation slab. For a large surface, a $(4 \times 4)$ reconstruction based on electron redistribution is an energetically stable configuration for a GaN(0001) stoichiometric surface. This identification was confirmed by simulations of Mg- and Si-doped slabs where the fraction of the $sp^3$ and $sp^2$ hybridized Ga topmost atoms was changed.

The driving force of this reconstruction is the difference in the energies of $Ga - s$ and $Ga - p$ orbitals. Therefore, this reconstruction cannot occur on the surfaces of typical III-V semiconductors such as GaAs(111). In the case of nitrides, Ga-terminated surfaces, such as the investigated GaN(0001) surface, can undergo this reconstruction. In contrast, nitrogen-terminated surfaces do not undergo this reconstruction at the N-terminated $GaN(000\bar{1})$ surface [12, 90]. Because of the large difference in the energies of the nitrogen $s$ and $p$ orbitals, the bonding of the solid does not involve $sp^3$ hybridization of nitrogen; therefore, such reconstruction is not observed.

In conclusion, these results demonstrate the charge-balance control of the energy optimization of semiconductor surfaces via reconstruction. This mechanism was proved in the case of a simple $sp^3$ to $sp^2$ bonding transition. In the future, more elaborate bonding, such as bridge formation, needs to be elucidated by investigating the charge contribution to energy optimization.

## VII. Charge role in adsorption

### 1. Basic description

During the adsorption of any species, a positively charged nucleus and negatively charged electrons are attached to the surface. In most cases, when no charged species are present, i.e. in the exclusion of plasma processes, the attached adsorbate is electrically neutral. Thus, the existence of the external and internal dipole layers does not affect the adsorbate, as the interactions between the field and the positive and negative charges cancel out. Nevertheless, other charge-related contributions are not perfectly balanced [146, 147].

Adsorption at semiconductor surfaces generally leads to the formation of new covalent bonds between the adsorbate and surface. Thus, the quantum states are modified such that their energies are lower. Additionally, some states in the adsorbate were added. In this process, the number of electrons also increases. Naturally, some states with energies higher than the Fermi level were also created, but these did not contribute to the system energy. The important ones are those located below the Fermi level. Thus, the contribution to the energy change of the system during adsorption stems from two different factors [146–148].

i) the change of the energy of the states occupied by the electrons,

ii) the change of the number of the electrons in the system.

Because the number of newly added electrons and the number of newly added states could be different, the second effect may involve a change in the occupation of the surface states by their shift to lower or higher energies. Depending on the number of electrons and new states created, both directions are possible. These additional electron transitions between the surface states are different from bond creation, as the effect involves electrons that are located on the surface states both before and after adsorption; therefore, the phenomenon was named intra-surface state electron transition [14].

This intra-surface electron transition contribution depends on the position of the Fermi level. These states, which are located close to the Fermi energy, donate or accept electrons, thereby contributing to the adsorption energy. This effect may exist in adsorption at any surface, causing a shift of the Fermi level by the coverage increase until the maximum possible value is reached. There are two scenarios.

i) There is no gap in the density of states (DOS) in the energy range between the initial and final positions of the Fermi level.

ii) there exist DOS gaps, one or more, in the above range.

The initial and final positions of the Fermi level are taken as the Fermi energy at the surface in the total absence, close to, or at full coverage of the adsorbate. Thus, in some cases, the Fermi level may be selected for coverage very close to zero or unity.

In the absence of the gap, the change in adsorption energy as a function of coverage is continuous. Thus, the adsorption energy change related to electron transition can be accommodated as a part of the bonding energy. This is the case for metal surfaces and, in some cases, for the surfaces of semiconductors or insulators. The gap case is different because in the full coverage range, there exists a critical coverage ($\theta = \theta_{cr}$) at which the fraction of the surface states is occupied and the remaining ones are empty (at $T = 0\ K$). At this coverage, the adsorption energy may undergo a finite jump to a new value as the electron transition contribution changes in a jump-like manner. This phenomenon can be observed only for nonmetallic surfaces, that is, semiconductors and insulators.

To determine this possibility, the procedure denoted as extended electron counting rule (EECR) was proposed in Ref. 149 [149]. First, the possibility of a Fermi-level jump across any energy gap should be determined. To do so, the adsorption process must be considered using the smallest periodic surface unit at which a single adsorbate may be attached. The unit must be defined such that the entire surface is the direct sum of such units. Thus, the surface coverage is zero or full. The EECR procedure was used to determine the Fermi level position for a jump across the gap at fractional coverage from the analysis of the surface states. In this analysis, all available additional information can be used, including the projected density of states (PDOS) and Crystal Orbital Hamilton Population (COHP) data [147, 149].

Technically, the EECR calculation is used to determine the average fraction of units with the adsorbate attached $g$ $(0 \le g \le 1)$. Assuming that the number of topmost surface atoms (adsorption sites) in the unit is $z$. Thus, the unit may be a single site in the case of adsorption at the on-top position ($z = 1$), a pair of sites in the case of the bridge configuration ($z = 2$), three sites in the case of H3 or T4 positions ($z = 3$), or even larger. The coverage $\theta$ is defined as the ratio of the adsorbates (molecules or atoms) to the number of topmost sites in the unit, that is, $\theta = g/z$. Assume that the surface electron charge in the unit before the adsorption event (i.e., for coverage $\theta$ equal to zero) is $Q_i = \sum_i q$ where the occupation of the broken bond surface states is calculated using the ECR prescription described above. The electron charge contributed by the adsorbate is $Q_a = \sum_a q_a$ where $q_a$ is the number of electrons in atom $a$ (index $a$ runs over all the atoms in the adsorbate). In the fractional occupation $g$ there exist surface states that were present before and after adsorption. However, they can be occupied in accordance to the Fermi level position after adsorption. The number of such states is $N_f$. The number of occupied states due to the attached adsorbate was $N_a$. Then extended electron counting rule (EECR) determines the critical fraction $g_{cr}$ of the units attached at which all states below the Fermi level are occupied. The EECR charge balance was obtained from the electron charge (number) conservation equation:

$$Q_i + \; g_{cr}\, Q_a \; = N_a\, g_{cr} + (1 - g_{cr})\, N_f \tag{3a}$$

This is calculated as follows: on the left is the charge in the system before adsorption, $Q_i$. This is the charge of the unit, including the fractionally occupied states, in accordance with the ECR of broken bonds. The second contribution is the charge added by the adsorbate, which is proportional to the critical occupation and charge of the adsorbate, that is, $g_{cr}\, Q_a$ . On the right hand, this charge is distributed into the states of the covered part, which is again proportional to the critical occupation, that is $N_a\, g_{cr}$. The second term is the number of states occupied by electrons in the non-covered fraction of the surface i.e. $(1 - g_{cr})\, N_f$. This balance was used to achieve full and zero occupation below and above the Fermi energy, respectively. Thus, the critical coverage at which the Fermi level jumps from occupied to empty states was determined. In the double spin degenerate case, that is,, the factor $f = 2$ is used for the calculation of the number of states $N_f$ and $N_a$ (those numbers are calculated after the adsorption). From this relation, the critical fraction and critical coverage can be obtained:

$$g_{cr} = z\theta_{cr} = \frac{Q_i - N_f}{(N_a - N_f) - Q_a} \tag{3b}$$

Alternatively, this relation can be formulated as a calculation of the covered fraction of the units by considering the number of donated and accepted electrons:

i) in the case of the donation of electrons from the initially occupied surface state ($Q_i > N_f$), that is, when the Fermi level decreases. The number of donated electrons from the surface states in the unit is: $Q_i - N_f$ as the initial number of electrons is $Q_i$ and the number of occupied states after (in ECR, all states are occupied or empty) is $N_f$. The number of accepted electrons in the final state is $\left[\left(N_a - N_f\right) - Q_a\right]$ because $\left(N_a - N_f\right)$ is the number of electrons in the final state from which the number of electrons contributed by the adsorbate, that is, $Q_a$, should be subtracted.

ii) in the case of the acceptance of electrons in the initial state, that is, initially empty ($Q_i < N_f$ ), that is, when the Fermi level increases. The number of accepted electrons is $N_f - Q_i$ as the number of electrons in the final state is $N_f$ and this number in the initial state is $Q_i$. The number of donated electrons from the final state is $Q_a - \left(N_a - N_f\right)$ as $(N_a - N_o)$ is the number of electrons that decreased (due to $N_a > N_o$) in the final states for which the number of electrons contributed by the adsorbate, that is, $Q_a$ should be added.

Thus, a fraction of the units $g_{cr}$ is occupied as determined from the number of available electrons.

Several examples are discussed to demonstrate the EECR calculation of the charge. First, the case of hydrogen adsorption on-top ($z = 1$) of the topmost Ga atom of the GaN (0001) surface (Figs 13 and 14). In this case, the initial charge is $3/4$ electrons associated with the broken Ga bond, that is, $Q_i = 3/4$. These broken bond states are located in the vicinity of the CBM (Fig. 13). Hydrogen brings one electron, that is $Q_a = 1$. In the final state, the Ga-H state is at the VBM; therefore, $N_a = 2$. After adsorption, the broken Ga bond state is above the Fermi level; thus, it is empty, that is, $N_f = 0$ (Fig. 14). From these data, it follows that from Eq. 3b, where $g_{cr} = \theta_{cr-H} = 3/4$.

The alternative determination based on argument (i) is as follows: the broken bond state has $3/4$ electrons initially ($Q_i = 3/4$), and zero in the final state $\left(N_f = 0\right)$. The adsorption of hydrogen brings one electron ($Q_a = 1$), thus in the final state, we have from the adsorption $1\frac{3}{4}$ electrons, i.e. $1/4$ electrons are missing. This must be borrowed from Ga broken bond states that are higher in energy. In fact, a single Ga broken bond state provides $3/4$ electrons, which is sufficient to fully occupy the three Ga-H states. Therefore, the ratio of covered to empty sites is $3:1$, for three covered sites, we have on site empty, and accordingly, the critical coverage is $\theta_{H-cr} = 3/4$.

The second example is the mixed $NH_2$-$NH_3$ coverage where GaN(0001) surface is covered by mixture of $NH_2$ radicals $\left(\theta_{NH_2} = 0.75\,ML\right)$ and $NH_3$ admolecules $\left(\theta_{NH_3} = 0.25\,ML\right)$ [148] presented in Fig. 20. The EECR condition can be determined as follows: the coverage is full, that is, all sites are covered either by $NH_2$ radicals or $NH_3$ molecules. Therefore, the change from the $NH_2$ to the $NH_3$ configuration may be treated as an attachment of single H atom. The surface unit can be selected to include a single Ga surface atom site, that is, $z = 1$. Thus, the initial configuration consisted of a broken Ga bond with an attached $NH_2$ radical. All $NH_2$ states are degenerated with the valence band: Ga-N states are located deep, while the N broken bond is at the VBM. This confirms the location of the Fermi level at the mid-gap and the absence of a field inside the slab. Moreover, the projected density of states (PDOS) and Crystal Orbital Hamilton Population (COHP) diagrams confirmed that the nitrogen and hydrogen s and p overlapped states were located in the valence band.

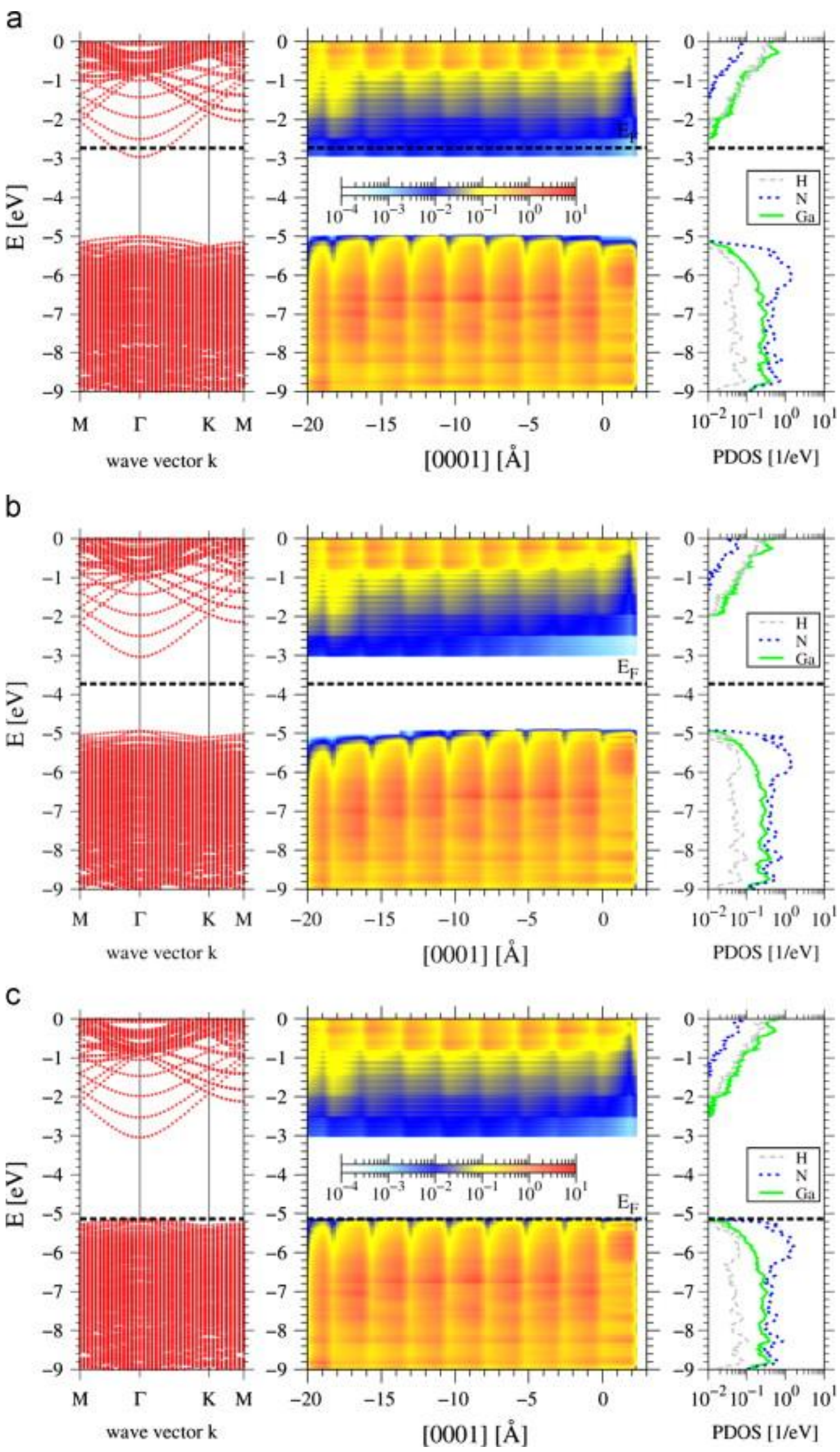

Fig. 20. Electronic properties of the (4×4) 8 GaN DALs slab representing GaN(0001) surface covered with 0.25 $NH_3$ admolecules and 0.75 $NH_2$ radicals. Left – the band diagram of the slab; middle – DOS projected on the atom quantum states (PDOS), showing the spatial variation of the valence and conduction bands in the slab; right – DOS of the topmost Ga atoms and $NH_3$/$NH_2$ admolecules. The states density scale (logarithmic) is included in the diagrams. Top, middle and bottom panels represent *n*-type, semi-insulating and *p*-type

material respectively, simulated by charge background procedure implemented in SIESTA package. Reproduced Fig. 3 from Ref. 148 [148].

This is in agreement with the data presented in Fig. 20. More interesting are the nitrogen *p* states of the $NH_2$ radicals without overlap. As indicated by the COHP diagram, these states are located at the top of the valence band; thus, they are occupied. Therefore, the occupation is full, but $3/4$ electrons are borrowed from valence band. Accordingly, the initial charge is $Q_i = 3/4 + 7$ in which $3/4$ in the Ga-broken bond contribution and seven stem from five electrons in nitrogen and two electrons from two hydrogen atoms. After adsorption, the number of states is $N_f = 8$. Thus, all $NH_3$ states are occupied, and $1/4$ electrons are shifted to the conduction band, as there are no states in the gap. The adsorbate, that is, the hydrogen atom charge, was $Q_a = 1$. From Eq. 3b, it follows that $g_{cr-H} = \theta_{cr-H} = \theta_{cr-NH_3} = 1/4$. Thus the critical coverage is the mixture of $NH_2$ radicals $\left(\theta_{NH_2-cr} = 0.75\ ML\right)$ and $NH_3$ admolecules $\left(\theta_{NH_3-cr} = 0.25\ ML\right)$.

Notably, the EECR critical state does not correspond to any energetically preferable state of the surface; it is merely the coverage at which the Fermi level is liberated, that is, located in the gap between one pinning state and the other. As such, it is mere termination of the coverage range in which the jump in the Fermi level corresponds to the jump in the adsorption energy. As argued above, this change usually corresponds to the absence of a subsurface dipole layer because the Fermi level is governed by its position in the bulk, that is, no subsurface charge is present. An example of the absence of an electric field in a slab is shown in Fig. 20.

Therefore, it is expected that in the vicinity of the EECR state, the adsorption energy depends on doping in the bulk. The band profiles presented in Fig. 20 confirm this prediction; the bands are flat, and the Fermi level is located in the CB, VB, and midgap for n-type, p-type, and semi-insulating bulk GaN, respectively. As claimed, the adsorption energy should be approximately constant for a wide range of coverage (it may depend on adsorbate-adsorbate interaction only) and undergo a considerable jump at the EECR state where the Fermi level is liberated to be shifted to the other pinning state. This was best observed for hydrogen adsorption on the GaN(0001) surface. In this case, molecular hydrogen is dissociated during adsorption, and the resulting hydrogen atoms are located at the top positions. Hydrogen is an exemplary case because it is small; therefore, the quantum overlap is extremely small, and these atoms are essentially noninteracting [150]. The surface unit consists of a single Ga site to which an H atom is attached. The Ga broken bond state is located 0.5 eV below the CBM; therefore, it is empty, that is, $N_i = 0$, and is fractionally charged by ¾ electrons, thus $Q_i = 3/4$ . The Ga-H two spin states are located at the VBM; therefore, it is occupied; thus, $N_f = 2$. Hydrogen brings about a single electron; therefore, $Q_a = 1$. From Eq. 2a, the critical EECR hydrogen coverage was $\theta_{H-cr} = 3/4$. At this coverage, the Fermi level is free; for $\theta < \theta_{H-cr}$ the Fermi level is pinned by the Ga-broken bond state below CBM; and for $\theta > \theta_{H-cr}$ it is pinned by the G-H state at the VBM. Therefore, it is expected that the adsorption energy will increase, a jump as shown in Fig. 15. The adsorption energy was obtained from *ab initio* calculations using the following formula:

$$\Delta E_{DFT}^{ads-GaN}(H) = E_{DFT}^{tot}(slab + H) - E_{DFT}^{tot}(slab) - E_{DFT}^{tot}(H) \quad (4)$$

and is plotted in Fig 15 for the entire range of hydrogen coverage, $(0 \leq \theta_H \leq 1)$. The adsorption energies were calculated for the atomic hydrogen adsorption $[H(v) \rightarrow H(s)]$ and the molecular adsorption $[H_2(v) \rightarrow 2H(s)]$ is [149]

$$\Delta E_{DFT}^{ads-GaN}(H) = \begin{cases} -3.40\ eV & \theta_H < 0.75\ ML \\ -1.10\ eV & \theta_H > 0.75\ ML \end{cases} \quad (5a)$$

and

$$\Delta E_{DFT}^{ads-GaN}(H_2) = \begin{cases} -2.24\ eV & \theta_H < 0.75\ ML \\ 2.36\ eV & \theta_H > 0.75\ ML \end{cases} \quad (5b)$$

The jump by about 2.30 eV is due to electron transfer from Ga broken bond to H bonding state located below VBM, present and absent for low and high coverage, respectively. The energy difference obtained for the dissociative adsorption of molecular hydrogen and the adsorption of atomic hydrogen was in accordance with the molecular hydrogen dissociation energy $E_{DFT}^{diss}(H_2) = 4.56\ eV$. This calculated value is slightly different from the experimental value $E_{exp}^{diss}(H_2) = 4.58\ eV$ [151]. As shown in Fig. 21, the adsorption energy was constant with a random variation below 0.1 eV.

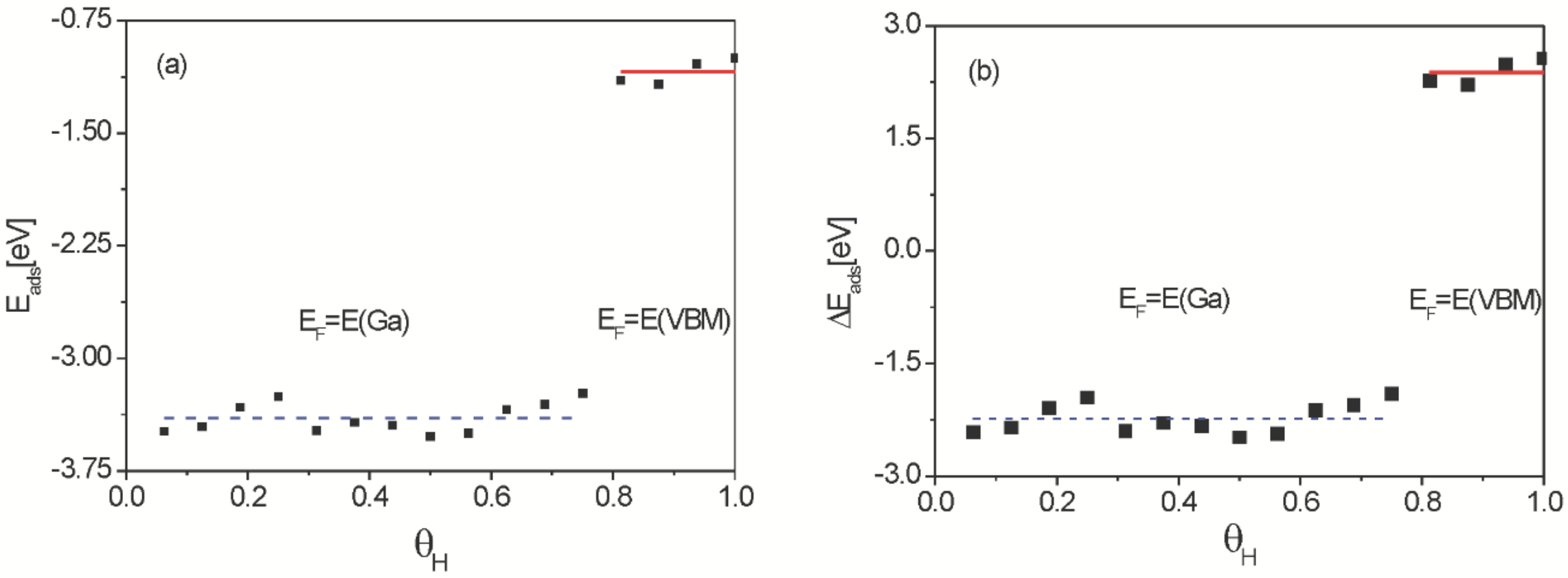

Fig. 21. Atomic (a - Eq. 3a) and molecular (b - Eq. 3b) hydrogen adsorption energy in function of hydrogen coverage of the GaN(0001) surface (black squares), calculated using 4 x 4 slab. The horizontal lines represent coverage independent energy values for low and high coverage regime, denoted by dashed blue and solid red lines respectively. Reproduced Fig. 8 from Ref. 149 [149].

An additional effect was observed for the above-presented case of $NH_2$/$NH_3$ coverage of the GaN(0001) surface [149]. The band structures of the surface covered by $\theta_{NH_2-cr} = 3/4$ and $\theta_{NH_3-cr} = 1/4$ are shown in Figs 20. As shown in Fig. 20, n-type, SI, and p-type doped crystals have a Fermi level descending from the CBM, midgap to the VBM. The $NH_2$ to $NH_3$ conversion is essentially the attachment of a single hydrogen atom. The adsorption energy presented in Fig. 21 shows the energy change during attachment of $H_2$ molecule at $(4 \times 4)$ i.e. 16 site slab. The adsorption in the case of a finite state above the critical value, that is, $\theta > \theta_{cr-NH_3} = 1/4$ leads to the shift of electrons to the other states, that is, above the Fermi energy. As shown in Fig. 22, the reduction in the energy gain occurs exactly at $\theta = \theta_{cr-NH_3} = 1/4$ for n-type, as these electrons are shifted up very high. Subsequently, this reduction is observed for SI and p-type materials, as in these cases, these electrons are located first in the partially empty VB states. A further increase in the coverage leads to an energy gain until the energy gain finally becomes negative, owing to electron transfer over the gallium nitride bandgap. This effect is the second confirmation of the electron character of the observed effect on the adsorption energy change.

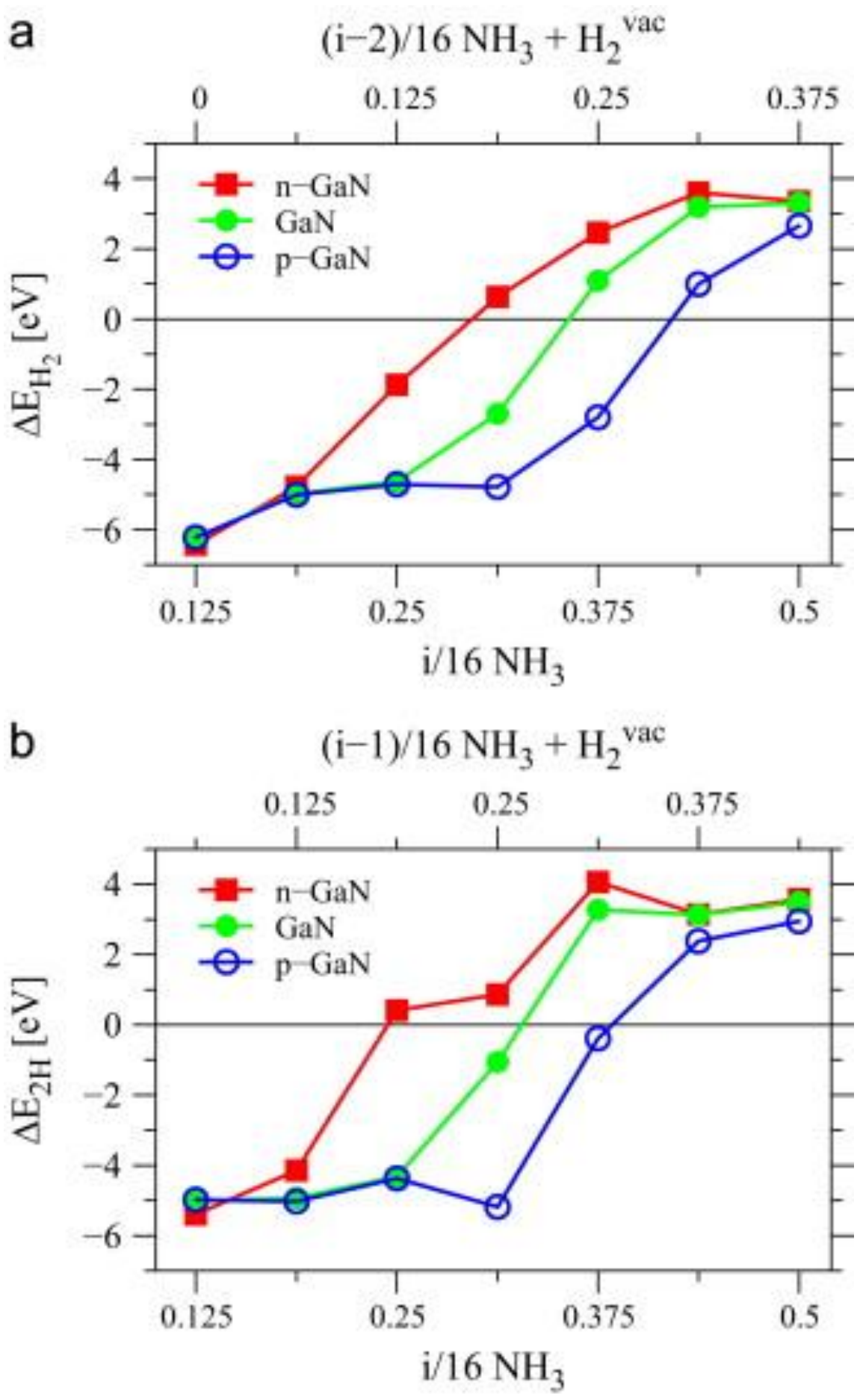

Fig. 22. Energy changes during adsorption/desorption of $H_2$ molecule on GaN(0001) surface covered with various $NH_3$/$NH_2$ mixture. The data are obtained for *n*-, *p*-type and semi-insulating bulk. The negative value of energy means that the surface configuration indicated on the bottom horizontal axis has a lower energy than the configuration marked on the upper axis. Reproduced Fig. 4 from Ref. 148 [148].

It is worth noting that the adsorption energy jump related to intra-surface electron transfer exerts a significant influence on the thermodynamic properties of two-phase semiconductor-vapor systems. From the Van t'Hoff relation, the equilibrium pressure of adsorbate $p$ depends exponentially on the evaporation energy $\Delta H$, that is, $p \approx \exp\left(-\frac{\Delta H}{kT}\right)$. In addition to *ab initio* calculated energy difference, the evaporation energy $\Delta H$ can also change owing to the thermal contribution to the enthalpy, that is, $\Delta H(T) = \Delta H(T_o) + \int_{T_o}^{T} \Delta C(T) dT$. The thermal contribution to this difference can be calculated by assuming that

i) for the vapor phase, the thermal contributions stem from the translational, rotational, and vibrational degrees of freedom [151-154],

ii) for the adsorbate, the thermal contribution to enthalpy, entropy, and free energy arises only from lattice vibrations [155].

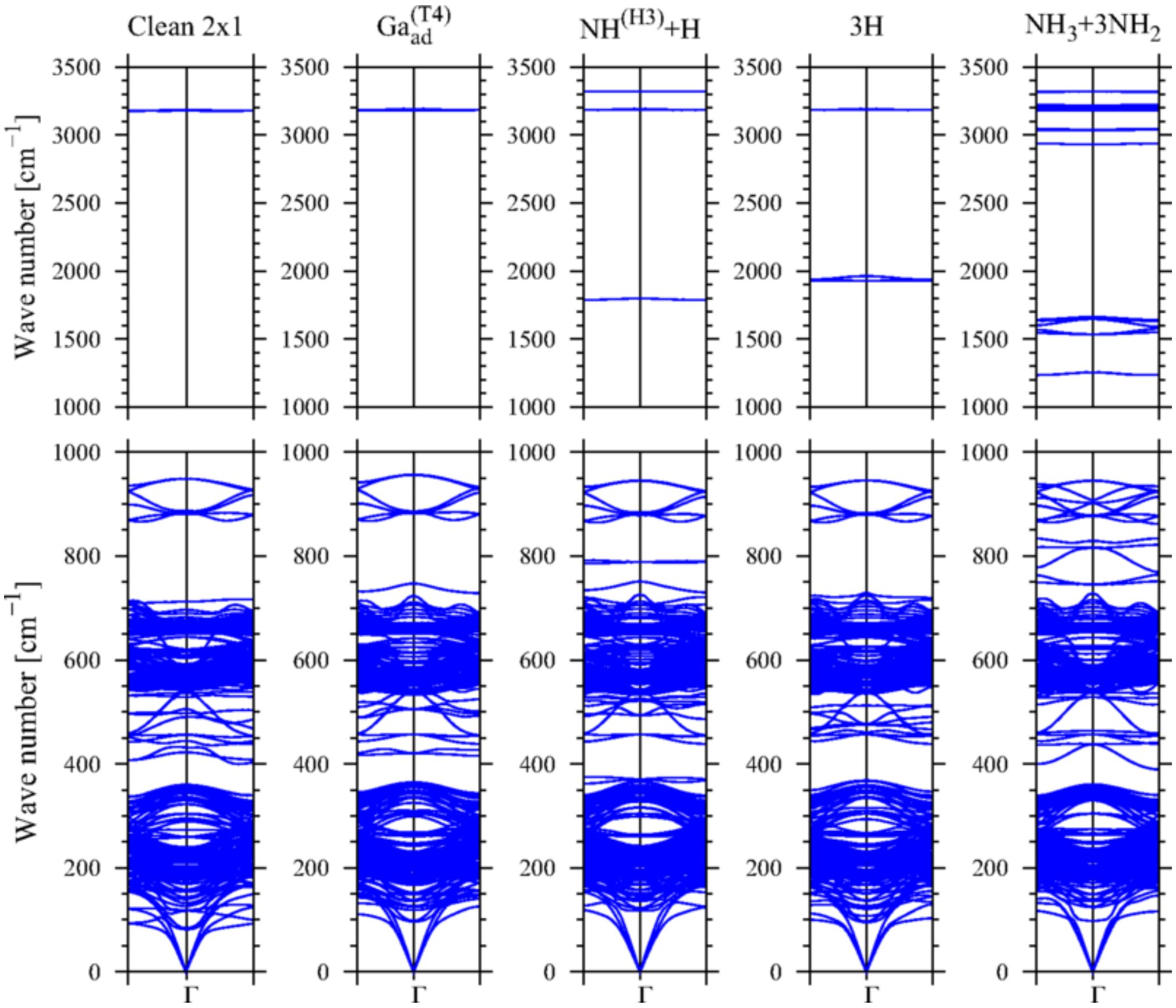

Fig. 23. Phonon dispersion relations of (2 x 2) slab representing GaN(0001) surface: lower row – full slab spectra, upper row – only spectra associated with different adsorbate: from the left: none, Ga adatom, NH-radical in H3 position plus H adatom, 3 H adatoms, $NH_3$ admolecule and 3 $NH_2$ radicals. In addition, the modes associated with pseudo-hydrogen termination atoms at the opposite surface are shown (approximately $3200\ cm^{-1}$). The bands in the range between $850\ cm^{-1}$ and $950\ cm^{-1}$ are related to N atom close to the termination. Reproduced Fig. 3 from Ref. 155 [155].

The adsorbate was treated in a simplified manner by employing only vibrational degrees of freedom (ii). In some sense, this is the reduction of the more general approach (i) used for the derivation of the data shown in the thermochemical tables for the molecules and atoms. Because the adsorbates were located in the lattice sites, only vibrational calculations were performed [155]. In this case, the adsorbate contribution was defined as the difference between the vibrational spectra of the slab with and without the adsorbate. An example of such calculation results is presented in Fig. 23 for a GaN(0001) surface with several adsorbates. The results showed additional modes associated with several adsorbates. In addition, by direct summation, the zero-point energy can be calculated as the difference in the energies of the ground quantum states of the adsorbed atoms and the molecule in the vapor: $\Delta E_{sv}^{ZPE} = 2E_{s}^{ZPE}(H) - E_{v}^{ZPE}(H_2)$

From these spectra, standard thermodynamic formulae may be used to derive thermodynamic quantities by direct summation [155-157]

$$E^{vib}(x) = k_B T \sum_j \frac{x_j}{exp(x_j)-1} \tag{6a}$$

$$C^{vib}(x) = k_B \sum_j \frac{x_j^2 exp(x_j)}{\left[exp(x_j)-1\right]^2} \tag{6b}$$

$$S^{vib}(x) = k_B \sum_j \frac{x_j}{exp(x_j)-1} - ln\left[1 - exp\left(-x_j\right)\right] \tag{6c}$$

$$F^{vib}(x) = k_B T \sum_j ln\left[1 - exp\left(x_j\right)\right] \tag{6d}$$

where $x_j \equiv \frac{\hbar\omega_j}{k_B T}$ and $\omega_j$ is the phonon frequency of the jth phonon mode. From these formulae, the temperature-dependent free energy vibrational contribution $\Delta F^{vib}(T)$ and zero-point energy contribution was $\Delta E^{ZPE}$ were calculated for several adsorbates on the GaN(0001) surface [155]. The results are shown in Fig. 24.

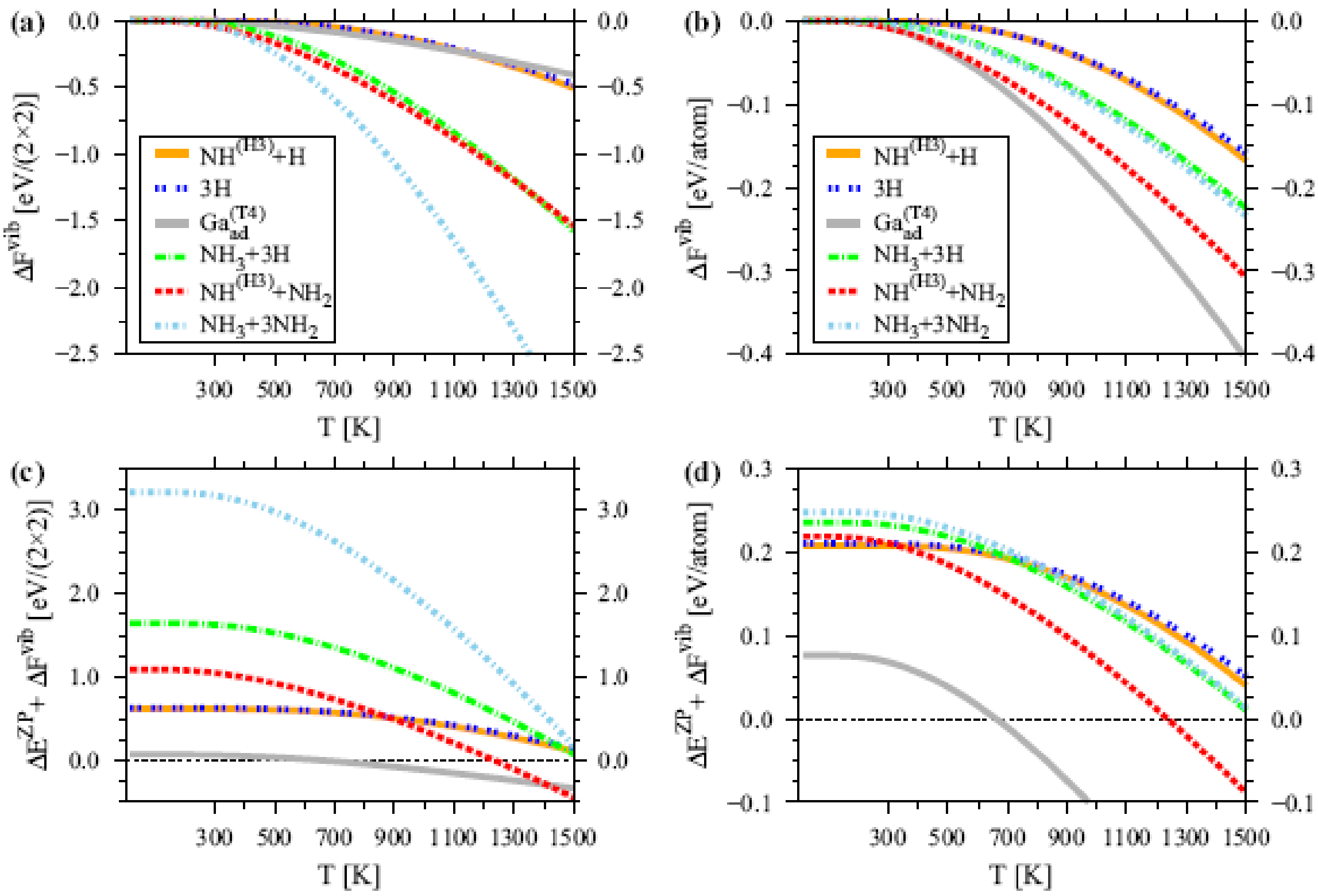

Fig 24. Changes in the vibrational free energy versus temperature for the GaN(0001) surface covered with different adsorbates, determined in relation to clean surface with 2 × 1 reconstruction; (a), (b) $\Delta F^{vib}(T)$ dependence; (c), (d) $\Delta F^{vib}(T)$ dependence including zero-point $\Delta E^{ZPE}$ offset . Graphs (a) and (c) show energies of the entire 2 × 2 surface cell, while graphs (b) and (d) show the average energy per atom. Fig. 4 from Ref 155 [155].

Alternatively, a simplified treatment may be used, based on the Debye theory, in which the acoustic phonon frequencies are approximated by linear dependence on the wavevector. Accordingly, the maximum phonon energy, known as the Debye energy, and its equivalents, the Debye frequency and Debye temperature, are related as follows: $E_D = \hbar\omega_D = k_B\theta_D$. Using this simplified representation, the complex summation can be replaced by integration [157]:

$$E^{vib}(T) = 9k_BT\left(\frac{T}{\theta_D}\right)^3 \int_0^{\theta_D/T} \frac{x^3dx}{exp(x)-1} \tag{7a}$$

$$C^{vib}(T) = 9k_B\left(\frac{T}{\theta_D}\right)^2 \int_0^{\theta_D/T} \frac{x^4exp(x)dx}{[exp(x)-1]^2} \tag{7b}$$

$$S^{vib}(T) = 9k_B\left(\frac{T}{\theta_D}\right)^3 \int_0^{\theta_D/T} \left\{\frac{x}{exp(x)-1} - ln[1-exp(-x)]\right\} x^2dx \tag{7c}$$

$$F^{vib}(T) = 9k_BT\left(\frac{T}{\theta_D}\right)^3 \int_0^{\theta_D/T} x^2 ln[1-exp(-x)]dx \tag{7d}$$

In the case of hydrogen adsorbed on the GaN(0001) surface, the Debye formulation provides a satisfactory approximation to the exact summation [157]. This formulation shows that the interaction between the adsorbate and the substrate may be complex, which is quite different from the simple increase in the number of degrees of freedom and the emergence of additional frequencies. The interaction may affect the elastic lattice properties such that the specific heat change may be negative at low temperatures owing to the lower excitement of a large number of lattice vibrations, which can offset the increase in the number of degrees of freedom during adsorption. Generally, this is reflected in the coverage dependence of the effective Debye temperature, which may be substantial for complex adsorbates. Thus, in the case of Al adsorption on the AlN(0001) surface [158],

$$\theta_D(\theta_{Al}) \cong \begin{cases} 1002.4 - 1254.5\theta_{Al} + 745.1\theta_{Al}^2 & \theta_{Al} \le 1 \\ 493 & 1 < \theta_{Al} \le 2 \end{cases} \tag{8}$$

This reflects the dramatic reduction in the stiffness for higher Al coverage, for example, $T_D(\theta_{Al} = 0.1) = 875\ K$ which is reduced to $T_D(\theta_{Al} = 0.75) = 550\ K$ [158]. Naturally, the zero-point energy is also coverage-dependent.

$$E_{Al}^{ZPE}(\theta_{Al}) \cong 0.097 - 0.101\theta_{Al} + 0.06\theta_{Al}^2 \tag{9}$$

This complicates the use of simplified formulations. It is worth noting that the presence of the adsorbate may lead to an increase as well as a decrease in the heat capacity. This is shown for the case of Al adsorption on the AlN(000-1) surface in Fig. 26.

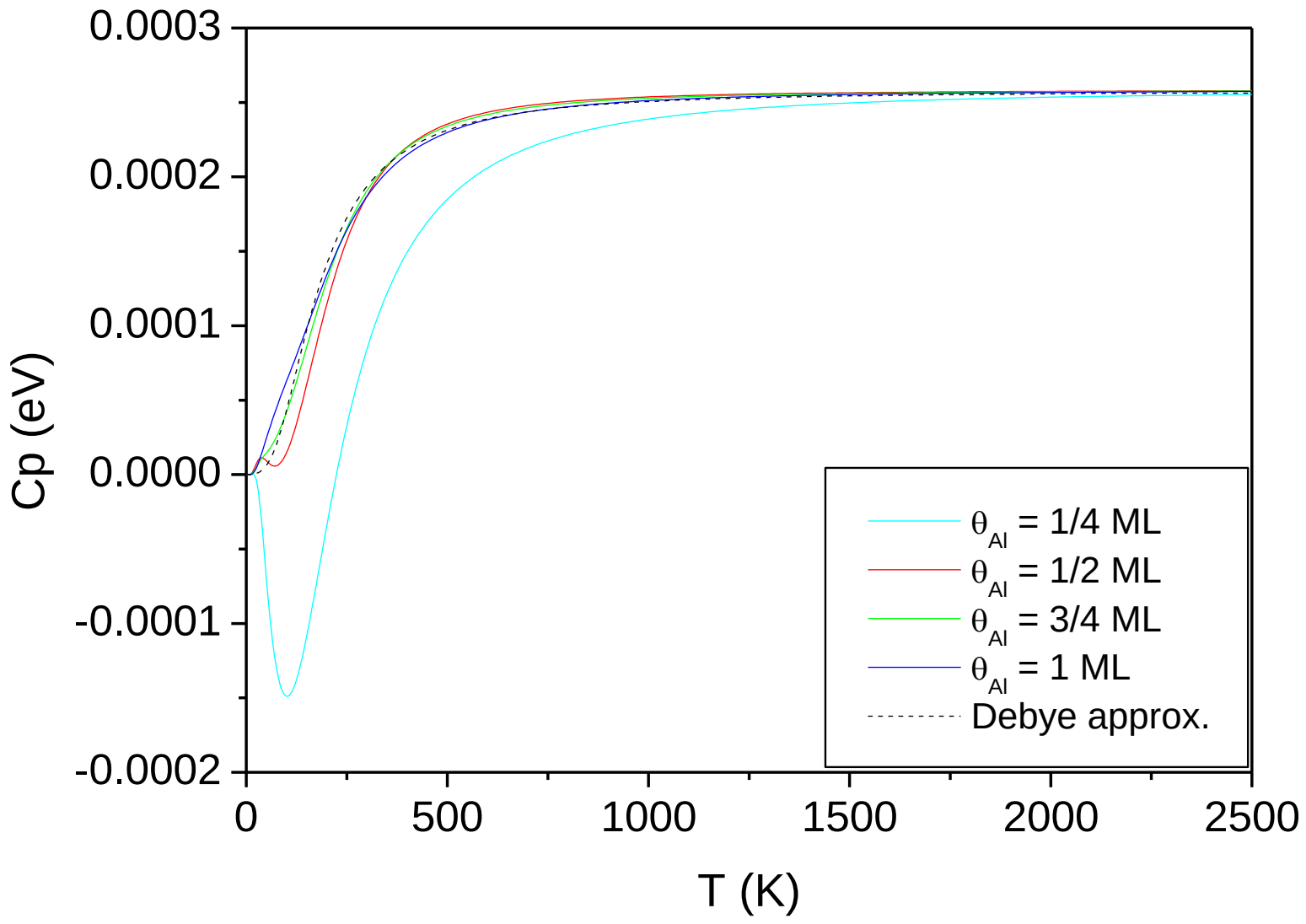

Fig. 25. Temperature dependence of the specific heat related to Al adatoms, attached to $4 \times 4$ slab representing AlN(0001) surface, obtained from phonon frequency summation (solid lines). For the comparison, the specific heat obtained for the Debye temperature $\theta_D = 740$ K is plotted using broken line. Reproduced Fig. 13 from Ref. 158 [158].

The differential specific heat was calculated as the difference between the specific coverage of the surface and the clean surface. This value becomes negative for a low Al coverage at low temperatures. The increase in the heat difference owing to the attachment of Al adatoms is compensated by the negative contribution related to the adsorption that stiffens the lattice, so that the overall contribution is negative. This is observed only at low temperatures; for higher temperatures, all degrees are excited, so the increment of the specific heat is positive.

The thermodynamic quantities of the vapor phase were calculated according to Formulation (i) and tabulated for a large number of substances that could be used to derive effective formulae for enthalpy, entropy, and free energy. These tables provide thermodynamic data for all elements and compounds in two basic states of matter: vapor at normal pressure $p = 1bar$ and the condensed phase (solid/liquid) up to 3000 K. Therefore, these precise data are sufficient for the estimation of the vapor phase contribution.

Accordingly, a full determination of the equilibrium is possible. This was first formulated for 3-d transitions by Jackson and Walsh [159] and later independently for surface adsorption (2-d) by Kempisty et al. [160]. Remarkably, the results formulated as the solid/surface-vapor chemical potential difference, $\Delta\mu_{sv}$, were identical. Here, we present the chemical potential difference (which is zero at equilibrium) formulae for the dissociative adsorption of molecular hydrogen ($H_2$) at the GaN(0001) gallium-terminated gallium nitride surface [160]:

$$\Delta\mu_{sv}(p,T) = \mu_{H(s)}(T,x) - \tfrac{1}{2}\mu_{H_2(v)} = \Delta H_{dis}^{DFT}(0) + \Delta G_{ads} + \Delta H_{therm} + \Delta G_{S-therm} + \Delta G_{pres} + \Delta G_{conc}(T,\theta_N) = \Delta H_{ads}^{DFT} - T_o\Delta s_{sv} + \int_0^{T_o}(C_s - C_v)dT - \int_{T_o}^{T}(s_s - s_v)dT + \int_{p_o}^{p}(v_s - v_v)dp + k_B T ln\left(\theta_N/1-\theta_N\right) = 0 \quad (10)$$

where the surface-vapor chemical potential difference $\Delta\mu_{sv}$ is expressed as sum of:

i) enthalpy change at adsorption $\Delta H_{ads}^{DFT} = \Delta E_{DFT}^{ads-GaN}(H_2) + \Delta E_{sv}^{ZPE}$, taken as the sum of the adsorption energy and zero-point energy difference. The former was obtained from *ab initio* calculations and is given by Eq. 4. Note that this quantity undergoes a jump at $\theta_H = \theta_{H-cr} = 0.75\,ML$. The latter is given by $\Delta E_{sv}^{ZPE} = 2E_s^{ZPE}(H) - E_v^{ZPE}(H_2)$. This value was obtained from *ab initio* phonon calculations $\Delta E_{sv}^{ZPE} = -0.21\,eV$ [158] and is much smaller than the enthalpy change.

ii) The free energy $\Delta G_{ads} = -T_o\Delta s_{sv}$ is related to the entropy change at adsorption, which could be obtained as the difference of entropies of nitrogen at the surface and in the vapor $\Delta s_{sv} = 2s_s - s_v$, where the solid phase entropy is due to phonon contribution only as the solid phase entropy is zero; thus, the free energy contribution is therefore $\Delta G_{ads} = 0.633\,eV$ [156].

iii) the thermal enthalpy change $\Delta H_{therm} = 2h_s(T_o) - h_v(T_o) = \int_0^{T_o}(2C_s - C_v)dT$ in which the enthalpy of the adsorbate is $h_s(T_o) = 1.71 \times 10^{-4}\,eV$ and $h_v(T_o) = 8.99 \times 10^{-2}\,eV$ that gives $\Delta H_{therm} = -0.051\,eV$ [158]. This component is relatively small.

iv) The thermal entropy change $\Delta G_{S-therm} = \int_{T_o}^{T}(2s_s - s_v)dT = \left[2[\mu_{H-s}(T) - \mu_{N-s}(T_o)] - \mu_{H_2}(T) - \mu_{H_2}(T_o)\right]$ can be obtained from the free energy dependence of ideal $H_2$ vapor at 1 bar, given as [160]:

$$\mu_{H_2}(T) = -1.22 \times 10^{-3}T - 0.26 \times 10^{-6}T^2 + 2.4 \times 10^{-11}T^3 \quad (10)$$

whereas the chemical potential of nitrogen at the surface can be approximated to the Debye thermal dependence as

$$\mu_{H-s}(T) = 7.56 \times 10^{-6}T - 2.38 \times 10^{-7}T^2 + 3.26 \times 10^{-11}T^3 \quad (11)$$

which combined $\Delta G_{S-therm}(T) = 2\mu_{H-s}(T) - \mu_{H_2}(T)$ to give

$$\mu_{H_2}(T) - 2\mu_{H-s}(T) = -1.23 \times 10^{-3}T - 2.82 \times 10^{-7}T^2 + 1.54 \times 10^{-11}T^3 \quad (12)$$

that determines this value at any T. At $T = 1300\ K$ this contribution is $G_{S-therm}(1300\ K) = -1.497\ eV$.

v) pressure-dependent term can be used in ideal gas approximations, as the typical nitride vapor growth conditions correspond to high temperature and relatively low pressure: $\Delta G_{pres} = k_B\ T\ ln(p_{N_2}/p_o)$ where $p_o = 1\ bar$.

vi) configuration term $\Delta G_{conc}(T, \theta_N) = k_B T\ ln\left(\theta_N/1 - \theta_N\right)$. The term is responsible for the Langmuir singularity at $\theta_N = 1$, therefore in the vicinity of full coverage, it can attain extremely high values.

Overall, several terms contributed to similar values. The pressure-dependent term can be used to determine the equilibrium pressure of hydrogen on the GaN surface at any temperature according to the following formula [156]:

$$\frac{p}{p_o} = \left(\frac{\theta_H}{1-\theta_H}\right)^2 exp\left\{\frac{-[\Delta H_{vap} + \Delta H_{therm} + \Delta G_{S-vap} + \Delta G_{S-therm}]}{k_B T}\right\} \quad (14)$$

The results of these calculations are given at Fig. 26.

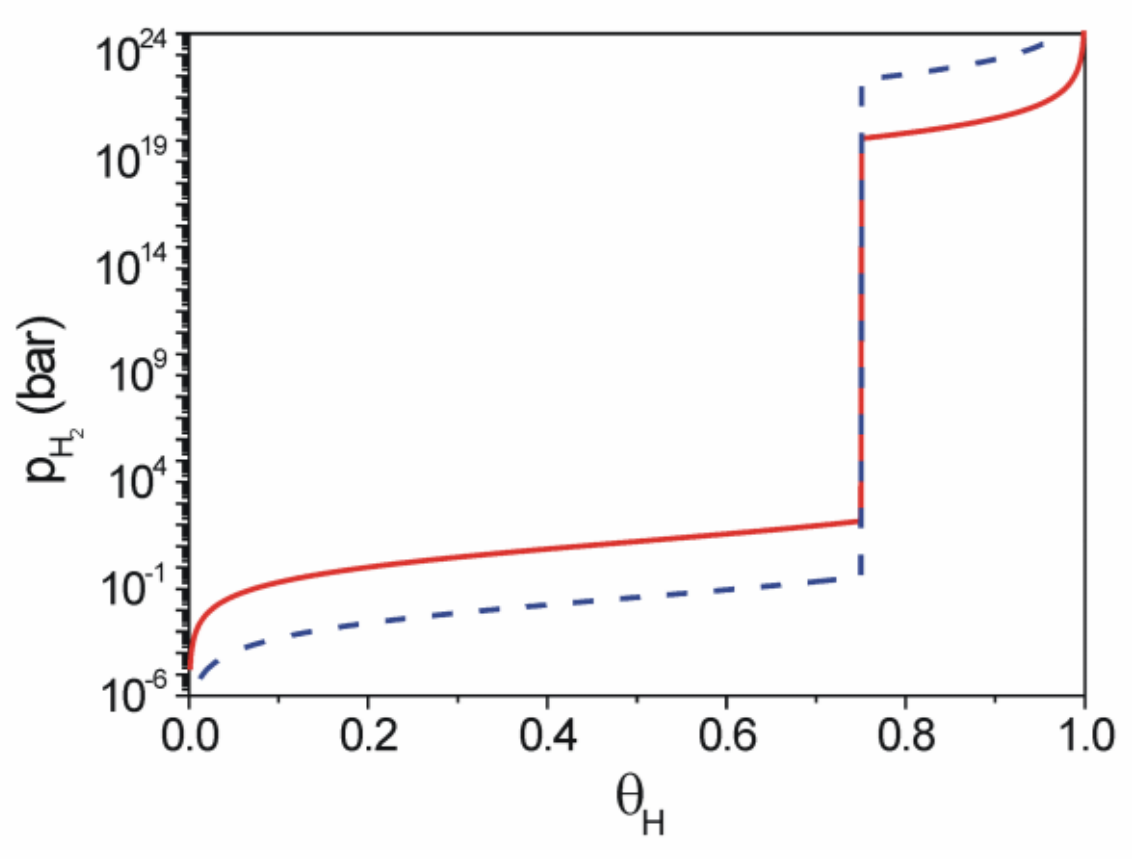

Fig. 26. Equilibrium pressure of molecular hydrogen at GaN(0001) in function of hydrogen coverage. The two lines: dashed blue and solid red correspond to 1000K and 1300K, respectively. Reproduced Fig. 2 from Ref. 156 [156].

From these data, it follows that the equilibrium pressure of the vapor strongly depends on the temperature, which is related to the thermal contribution to the chemical potential difference (ii-v). Nevertheless, the most important is a jump-like change at the critical coverage $\theta_H = \theta_{H-cr} = 0.75\ ML$ where the pressure change is the greatest. Additionally, at both ends, that is, $\theta_H = 0$ and $\theta_H = 1\ ML$, the entropy contribution diverges asymptotically, as shown in Fig. 26. In summary, the charge-related term is the most important factor affecting the equilibrium between molecular hydrogen and GaN(0001) surface. In the following, we discuss the vapor-solid equilibria historically, showing that this phenomenon was identified in many systems, changing the description of these systems.

## 2. Historical resume and material review

Historically, the contribution of intra-surface electron transfer was first obtained in a study devoted to the determination of the adsorption energy of hydrogen at the GaN(0001) surface, published in 2012 [150]. Specifically, the adsorption energy of single hydrogen atom at the surface was about 3.4 eV for hydrogen coverage below 0.75 monolayer ($\theta_H < 0.75\ ML$), and was reduced to 1.4 eV for higher coverage ($\theta_H = 0.75\ ML$) [150]. The first diagram showing the change in the adsorption energy is shown in Fig. 27. In the following years, far more precise and extensive calculations of hydrogen adsorption have been published, proving that the energy gain follows Eq. 4a and Eq. 4b for atomic and molecular hydrogen atoms, respectively [149]. It was shown that far away from the critical coverage (i.e., $\theta_H = 0.75\ ML$, the adsorption energy is essentially constant and independent of the doping, that is, H adatoms effectively do not interact with the neighbors. In the vicinity of the critical coverage, that is, for the Fermi level not pinned at the surface, the adsorption energy depends on the doping in the bulk, finally confirming the electron transfer contribution.

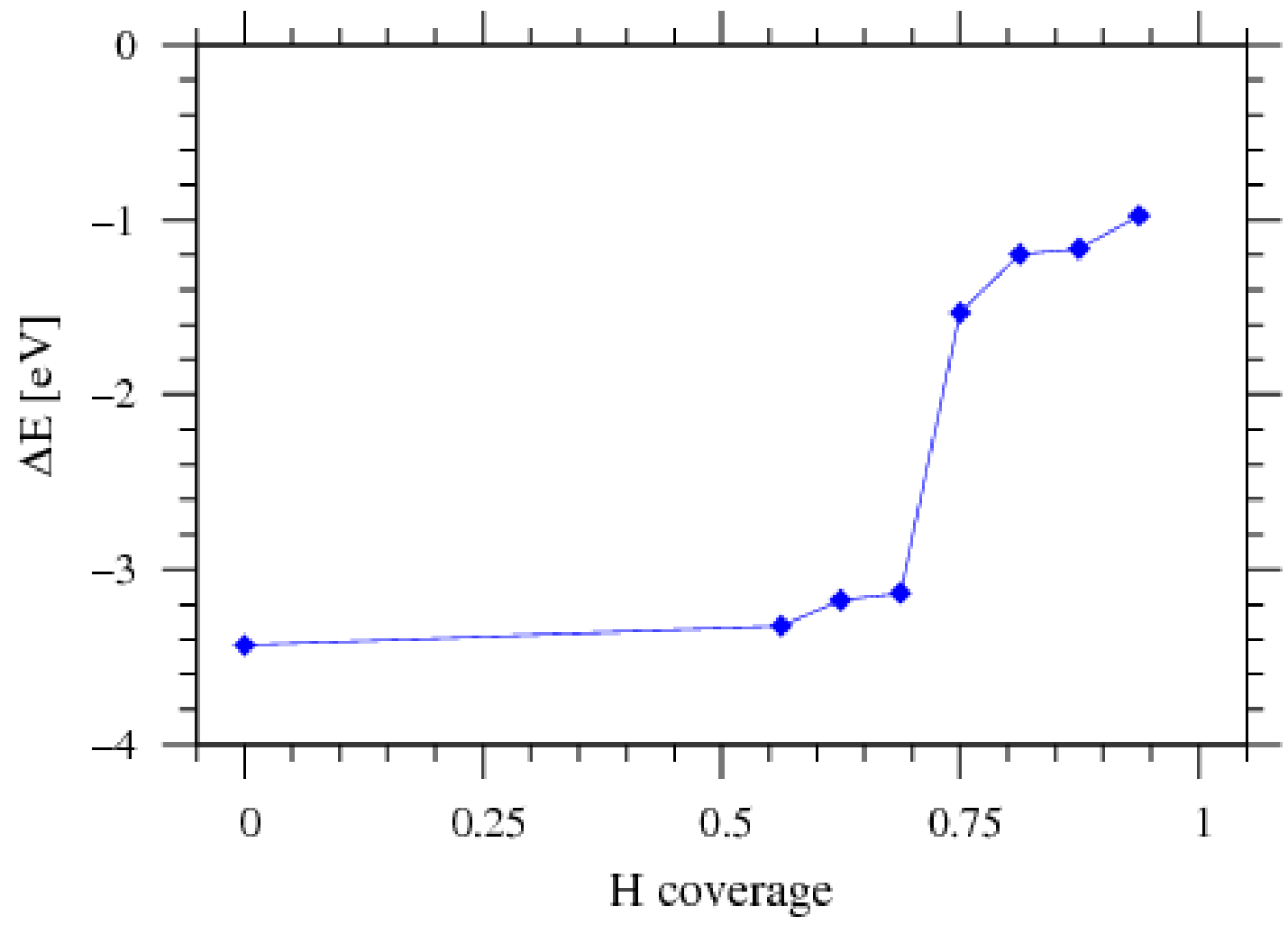

Fig. 27. Adsorption energy of single hydrogen atom, at differently covered GaN(0001) surface, obtained using 4 x 4 supercells. Reproduced Fig. 5 from Ref. 150 [150].

Somewhat later, a phonon-related contribution was introduced, so that the free energy of the hydrogen adsorbate was temperature-dependent [155]. Subsequently, the full contribution to the chemical potential of both the adsorbate and vapor was determined to obtain a complete equilibrium [156]. Therefore, hydrogen vapor adsorption on the Ga-terminated GaN(0001) surface is fully determined by the surface coverage, temperature, and pressure of the vapor.

Several months after the first publication of hydrogen adsorption energy dependence on coverage [150], Chen and Kuo published an investigation into the wetting of GaN(0001) surfaces [144]. They calculated the energy change during the adsorption of water molecules on the surface in three different forms:

i) molecular,

ii) fully dissociated water,

iii) molecular hydrogen plus O adatom.

They observed some changes in the adsorption energy during the adsorption of four water molecules. These changes were relatively small and of different natures, and were generally smaller for the set of four molecules. No clear profound differences were identified. The authors speculated that this was due to the reconstruction contribution, which was discussed earlier. This identification is related to the reconstruction contribution, and the magnitude of the effect is different from the later confirmed contribution from charge transfer. Therefore, a clear identification of this effect was not obtained. Subsequently, the reconstruction hypothesis was shown to be incorrect.

Molecular hydrogen adsorption on the opposite N-terminated GaN (000-1) surface has also been investigated, showing that the hydrogen adsorption process is dissociative and barrierless [161]. Hydrogen atoms are adsorbed at the on-top position; therefore, the EECR unit has a single N topmost atom ($z = 1$). Their fractional charge is $q_i = 5/4$ thus, $Q_i = q_i = 5/4$. As shown, the N-terminated surface is different from the Ga-terminated surface because the N-broken bond state energies are located close to the VBM [90]. Thus, the Fermi level was pinned at the clean GaN(000-1) surface. The number of states before and after adsorption was identical, $N_a = N_f = 2$. Thus from Eq. 2b we obtain $g_{cr} = \theta_{cr} = \frac{3}{4}$. In other words, the addition of a hydrogen charge $Q_a = 1$ creates a surplus of $1/4$ electrons that are donated to the other N-broken bond states that are not saturated. As the N-broken bond state can accept $3/4$ electrons, three H-adsorbed sites are required for full saturation of a single N-broken bond, so that the critical condition for filling all broken states is at $\theta_H = \theta_{H-cr} = 0.75\, ML$ [161]. Thus, the energy change in the adsorption of hydrogen molecules on the fractionally H-covered GaN(000-1) surface is

$$\Delta E_{DFT}^{ads-GaN}(H_2) = \begin{cases} -7.04\, eV & \theta_H < 0.75\, ML \\ -1.60\, eV & \theta_H > 0.75\, ML \end{cases} \tag{15}$$

The jump is considerable; for high coverage, the $1/2$ electrons are shifted to the conduction band across the bandgap. These data were used to estimate hydrogen equilibrium pressure at the benchmark temperature $T = 1300\, K$ via simple van t'Hoff approach to be $p \cong 10^{-21} bar$ and $p \cong 1\, bar$ for $\theta_H < 0.75\, ML$ and $\theta_H > 0.75\, ML$, respectively. Thus GaN(000-1) surface rapidly adsorbs molecular hydrogen, which can lead to decomposition via the desorption of ammonia. The high coverage of hydrogen is therefore not accessible for the experimental investigation of hydrogen at the GaN(000-1) surface, unless ammonia ambient is used. To date, no further theoretical investigations of this process have been conducted. Nevertheless, scarce data confirm the decisive role of charge transfer in this process.

Epitaxial processes, widely employed for the creation of nitride device structures, use either plasma-activated nitrogen [162,163] or ammonia [164-166] as active nitrogen source. The growth is conducted almost exclusively on the GaN(0001) surface by metal organic vapor phase epitaxy (MOVPE) because the layers grown on this surface have superior optical properties and are useful for optoelectronic devices. Therefore, the adsorption of ammonia onto this surface is of considerable interest and has been investigated by many researchers. The early stages of these investigations were summarized by Van de Walle and Neugebauer in their Letter [167]. They presented a phase diagram for the GaN(0001) surface, in which the nitrogen-hydrogen-rich region of the surface can be divided as follows:

i) hydrogen covered at $\theta_H = 0.75\ ML$

ii) mixed $HN_3 - NH_2$ covered at $\theta_{NH_3} = 0.25\ ML$ and at $\theta_{NH_2} = 0.75\ ML$

iii) mixed $HN_3 - H$ covered at $\theta_{NH_3} = 0.25\ ML$ and at $\theta_H = 0.75\ ML$.

The remaining part of the diagram corresponds to a Ga-rich diagram, which is not relevant to ammonia adsorption. This diagram was obtained using $(2\ \times 2)$ slab; therefore, these data indicate only the basic features. In fact these values correspond to the critical coverages at which the Fermi level becomes free as it was proved above for the case of hydrogen $(\theta_{H-cr} = 0.75\ ML)$ and mixture of $NH_2$ radicals $\left(\theta_{NH_2-cr} = 0.75\ ML\right)$ and $NH_3$ admolecules $\left(\theta_{NH_3-cr} = 0.25\ ML\right)$. As it will be shown below, from the point of the charge $NH_2$ radicals and equivalent to H adatoms, therefore the mixture $(\theta_{H-cr} = 0.75\ ML)$ and $NH_3$ admolecules $\left(\theta_{NH_3-cr} = 0.25\ ML\right)$ is also critical. In accordance with the above arguments, the coverage indicated in the phase diagram is therefore the point at which the change in electron transfer causes a jump-like change in the adsorption energies.

The following investigations prove that the data presented in Refs. 167 are basically correct, but overly simplify the issue. In fact, they correspond to relatively high pressures of ammonia and hydrogen in vapor. Several years later, detailed investigations began on ammonia adsorption on a clean GaN(0001) surface. Because ammonia decomposition leads to hydrogen release, the incorporation of hydrogen coverage into $NH_3$ investigations is indispensable. It was shown that two different adsorption scenarios are possible: ammonia can be adsorbed molecularly with the energy $\Delta E_{DFT}^{ads-mol}(NH_3) = 2.0\ eV$ or dissociatively with $\Delta E_{DFT}^{ads-mol}(NH_3) = 2.6\ eV$ for hydrogen coverage $\theta_H\ <\ 0.75\ ML$ [148]. For higher coverage, the molecular adsorption energy remained unchanged, whereas the dissociative adsorption energy became negative [168]. As shown in Fig. 21, all states of the $NH_3$ admolecule degenerate with the valence band, so they are occupied. Ammonia molecular adsorption is neutral because it brings eight electrons to occupy all states. Therefore, $3/4$ electrons were shifted to the conduction band. Thus, the EERC condition is not possible, and the adsorption energy is reduced by electron transfer for any hydrogen coverage. This was confirmed by *ab initio* adsorption energy presented in Fig 28. As shown, the molecular ammonia adsorption energy was independent of the coverage for the entire simulated hydrogen coverage range of up to 1 ML.

Dissociative adsorption creates $NH_2$ admolecules and H adatoms. Both cases were equivalent thus $1/4$ electrons were missing. Therefore, the number of electrons transferred from the Ga broken bond state is doubled, which substantially contributes to the adsorption energy. The energy difference at the critical coverage $\theta_{cr-H} = 3/4$ is close to 5 eV, as shown in Fig 28. [170].

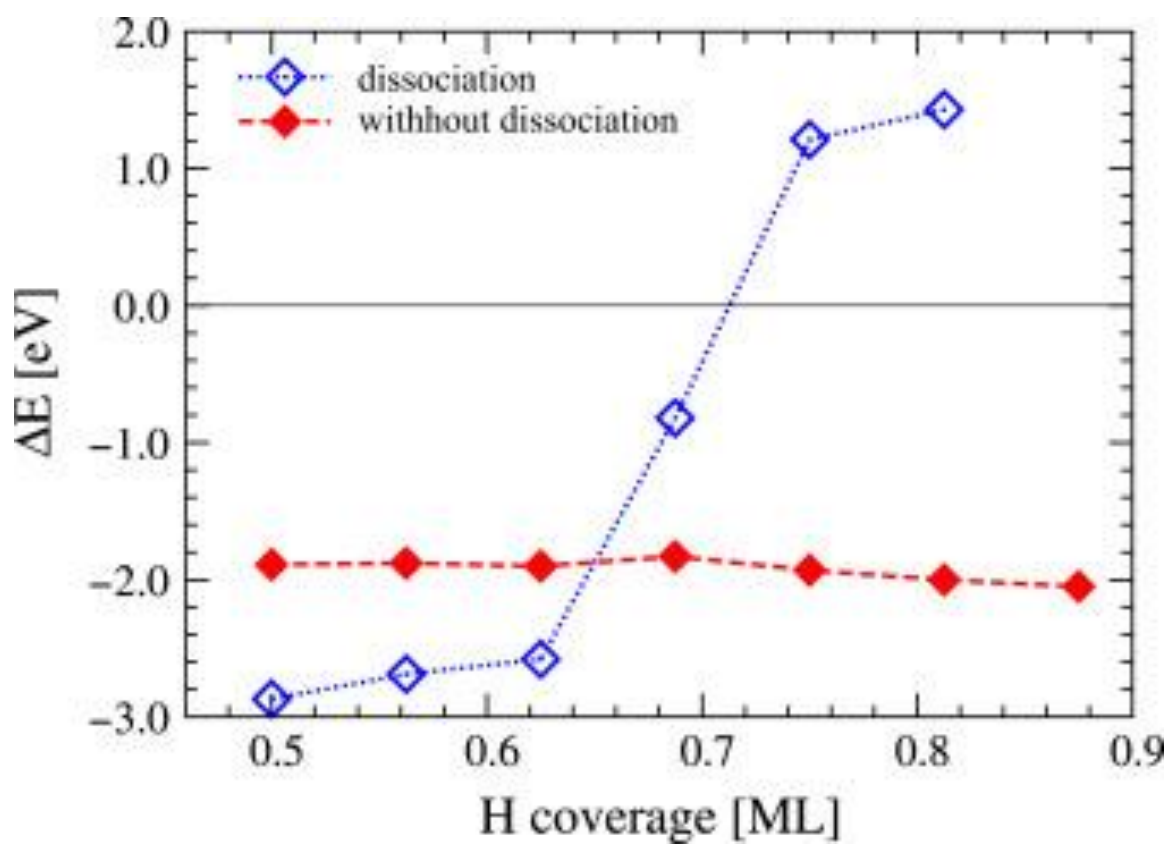

Fig. 28. Adsorption energies of ammonia molecule on relatively highly hydrogen covered GaN(0001) surface: blue – dissociative and red – molecular process. Reproduced from Fig. 3 Ref. 168 [168].

These data prove that the variation in adsorption energy is not due to adsorbate interactions, as this is similar for both cases. This difference is related to the electron contribution being controlled by the Fermi level position.

More detailed investigations revealed a complex pattern of ammonia dissociative adsorption at the GaN(0001) surface, particularly with low surface coverage [134, 149]. Several configurations are possible.

i) NH radical in H3 position
ii) NH radical on-top
iii) $NH_2$ radical – bridge
iv) $NH_2$ radical – on top

Based on the data in Ref. 149, all the states of these radicals degenerate with the valence band [149]. In the case of the on-top position, the unit was a single site, $(z = 1)$. Thus, charge is $Q_i = 3/4$. In the case of the NH radical, the adsorbate charge was $Q_a = 6$. The number of initial occupied states after adsorption is $N_f = 0$, and the configuration number of states after adsorption $N_a = 8$. This leads to NH critical coverage at coverage $\theta_{cr-NH-top} = 3/8$. And accordingly, the adsorption energy gain of dissociative adsorption of ammonia at $\theta_{NH-top} < 0.375\ ML$ is $\Delta E_{NH-top}^{DFT} \cong 1.42\ eV$. For a higher coverage, this gain is negative.

In the case of the NH radical located at the H3 position, the unit contains three Ga sites, that is, $(z = 3)$, so that the initial charge $Q_i = 9/4$. The adsorbate charge was $Q_a = 6$. The number of occupied initial states after adsorption was $N_i = 0$, configuration number of occupied states after adsorption $N_a = 8$. This led to a critical fraction, $g_{cr-NH-H3} = 9/8$. Accordingly, the critical coverage is $\theta_{cr-NH-H3} = 3/8$. Therefore, this energy jump should be observed $\theta_{NH-top} \cong 0.375\ ML$. This is in accordance of general result, the adsorption energy gain of dissociative adsorption of ammonia at $\theta_{NH-top} < 0.375\ ML$ is $\Delta E_{NH-top}^{DFT} \cong 1.42\ eV$. For a higher coverage, this gain was negative.

In the case of the $NH_2$ radical at the bridge position, the unit contains two Ga sites, that is, $(z = 2)$, so that the initial charge $Q_i = 3/2$. The adsorbate charge was $Q_a = 7$. The number of occupied initial states after adsorption was $N_f = 0$, configuration number of occupied states after adsorption was $N_a = 8$. This leads to a critical fraction $g_{cr-NH2-bridge} = 3/2$. Accordingly, the critical coverage was $\theta_{cr-NH2-bridge} = 3/4$. This was not possible because the top coverage was $\theta_{cr-NH2-bridge} = 1/2$.

Finally, in the case of the $NH_2$ radical at the on-top position $(z = 1)$, the initial charge is $Q_i = 3/4$, and the adsorbate charge is $Q_a = 7$. The number of occupied initial states after adsorption was $N_f = 0$, configuration number of occupied states after adsorption was $N_a = 8$. This led to a critical fraction, $g_{cr-NH2-on-top} = 3/4$. Accordingly, the critical coverage is $\theta_{cr-NH2-on-top} = 3/4$. Therefore this energy jump should be observed $\theta_{NH2-on-top} \cong 0.75\ ML$. However, no such investigation has been conducted because this configuration is unstable.

These investigations were completed before a full thermodynamic procedure for the determination of vapor-surface equilibria was formulated [155,156]. Therefore, the thermal contribution was limited to the configurational entropy. Using this, an estimate of the mixed hydrogen-ammonia pressure in equilibrium with the specified coverage of GaN(0001) was obtained [149]. The coverage by both ammonia-derived radicals, NH (on-top) and $NH_2$ (on-top and bridge), is limited to hydrogen pressures below $p_{H_2}(min) = 10^{-5}\ bar$ and ammonia pressures below $p_{NH_3}(min) = 10^{-2}\ bar$. These conditions cannot be achieved separately in ammonia growth of GaN due to thermal decomposition of ammonia; therefore, this suggests that the standard coverage of the GaN(0001) surface is full $NH_2 - NH_3$ mixed coverage. The relative ratio of the radicals and molecules depends on the ammonia-to-hydrogen pressure ratio. It must be stressed that this is a rough estimate, and precise determination of the coverage at given temperature–pressure conditions requires a number of simulations in the future. These investigations include phonon simulations, which are extremely demanding. On the other hand, the complex atomic structure of the adsorbate indicates a much larger phonon contribution to the free energy, which affects the equilibrium much more than in the case of hydrogen. Thus, it is expected that the phonon contribution is comparable to the adsorption energy. Therefore, the obtained equilibria may be modified considerably. Because the MOCVE and HVPE methods are very important technologically, these calculations are needed in the nearest possible future.

An important example of the influence of the electronic charge on the results of the MOVPE growth of nitrides is the well-known problem of the growth of GaInN necessary for visible-range active LED's and LDs [169-177]. The growth was conducted at a high V-III ratio of approximately $p_{NH_3}/p_{Ga} = 4000$ for GaN and $p_{NH_3}/p_{In} = 10000$ for InGaN. In addition, the incorporation of indium into the nitride layers was completely blocked by the minimal presence of hydrogen in the vapor. Thus, in InGaN epitaxy, molecular nitrogen is used as a group III metal carrier gas and ammonia is the second main component. The typical pressure ratio of ammonia to the carrier gas is $p_{NH_3}/p_{H_2} = p_{NH_3}/p_{N_2} = 1$. Nitrogen is essentially inert; therefore, the vapor consists of two active components: ammonia and molecular hydrogen, which are generated by thermal decomposition of ammonia. The total vapor pressures used in the MOVPE processes is in the $(0.1\ bar \div 1\ bar)$ range. Thus, the ammonia partial pressure was within the same range. In the case of the nitrogen carrier gas, the hydrogen pressure is a small fraction of this value.

In the *ab initio* calculations, it was assumed that the mixed $NH_2 - NH_3$ coverage of the GaN(0001) surface is full, that is, all topmost Ga atoms have broken bonds saturated either by $NH_2$ radicals or $NH_3$ admolecules located in the on-top positions. The critical $NH_3$ fraction is 0.25 ML $\left(\theta_{NH_3} = 1/4\right)$ for which the Fermi level is free in the gap. For lower $\left(\theta_{NH_3} < 1/4\right)$ and higher $\left(\theta_{NH_3} > 1/4\right)$ ammonia fraction the Fermi level was located at the VBM and CBM.

The indium carrier trimethylindium (TMI) is highly unstable at temperatures $T > 600\ °C$. Therefore in the typical GaInN growth conditions at the temperature $T \cong 800\ °C$ it is decomposed and single In atom is adsorbed. For low coverage, the $\left(\theta_{NH_3} < 1/4\right)$ In atom is located at the T4 position. In *sp*[3] orbitals, overlap with $NH_2$ radical orbitals creates states in the valence band, that is, they are occupied. The upper *sp*[3] orbital was not occupied. An increase in the ammonia ratio leads to a shift in the In position to a bridge, the shift of the Fermi level to the conduction band, and a drastic reduction in the In bonding energy. Therefore, the adsorption energy of indium is dependent on GaN(0001) surface coverage. This contribution is important, as the number of electrons shifted may reach three. The adsorption energy as a function of coverage is shown in Fig. 29. It should be noted that this relation is jump-like, but owing to the limited size of the slab, these jumps were not obtained.

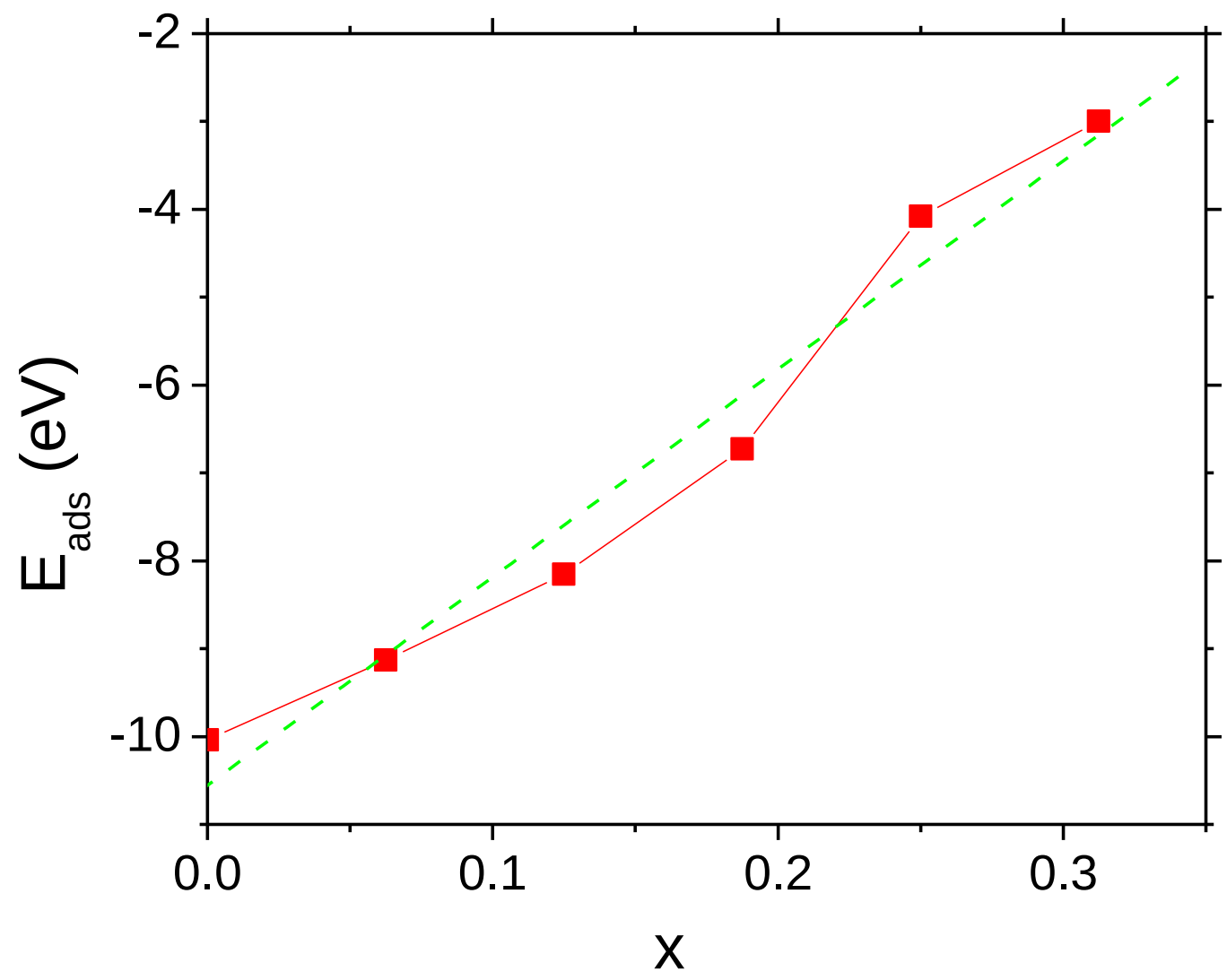

Fig. 29. Indium adsorption energy in function of the $NH_3$ molecules $\left(n_{NH_3}\right)$ to $NH_2$ radicals $\left(n_{NH_2}\right)$ ratio (x), simulated using $4 \times 4$ GaN slab. All sites on-top of top Ga atoms of GaN(0001) surface are occupied by $NH_3$ molecules or by $NH_2$ radicals. Green dashed line represent the linear approximation in Eq. 15. Reproduced from Fig. 8 Ref 169 [169].

These data were approximated by the following expression

$$E_{ads}^{DFT}(x) = -10.55 + 23.68\,x \tag{16}$$

where the energy is expressed in electronvolts. The difference between the n- and p-type bulk material was negligible. As shown, the dependence is steep; thus, the energy gain is small for a higher ammonia coverage. The thermodynamics of In adsorption were also investigated, showing a drastic reduction in the probability of In adsorption for increased hydrogen pressure.

This was confirmed by MOVPE growth experiments, where a slight presence of hydrogen blocks the incorporation of In into the growth layers [174–176]. Moreover, this effect was used in the technology because the growth of In-free GaN does not require the termination of In flow. The addition of hydrogen to the flow is much faster, which leads to growth of GaN layers [174–177].

In recent years, an important part of nitride investigations has been redirected towards shorter wavelength devices that are active in the UV range [178-181]. This process entails the growth of Al-rich structures and AlN substrates. AlGaN layers crack under tensile strain when grown on GaN substrates; therefore, the growth of high-quality AlN single crystals is indispensable [182-184]. The most promising method is physical vapor transport growth of Al and $N_2$ [185]. These crystals were grown on two polar AlN surfaces: Al-terminated AlN(0001) and N-terminated AlN(000-1). Therefore, the adsorption of both Al and $N_2$ was investigated by *at initio* methods [185-187]. It is important to note that these processes are hydrogen-free.

The adsorption of nitrogen on an AlN(0001) surface was first investigated [185]. The bare AlN surface had no reconstruction with a partially filled Al-broken bond state located below the CBM. Molecular nitrogen is adsorbed dissociatively without any energy barrier at the clean AlN(0001) surface in the H3 position by a very complex path, as shown in Fig. 30.

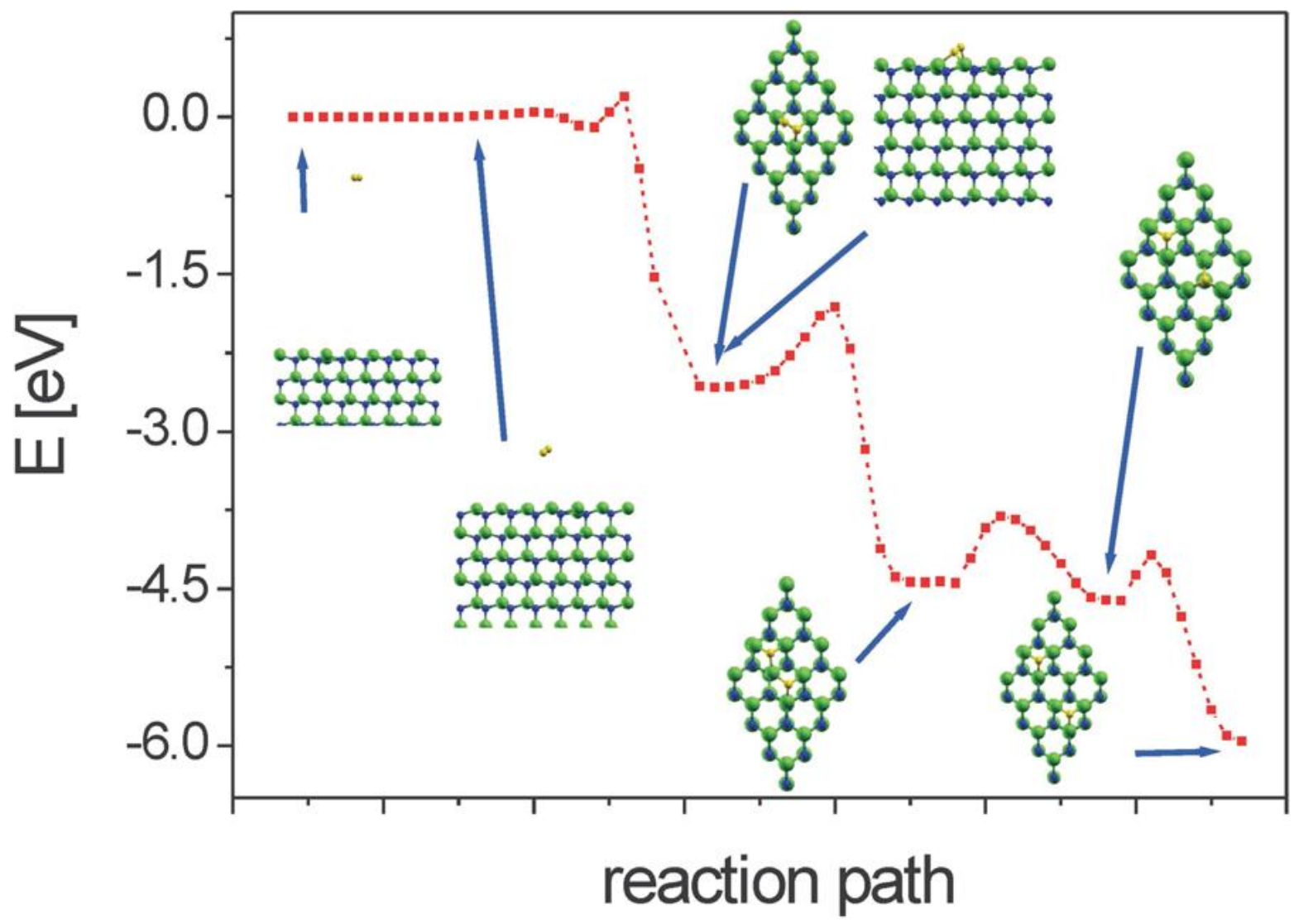

Fig. 30. $N_2$-AlN(0001) clean surface slab interaction energy in function of the molecule reaction path. The configuration of the molecule is presented in the inserted diagrams. The molecule undergoes dissociation before reaching the minimum energy point. The energy level is set to zero for asymptotically far distance. Reproduced from Fig. 8 Ref. 185 [185].

The adsorbed nitrogen atom *Ns* state is located deep in the valence band, whereas *Np* states are located in the bandgap, below the Al-broken bond state. Thus, at low coverage, these states are occupied by electrons drawn from the broken Al bond state. The EECR unit may encompass three Al sites: i.e. $(z = 3)$. The broken Al bond charge is $q_{Al-br} = 3/4$ thus, the initial electron charge of the unit is $Q_i = 9/4$. The adsorbed N atoms bring five electrons, that is $Q_a = 5$. The number of initial states occupied after adsorption was $N_f = 0$. The number of occupied states after adsorption is $N_a = 0$. Therefore, the EECR condition is $g_N = 3/4$ from which the critical N coverage is $\theta_{N-cr} = 1/4$. The *ab initio* data confirm this prediction; at $\theta_{N-cr} = 1/4$ the Fermi level is at the top of the highest-energy N state [185]. Accordingly, the adsorption energy of molecular nitrogen jumps at this coverage, as shown in Fig. 31.

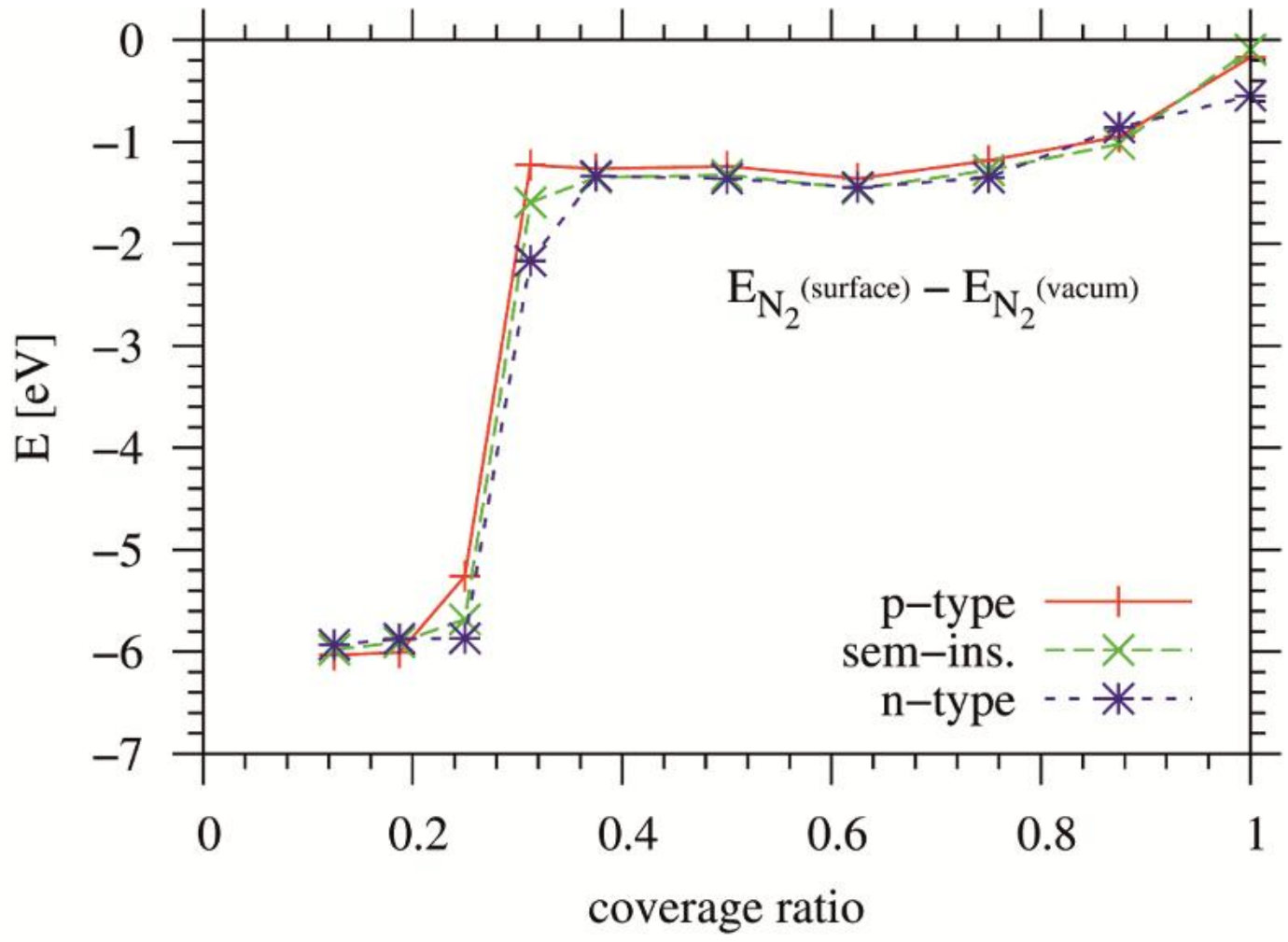

Fig. 31. Adsorption energy of $N_2$ molecule in function of N adatom coverage ratio, calculated using an energetically stable configuration: for $\theta_N < 0.25\ ML$ - dissociative process leading to concentration of separate N atoms in H3 position, for $\theta_N > 0.25\ ML$ - the excess nitrogen is in the form of $N_2$ admolecule in a skewed-vertical position located on top of an Al surface atom. The colors present data for a slab representing the following dominant doping in the bulk: *p*-type, *semi*-insulating , and *n*-type. Carrier concentration was set to 2.4 x $10^{20}$ $cm^{-3}$. Reproduced from Fig. 9 Ref. 185 [185].

The obtained adsorption energy change for $N_2$ molecule could be summarized as follows:

$$\Delta E_{DFT}^{ads-AlN}(N_2) = \begin{cases} -6.0\ eV & \theta_N < 0.25\ ML \\ -1.0\ eV & \theta_N > 0.25\ ML \end{cases} \tag{17}$$

For $\theta_N < 0.25\ ML$ the adsorption is dissociative; for the remaining part, the adsorption has no electron transfer contribution, which lowers the energy gain by 5.6 eV which is much lower than the dissociation energy of the nitrogen molecule. Therefore, for $\theta_N > 0.25\ ML$ adsorption is molecular with an energy gain of 1.0 eV. Thus, the drastic difference in the adsorption type and energy is related to the electron transfer. From this, the equilibrium pressure of nitrogen was estimated without full thermodynamic analysis. The result the saturation nitrogen pressure below 0.01 bar for the nitrogen coverage below critical ($\theta_N < 0.25\ ML$) at PVT AlN growth temperature $T = 2300\ K$. For molecular adsorption ($\theta_N > 0.25\ ML$) the required nitrogen pressure is prohibitively high, of order of $10^5$ bar.

The molecular nitrogen adsorption on the opposite N-terminated AlN(000-1) surface is drastically different [186]. The Clean N-terminated AlN(000-1) surface did not undergo any reconstruction. Molecular nitrogen interacts with the surface via weak van der Waals forces, which are typical for closed-shell systems. Therefore, it is adsorbed molecularly without localization, with an energy gain of approximately $\Delta E_{DFT}^{ads-AlN}(N_2) = 1.0\ eV$. In the case of an Al-covered surface, molecular nitrogen is dissociated with the energy gain given by the relation (in eV):

$$\Delta E_{DFT}^{ads-AlN}(N_2, \theta_N) = 2\Delta E_{DFT}^{ads-AlN}(N, \theta_N) - \Delta E_{DFT}^{diss}(N_2) = -8.32 + 4.88\ \theta_N \quad (18)$$

This reflects the transformation of the metallic surface into semiconducting, gradual downward motion of the Fermi level, and reduction of the electron transfer contribution to the adsorbed N adatom. A single N adatom has eight atomic states, while the number of electrons is $N_e = 7\frac{1}{4}$ thus, $\frac{3}{4}$ electrons are transferred from the top energy states of the Al coverage. Therefore, the change in the adsorption energy is considerable. The thermodynamics of nitrogen adsorption were formulated based on the contribution of vibrational energy to the chemical potential of adsorbed nitrogen at high temperatures. The results show considerable variation in the pressure for limited coverage, which is in agreement with the propensity for the creation of metallic Al inclusions during growth.

The complementary process of Al adsorption at principal AlN surfaces was also investigated [158, 187]. Al was adsorbed on the Al-terminated AlN(0001) surface without any energy barrier. The adsorption energy depends on the Al coverage  as follows:

$$\Delta E_{DFT}^{ads-AlN}(Al) = \begin{cases} -5.06 + 0.02\,\theta_{Al} & \theta_{Al} < 1 \\ -6.02 - 8.35\,(\theta_{Al} - 1) + 47.45\,(\theta_{Al} - 1)^2 & 1 \le \theta_{Al} \le 1.2 \\ -3.43 - 1.64(\theta_{Al} - 1) & \theta_{Al} > 1.25 \end{cases} \tag{19}$$

The energy gain has maximum of about 6.1 eV for $\theta_{Al} = 1.1$. Thus, the contracted layer model formulated for Ga coverage of the GaN(0001) surface is also applicable to AlN(0001) surfaces [158]. No direct confirmation of the role of the electron transfer was observed. Thus, this process is similar to the typical adsorption processes at metal surfaces.

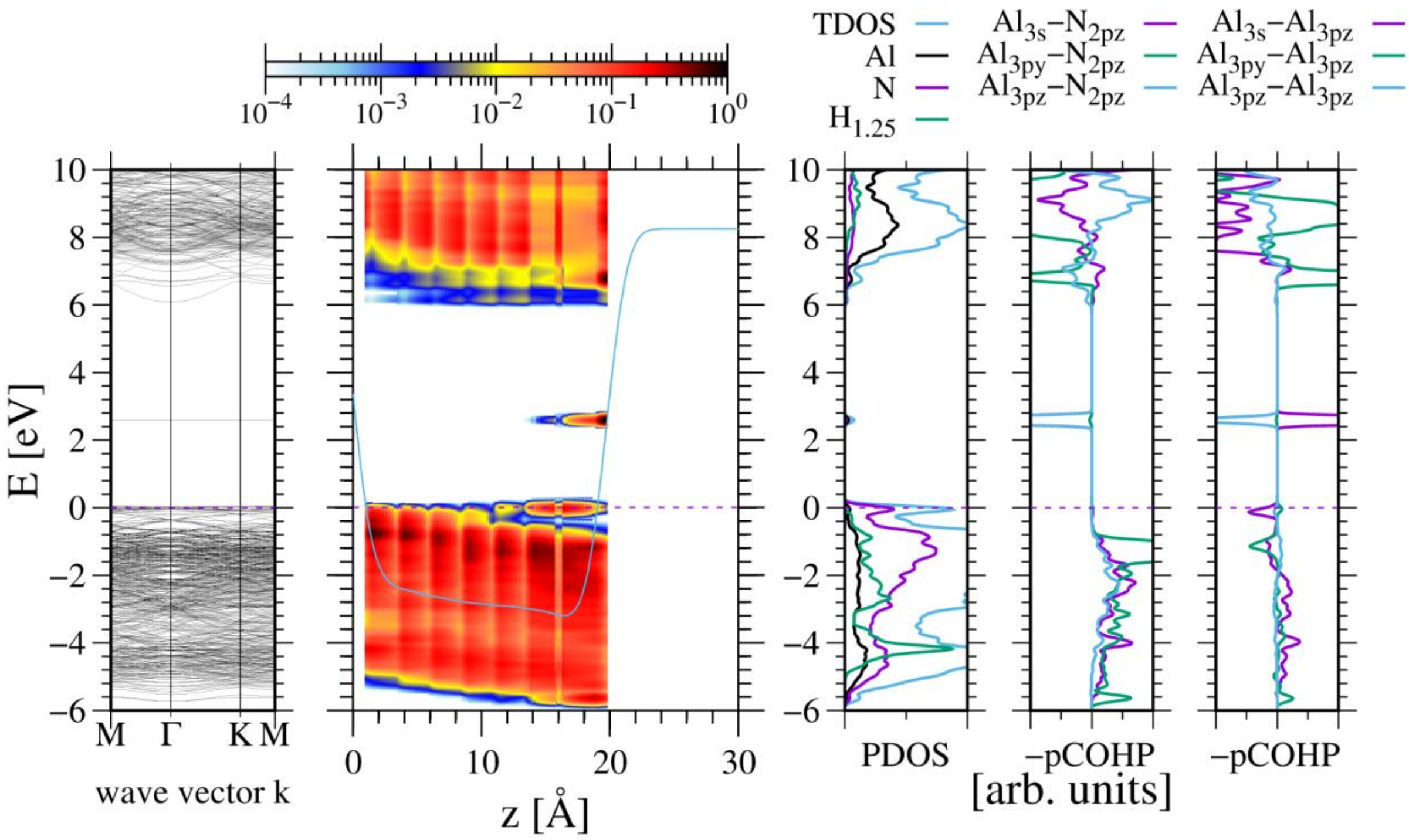

Fig. 32. Electronic properties of 4 x 4 AlN slab with single Al adatom, representing AlN(000-1) surface with $\theta_{Al} = 0.0625\ ML$ coverage. The panels from the left present: electron energy in momentum space (band diagram); electron energy in real space (obtained by projection of the density of states of each atom on the atomic eigenstates and associated with position of the atom), plotted in the direction perpendicular to the surface; projected density of states (DOS), crystal orbital Hamilton population (COHP) [113,114] of Al adatom and neighboring N and Al surface atoms, respectively. The surface is located at the right hand edge of real space diagram, close to 20 Å. Reproduced from Fig. 1 Ref. 187 [187].

Different, complex pictures were obtained in investigations of the adsorption of Al on the opposite, N-terminated AlN(000-1) surface [187]. The electron contribution plays an important role in adsorption. As shown in Fig. 32, Al atoms create bonding states with the three topmost N atoms; thus, the unit has three sites, $(z = 3)$. *Al3s*, *Al3p$_x$* (and equivalently *Al3p$_y$*), and *Al3p$_z$* overlap with the nitrogen atom *N2p$_z$* state, creating bonding states. The first three are degenerate with the VB; thus, they are occupied, but *Al3p$_z$-N2p$_z$* state is located in the bandgap; thus, it is empty. Therefore, the six Al states occupied $(N_a = 6)$. The Al atom contributed three electrons; hence, $Q_a = 3$. The number of electrons in each N surface atom is $q_N = 5/4$ which yields $Q_i = 15/4$. The nitrogen states and full after adsorption $N_f = 6$. Therefore, the EECR critical fraction for the occupation of the six states is $g_{Al-cr} = 3/4$ and accordingly, the critical Al coverage is $\theta_{Al-cr} = 1/4$. Naturally, Al adsorption creates a surplus of $3/4$ electrons, which are shifted to other N-broken bond states that are located close to the VBM. At critical coverage, this charge is sufficient to fill a single N atom broken bond state to two electrons. Thus, both Al states and N broken bond states are occupied, and a further increase in Al coverage leads to electron transfer to higher energy states. This is a considerable energy cost; therefore, for coverage above the critical $\theta > \theta_{Al-cr} = 1/4$ the Al adsorption energy should be lower. The Al adsorption energy as a function of the Al coverage is shown in Fig. 33.

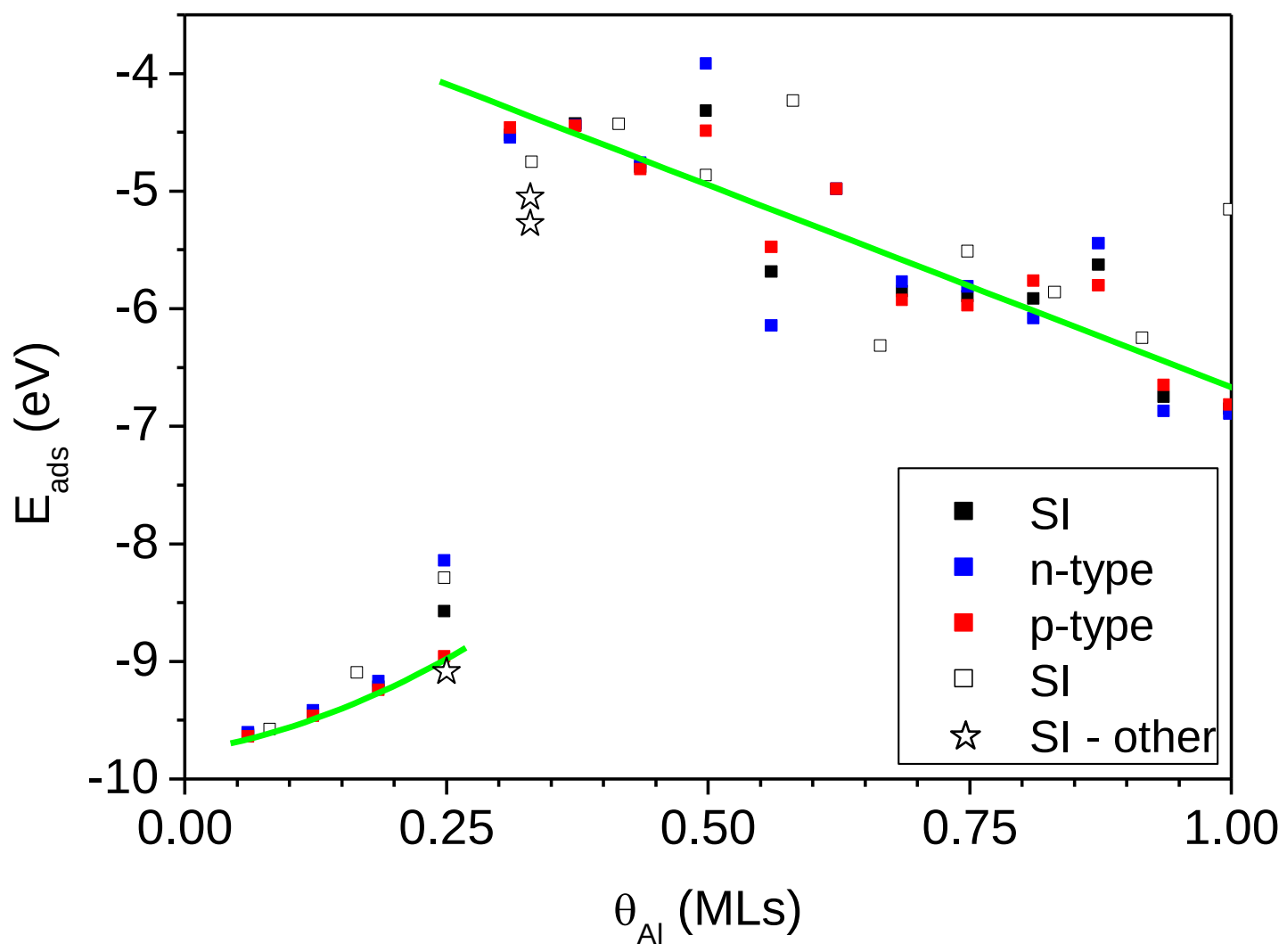

Fig. 33. Adsorption energy of Al atom at AlN(000$\underline{1}$) surface, in function of Al coverage. Full squares represent data obtained using $4 \times 4$ slab, empty squares - $2\sqrt{3} \times 2\sqrt{3}$ slab, empty stars – additional configurations in $2\sqrt{3} \times 2\sqrt{3}$ slab. The coverage on the horizontal axis refers to the value after the process. Green dashed line presents polynomial approximation to ab initio data. Reproduced from Fig. 3 Ref. 187 [187].

The obtained energy may be approximated by the formula:

$$E_{ads}(Al) = \begin{cases} -9.776 + 1.470\,\theta_{Al} + 6.856\theta_{Al}^2 & \theta_{Al} < \frac{1}{4} \\ -4.089 - 3.438\,\times\left(\theta_{Al} - \frac{1}{4}\right) & \theta_{Al} \geq \frac{1}{4} \end{cases} \quad (20)$$

As predicted at the critical coverage, the adsorption energy increased by approximately 3.5 eV. The high coverage dependence is due to the transition of the surface to the metallic state and closing of the gap. Full thermodynamic analysis was performed to obtain extremely low-pressure values for coverage below the critical value. The second stability region corresponds to coverage close to full, which is stable at a somewhat higher, but still relatively low, Al pressure. Thus, the N-terminated surface had a strong tendency to attain a metallic state.

These investigations were predominantly focused on nitride semiconductor surfaces. This is related to the rapid development of nitride-based optoelectronic devices, mostly light-emitting diodes (LEDs) and laser diodes. This is fortunate because GaN and AlN have wide bandgaps. This is essential because gas pressure jump is prerequisite to reveal the electron transfer contribution during adsorption as a result of the jump-like change in the Fermi energy. The other wide gap semiconductor, outside nitride family, suitable for these investigations, is silicon carbide SiC [190]. The clean Si-terminated SiC(0001) surface had a Fermi level pinned by a broken bond state located 0.8 eV below CBM. The Si atom is adsorbed at the H3 site, saturating three broken bonds of the topmost Si atoms ($z = 3$). Four bonds were created; thus, eight states occupied ($N_a = 8$). The topmost Si atom contributes one electron; thus, for three atoms, $Q_i = 3$. The Si-adsorbed atom had four electrons; therefore, $Q_a = 4$. The Si broken bond states are empty after adsorption; thus, $N_f = 0$. Therefore, the EECR critical fraction for the occupation of the six states is $g_{Si-cr} = 3/4$ and accordingly, the critical Si coverage is $\theta_{Si-cr} = 1/4$. The Si adsorption energies are shown in Fig. 34. As shown by the coverage in the neighborhood of the critical value, the Si adsorption energy is reduced.

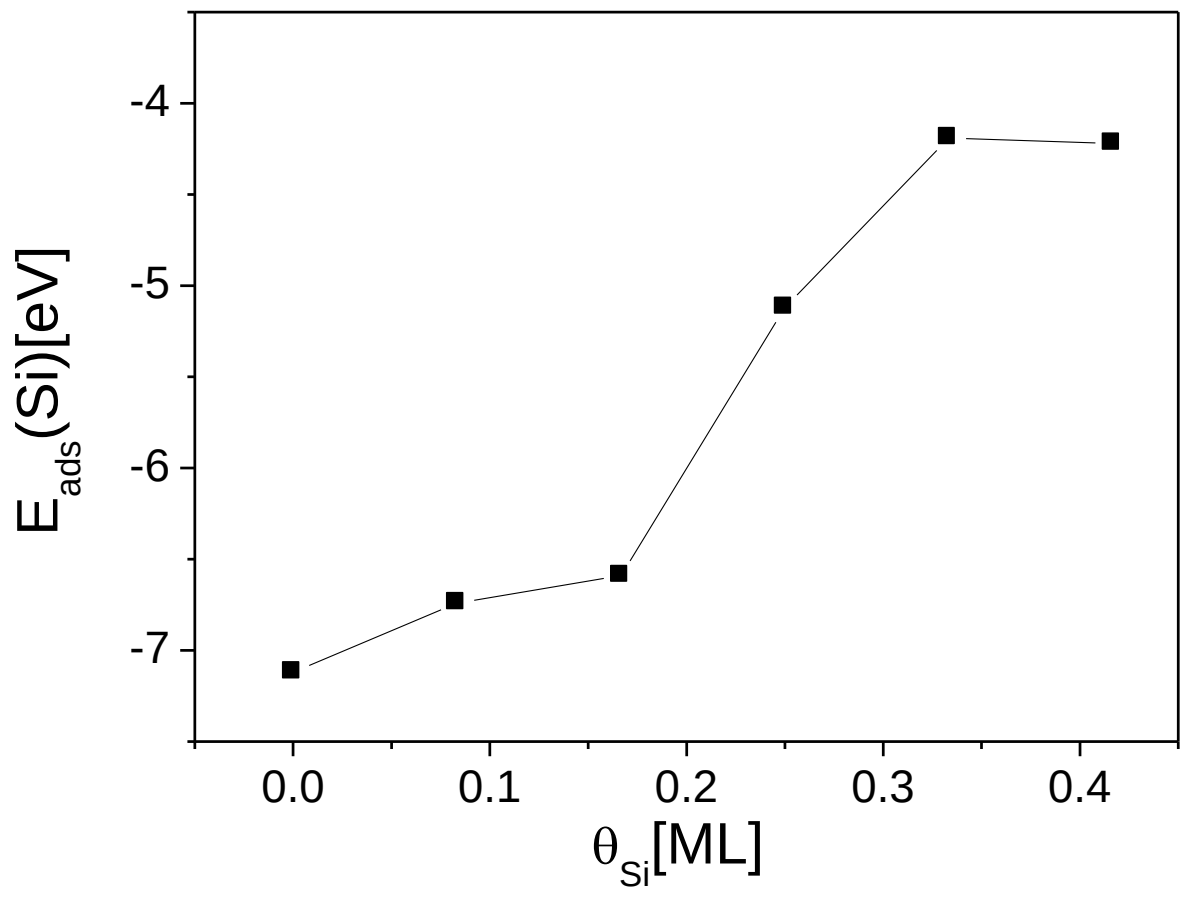

Fig. 34. Adsorption energy of Si adatoms located in H3 sites at the SiC(0001) surface simulated using $\left(2\sqrt{3}\ \times 2\sqrt{3}\right)$ slab. The coverage is defined as the ratio of the Si adatoms before adsorption to the number of the Si topmost atoms in the SiC lattice. Thus in case of finite coverage this points should be shifted to the right to the neighboring value. Reproduced from Fig. 10 Ref. 188 [188].

As shown again, Si adsorption is reduced by the fact that for coverage above the critical $\theta_{Si} > \theta_{Si-cr} =\ 1/4$ one electron should be shifted to a higher energy state. This causes the adsorption energy to be in the neighborhood of the critical value. This behavior indicates that the process is more complex and requires further investigation.

Recently, more extensive simulations of the interactions between polar GaN surfaces and mixed $NH_3/H_2$ vapors have been conducted [189]. They confirmed the dependence of the adsorption energy on the Fermi level pinning at the surface. They identified three different adsorption regimes  for Ga on both Ga- and N-terminated surfaces. As shown in Fig. 35, the adsorption energies varied considerably.

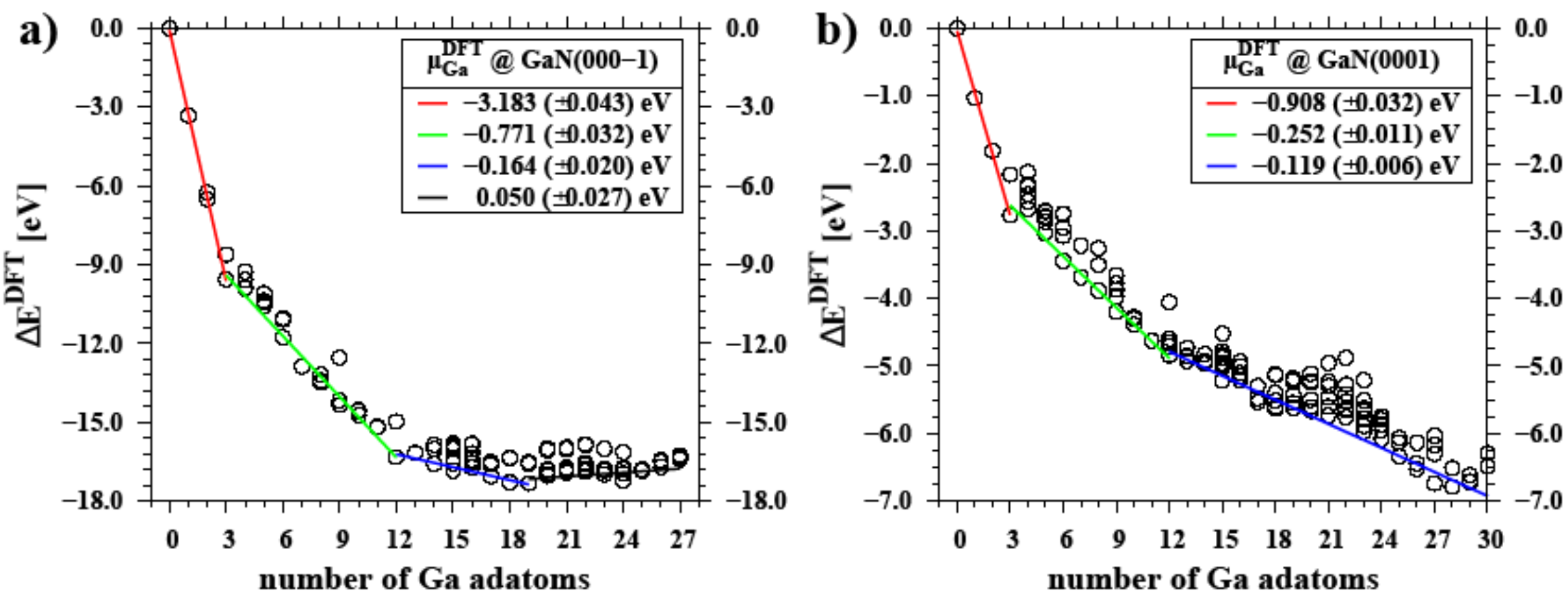

Fig. 35. Change of DFT total energy as a function of the number of Ga adatoms on polar GaN surfaces, (**a**) GaN(000-1), (**b**) GaN(0001). The reference level is the sum of the energies of the clean surface and corresponding number of Ga atoms bound in the crystal. All data points (circles) represent the energies obtained from the DFT calculations for different adatom distributions on the surface. Linear relationships were fitted to the points with the lowest energies; these reviews of adsorption processes show that the above-listed cases show jump-like changes in the adsorption energy at the critical coverage for many systems. In some cases, this process is gradual, and in these cases, other factors may be involved. In the case of jump-like changes, such additional factors were not found; thus, we could conclude that the effect was proven. Reproduced from Fig. 3 in Refs. 189 [189].

The data in these diagrams correspond to the surface free energy change calculated as the difference between the surface with the Ga adatom attached and the surface and G atom separated as a function of the attached atom. The dependence is linear in three separate parts. The slope of this line defines the adsorption energy at these coverages:

Using these data, the authors were able to determine the chemical potential of the Ga adatom on both surfaces. These data were projected onto the MOVE and HVPE growth conditions to determine the basic thermodynamic state of the surfaces during the growth of GaN layers using these methods.

To show the universal character of the above-described phenomenon of electron redistribution among the surface states during adsorption that leads to a jump-like change in the adsorption energy at some critical coverages, the identified jump-like changes are listed in Table 3.

*Table 3. The semiconductor surfaces in which critical adsorption state was identified*

| Surface adsorbate | Critical coverage ML | Fermi level below[a] | Fermi level above[a] | Adsorption energy below[b] | Adsorption energy above[b] | Ref.[c] |
|---|---|---|---|---|---|---|
| H on GaN(0001) | 0.75 ML H | 3.00 | 0.0 | -3.40 | -1.10 | [150] |
| Si → Si on SiC(0001) | 1 ML H | 3.54 | 0.00 | -4.54 | -3.90 | [188] |
| $NH_3$ → H & $NH_2$ (on-top) on GaN(0001) | 0.75 ML H | 3.00 | 0.0 | -2.86 | -0.72 | [149] |
| $NH_3$ → H & $NH_2$ (bridge) on GaN(0001) | 0.75 ML H | 3.00 | 0.0 | -3.05 | -2.02 | [149] |
| $NH_3$ → $NH_2$ & H on GaN(0001) | 0.5 ML $NH_3$ 0.125 ML $NH_2$ 0.125 ML H | 3.00 | 3.47 | -2.80 | 0.00 | [134] |
| $NH_3$ → $NH_3$ on GaN(0001) | 0.5 ML $NH_3$ 0.125 ML $NH_2$ 0.125 ML H | 3.00 | 3.47 | -2.00 | 1.00 | [133] |
| H→ H on GaN(000-1) | 0.75 ML H | 0.00 | 3.47 | -5.80 | -3.10 | [161] |
| Si → Si on SiC(0001) | 0.25 ML Si | 1.40 | 0.00 | -6.70 | -5.00 | [188] |
| N → N on AlN(0001) | 0.25 ML N | 5.50 | 4.00 | -7.20 | -4.00 | [186] |
| O → O on GaN(0001) | 0.3125 ML O | 3.00 | 3.47 | -3.74 | -3.67 | [190] |
| O → O on GaN(0001) | 0.375 ML O | 3.47 | 0.00 | -3.67 | 0.00 | [190] |
| Al → Al on AlN(000-1) | 0.25 ML Al | 0.40 | 2.50 | -8.16 | -4.09 | [187] |
| H → H on AlN(0001) | 0.25 ML H | 0.00 | 5.70 | | | [189] |

[a] *Fermi energy level is expressed in eVs above valence band maximum (VBM),*

[b] *Adsorption energy is expressed in eVs according to Eq. 4, i.e. the difference of the system energy after and before adsorption. Thus the energy is negative for the adsorption state energetically stable,*

[c] *Reference is given to the literature position listing the most precisely determined value.*

An important technological aspect of nitride-based devices is the preparation of substrates for epitaxy. This requires mechano-chemical polishing of the chemically resistant GaN(0001) surface [192,193]. The *ab initio* simulations proved that the exceptional chemical stability of the GaN(0001) surface is related to the specific dependence of the adsorption energy, which has a singular point of maximum energy gain at $\theta_O = \frac{17}{48}\ ML$. As the O adatom is located at the H3 sites, this corresponds to the total coverage of the GaN(0001) surface exposed to oxygen in a standard atmosphere [190].

An additional aspect is the synthesis of ammonia from its constituents [194] and the reaction known as Haber-Bosh process [195]. The process is possibly one of the most beneficial contributions of science to humanity, as it allowed for an uninterrupted increase in the world's human population without the danger of famine on a global scale. The exceptional significance of the invention was appreciated by the Royal Swedish Academy of Sciences, which awarded three Nobel Prizes in Chemistry to Fritz Haber in 1919 [196], Carl Bosh in 1931 [197] and Gerhard Ertl in 2007 [198]. The process was implemented on an industrial scale long ago; nevertheless, there are considerable costs of ammonia synthesis, so it is still actively investigated. These investigations are mostly related to the development of new catalysts such as Fe or Mo. The application of AlN as a catalyzer was also investigated, as it is inexpensive, durable, and possibly stimulates the decomposition of nitrogen molecules into N adatoms, which is a necessary step in the synthesis of ammonia [194]. These results confirm earlier results, indicating that molecular hydrogen and molecular nitrogen decompose during adsorption on the AlN(0001) surface [186, 194]. The decomposition is energetically favorable for molecular hydrogen and nitrogen for the coverage $\theta_H < 0.75\ ML$ and $\theta_N < 0.25\ ML$, respectively. The coverage limit was shown to be directly related to the position of the Fermi level at the AlN(0001) surface at these critical coverages [194]. As these atoms can migrate to the surface, they can react to synthesize ammonia molecules via the Langmuir-Hinshelwood mechanism [199]. However, these preliminary investigations showed the possibility of applying AlN as a catalyst for the Haber-Bosch process.

Another interesting application is the use of AlN as a catalyst for the synthesis of nitric oxide (NO) [200]. The implementation of this process would be as beneficial as the Haber-Bosh process because it would remove the use of hydrogen in the intermediate stages of nitride fertilizer production, resulting in enormous savings. Therefore, the adsorption of molecular nitrogen and oxygen on an AlN(0001) surface was also investigated [200]. The results do not exclude the possibility of the simultaneous adsorption of both gases and the creation of NO. Nevertheless, this is a far more tricky path as it would require that oxygen would not dominate the picture converting the surface into aluminum oxide, which would probably poison the surface [201]. A possible way to achieve a positive result would be to design a catalyst that attaches nitrogen from the vapor phase only. This amounts to the design of the reaction via the Eley-Rideal mechanism, in which only nitrogen is attached to the surface, while oxygen is directly captured from the vapor [201]. This process avoids oxidization and poisoning of the catalyzer. This would probably require the application of a mixed AlGaN crystal and precise control of nitrogen and oxygen partial pressures. The proposed scenario opens the route to difficult and cumbersome research in the future; nevertheless, the rewards would be enormous. Therefore, such investigations could be very attractive to many researchers.

## VIII. Charge role in diffusion

Surface diffusion is related to the surface mobility of localized species, that is, the jumps between the minimal energy positions at the surface. Consequently, this is most frequently simulated using the NEB procedure [202,203]. This approach leads to a single value for the diffusion energy barrier, obtained as the energy difference between the activated complex and the initial value [204]. It is widely used in many cases related to surface diffusion [205].

Recently, it was recognized that this phenomenon has to be investigated with the explicit incorporation of quantum effects in bonding and statistics [11,12]. Several possible scenarios were formulated, depending on the bonding type and statistics and the Fermi level. These cases are illustrated in Fig. 36. As shown, the diffusion barrier may be affected by the initial energy by: (i) the standard bonding picture (Fig 36(a)), (ii) bonding by resonant states (Fig 36(b)), (iii) statistical influence via change of the activated complex quantum states close to the Fermi energy level, and (iv) electron transfer to other states pinning the Fermi level in the case of the activated complex quantum states attaining energy above the Fermi energy and, in consequence, empty.

As an exemplary case for the investigation of the diffusion mechanism, the diffusion of nitrogen adatoms at the Ga-terminated GaN(0001) surface was investigated. This case is also important from a technological point of view as one of the growth methods for nitride structures used in devices in plasma-activated molecular beam epitaxy (PA-MBE) [162,163]. The layers were grown under metal-rich conditions; thus, it was assumed that the metal layer covered the surface. Accordingly the diffusion model was initially proposed in which the nitrogen atom diffuses underneath, known as “new diffusion channel”[206, 207]. First, the energy barrier for the diffusion of N adatoms on the clean GaN(0001) surface was found to be $\Delta E_{bar} = 1.3\ eV$ [206]. This was corrected later to a higher value $\Delta E_{bar} \geq 1.5\ eV$ [207]. In the case of the GaN(0001) surface fully covered by the indium adlayer, this barrier was dramatically lower, $\Delta E_{bar} = 0.5\ eV$. The high barrier for the clean GaN(0001) surface was attributed to the breaking of two Ga-N bonds, which was supposed to cost approximately $3.0\ eV$ while in the case of indium coverage, a strong In-N bond compensates for this effect. It was concluded that, in the case of gallium coverage, this effect is similar, which leads to the emergence of a fast diffusion channel. Subsequently, this effect was used in the interpretation of the PA-MBE growth experiments [208, 209].

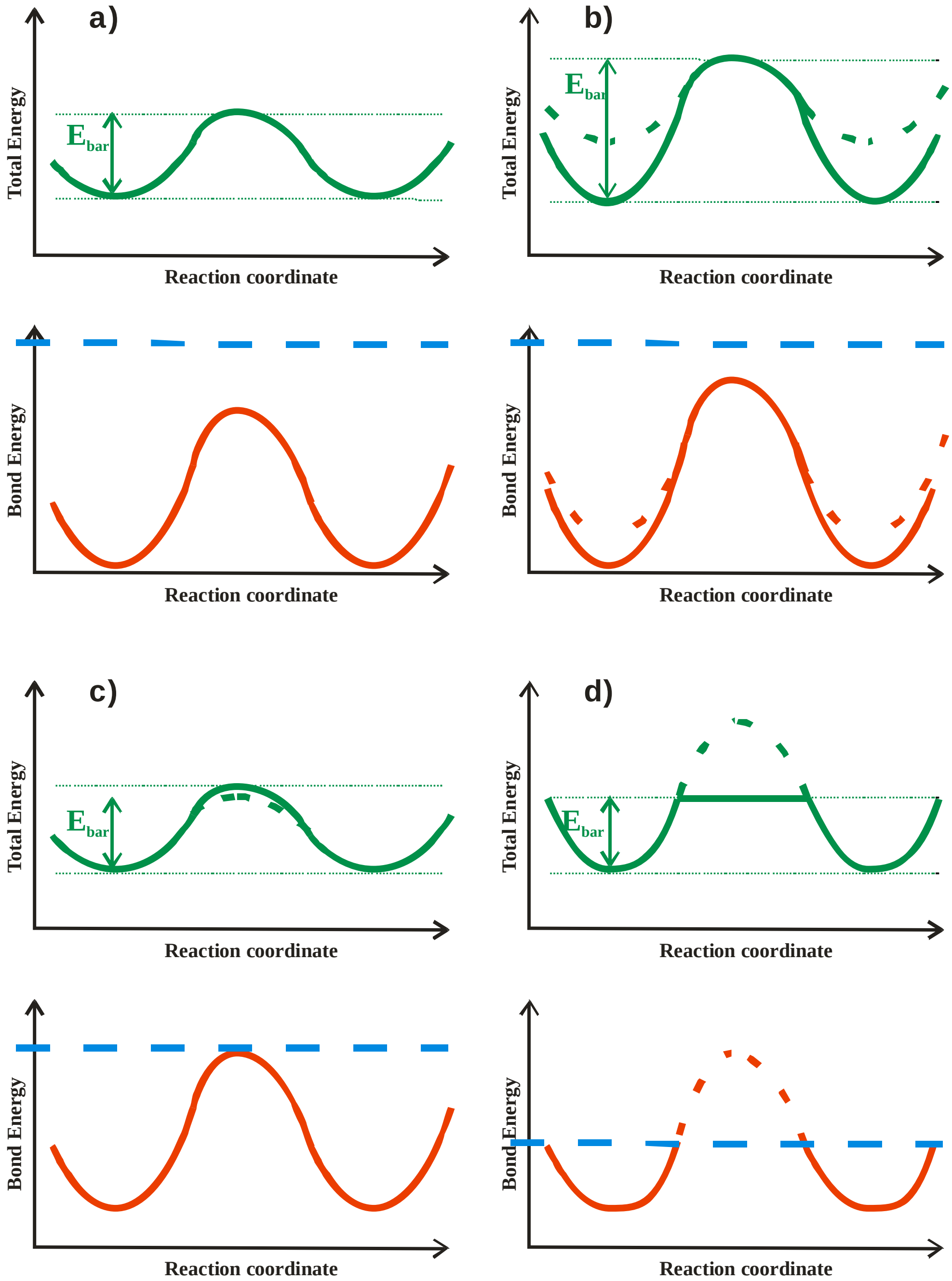

Fig. 36. The factors affecting the effective barrier for diffusion via modification of the energy of the quantum states (red line) and the resulting total energy change (green line) in the course of the jump between two sites: a) standard case: the energy of quantum state remains far below the Fermi energy, b) the energy and the occupation of quantum states, standard and resonant in initial state is changed affecting the barrier c) the energy of quantum states approaches the Fermi energy – the barrier affected via occupation of the states, d) the energy of quantum states extends over the Fermi energy – the energy of the barrier is determined by the Fermi level. The energy of quantum state pinning the Fermi energy (and the Fermi energy itself) is denoted by blue line. Reproduced from Fig. 1 Ref. 14 [14].

Therefore, the case of N adatom diffusion at the GaN(0001) surface under various Ga coverages was investigated again [14]. It was shown that for low Ga coverage, the N adatom diffusion path begins at the lowest energy point at the H3 site via the saddle point in the bridge configuration to the T4 site and repeats again in the inverse order in the symmetric part. The change in the energy of the N adatom was investigated along the NEB surface diffusion path. The results of these investigations on the fractional coverage of the GaN(0001) surface by Ga adatoms are plotted in Fig. 37.

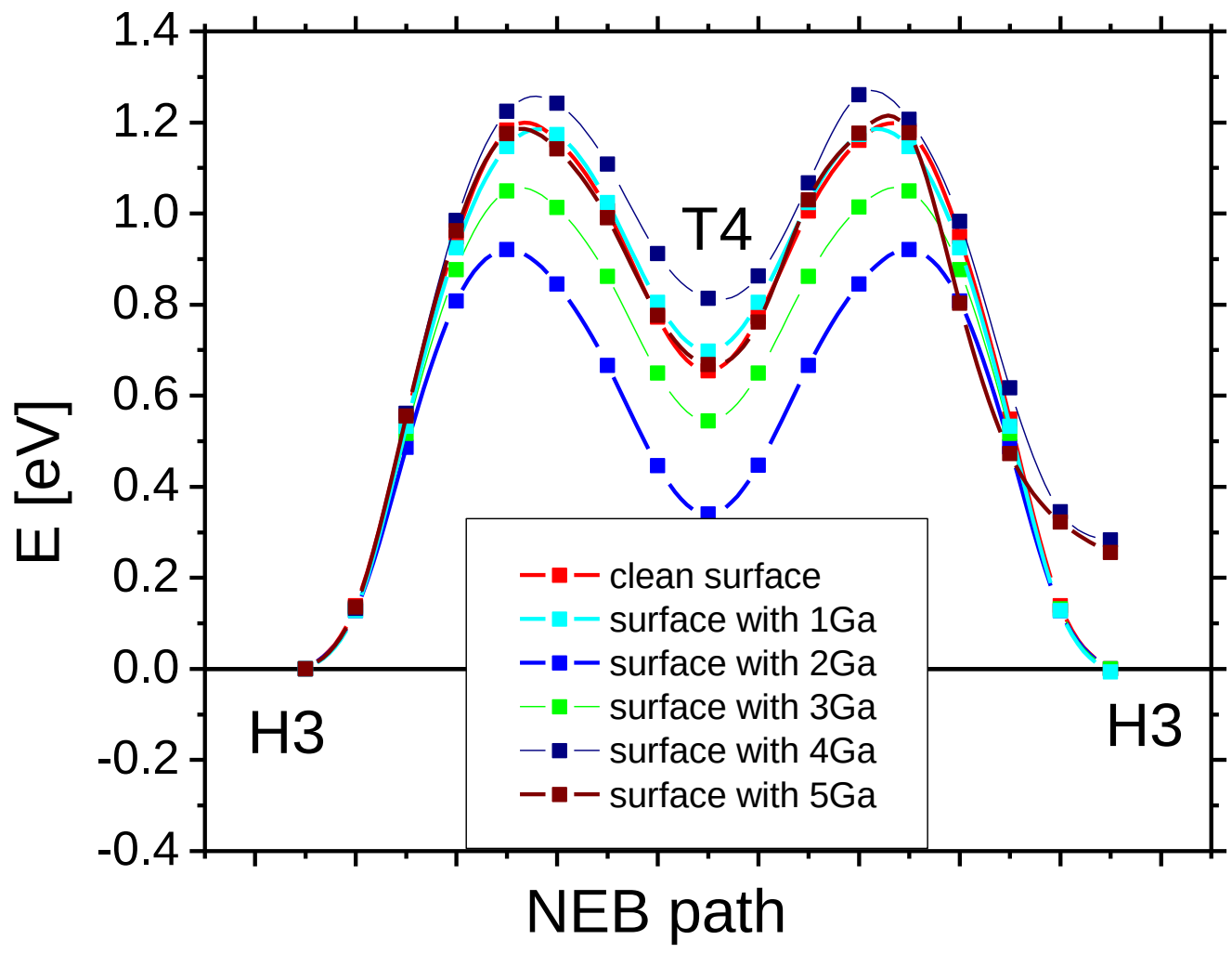

Fig. 37. System energy change during motion of the nitrogen adatom along the jump path over GaN(0001) surface partially covered with Ga adatoms. The number of Ga atoms refer to $(2\sqrt{3}\ \times 2\sqrt{3})$ slab representing GaN(0001) surface, thus a single Ga adatom refer to $1/12\ ML$ Ga coverage. The "clean" surface corresponds to the absence of Ga adatoms, i.e. with N adatom present. Reproduced from Fig. 5 Ref 14 [14].

As it is shown the diffusion energy barrier for clean GaN(0001)surface is $\Delta E_{bar} = 1.18\ eV$ which is good agreement with the results in Ref. 206 and basically also with reference 208. The barrier was determined using a $(2\sqrt{3}\ \times 2\sqrt{3})$ slab representing the GaN(0001) surface; thus, the coverage change was $(1/12)\ ML$ [14]. The lowest barrier obtained was $\Delta E_{bar} =\ 0.92\ eV$ for $(1/6) ML$ Ga coverage. For higher coverage, the barrier increases again.

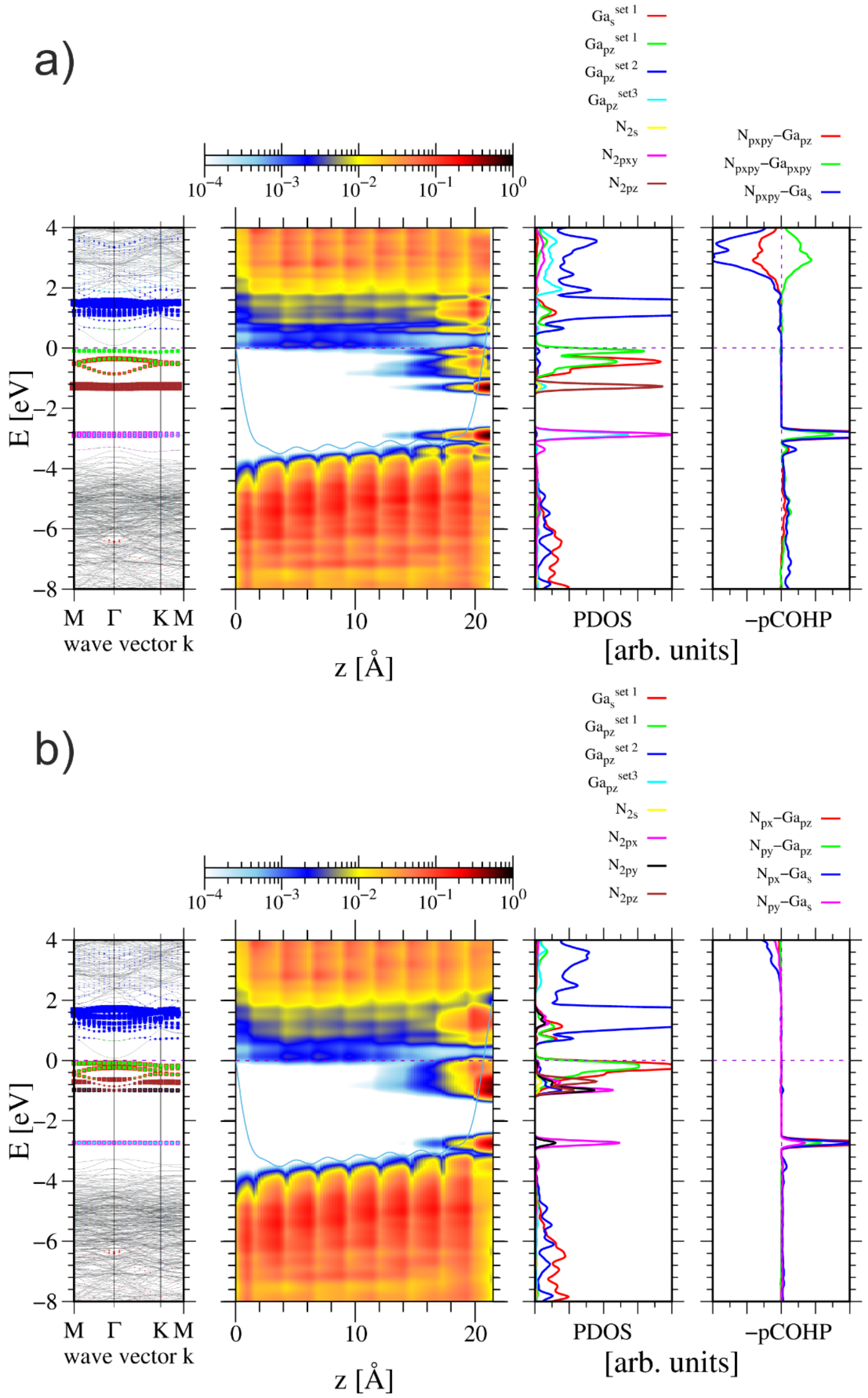

Fig 38. Energies of the quantum states of the $\left(2\sqrt{3}\ \times 2\sqrt{3}\right)$ slab representing clean GaN(0001) surface with single N adatom located in: (a) – H3 site, (b) MAX - max energy position. The panels represents, from the left: energy of the quantum states in the momentum and the position space, projected density of states (PDOS) of the top Ga atoms and N adatom, and the rightmost panel - Crystal Orbital Hamilton Population (COHP) of the nearest atoms [111, 112]. Thus the COHP data correspond to N adatom and the closest topmost Ga atoms. The Fermi energy is set to zero. COHP positive values correspond to the bonding overlap. Reproduced from Fig. 7 Ref 14 [14].

Therefore, detailed studies on the bonding properties of the path were undertaken. As shown in Fig. 40, the bonding of the N adatom at the H3 site was related to the existence of the three resonating bonds discussed earlier. They are located 2.4 eV below the Fermi level, which is pinned by the broken bond $sp^3$ hybridized state of the top-layer Ga atoms. Thus, they are fully occupied owing to their resonating character with a probability $P = 2/3.$ In addition the fourth N bond is broken having its energy about 0.5 eV below the Fermi energy and is also occupied. In the bridge configuration, one of these bonds is broken with energy located 0.68 eV below the Fermi levels. Therefore, it is still occupied. The barrier energy difference between H3 and the bridge configuration was related to the change in the energy of the fully occupied state. No N state is empty, which is related to the generally low energy of the nitrogen atom quantum states.

The increase in Ga coverage to $(1/6)\ ML$leads to a change in the Fermi level pinning, as these top layer Ga atoms have Ga adatoms attached. Somme others still preserve $sp^2 - p_z$ configuration in which $p_z$ state is empty. The Fermi level is pinned to the bonding $\left|Ga_{4p_z}\right\rangle - \left|Ga_{sp^3}\right\rangle$ states and, therefore, lower. Therefore, the broken nitrogen bond state changed its occupation, which led to a decrease in the energy barrier to $\Delta E_{bar} = 0.92\ eV$ for $(1/6)ML.$

The energy barrier for full Ga coverage (13 adatoms in $\left(2\sqrt{3}\ \times 2\sqrt{3}\right)$ slab) was also investigated. As shown, the minimal-energy N adatom position is changed to the on-top. This has important consequences for the general understanding of the atomic GaN growth mechanism in MBE. In fact, an on-top position is required to build a new crystalline layer of GaN. The N adatom is bonded to three Ga atoms from the top layer of the GaN lattice and two from the adlayer. The other two Ga neighbors belong to the Ga adlayer. The quantum states created by overlap between $\left|N_{2p}\right\rangle$ and $\left|Ga_{sp^3}\right\rangle$ hybridized states are identified at $E_1 = -7.97\ eV$ and $E_2 = -7.68\ eV$. These states and their overlaps are clearly shown in Fig. 38 (a) and Fig. 38 (b).

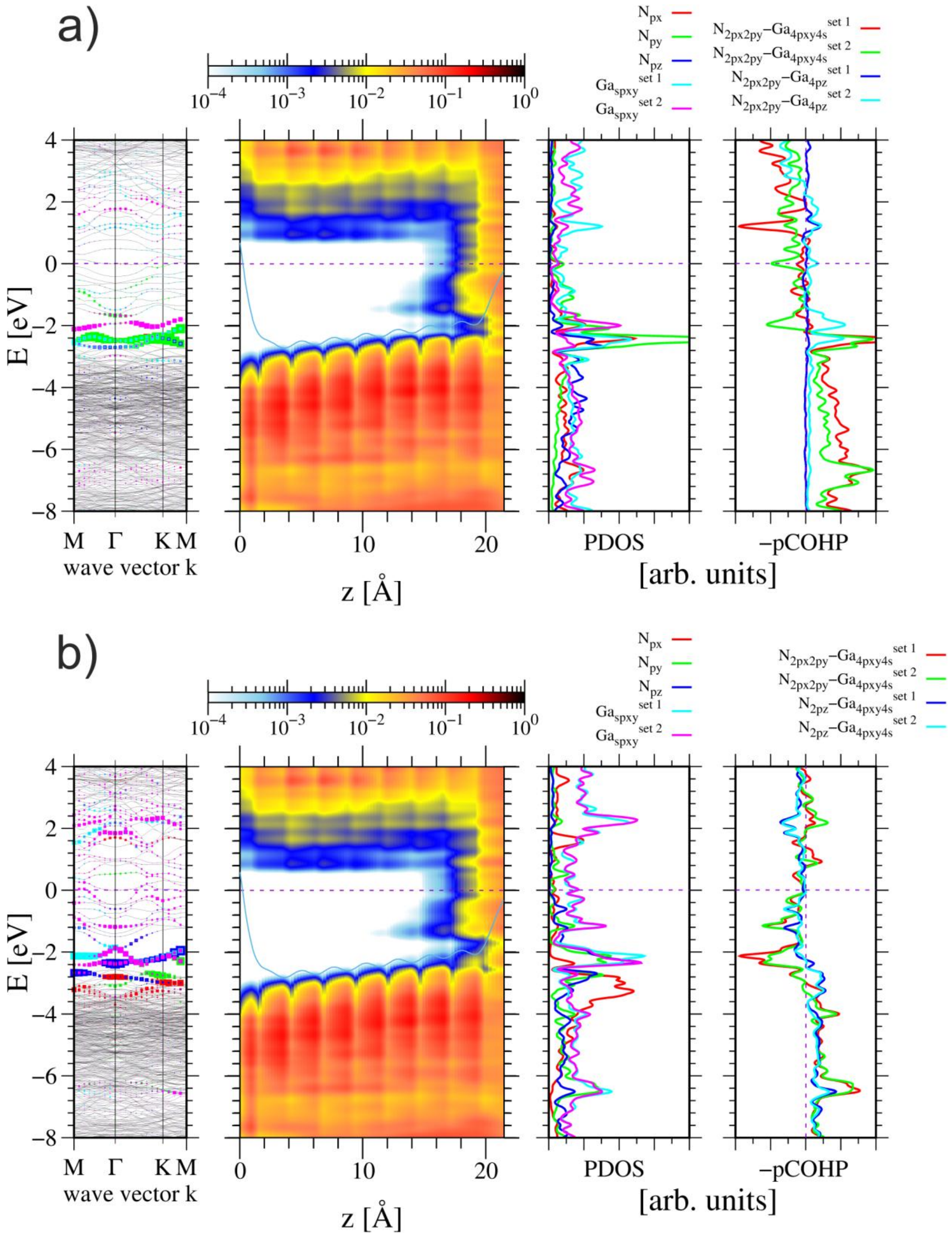

Fig 39. Energies of the quantum states of the $(2\sqrt{3}\times 2\sqrt{3})$ slab representing GaN(0001) surface fully covered by Ga adatoms, a single N adatom located in: (a) – on-top, (b) max energy position (H3). The symbols are analogous to Fig. 20. Reproduced from Fig. 14 Ref 14 [14].

The motion along the diffusional path encounters an energy maximum located at the H3 site. In this case, the path avoids the bridge location; therefore, the overlap is different from earlier results. Because of the coverage by excess Ga adatoms, no broken nitrogen bond states were created. In fact, the N adatom creates full bonding with the Ga neighbors. Nevertheless, the energies of these states change considerably to $E_1 = -7.54\ eV$ and $E_2 = -7.33\ eV$. Thus the energy change is substantial at $\Delta E_1 = 0.43\ eV$ and $E_2 = 0.35\ eV$. The change in energy of the extended state could not be assessed. Nevertheless, because these changes involve four electrons, they fully explain the existence of the barrier $\Delta E_{bar} = 1.23\ eV$. Thus, such a barrier is associated with a change in the energy of quantum states. The energy profiles are presented in Fig. 40.

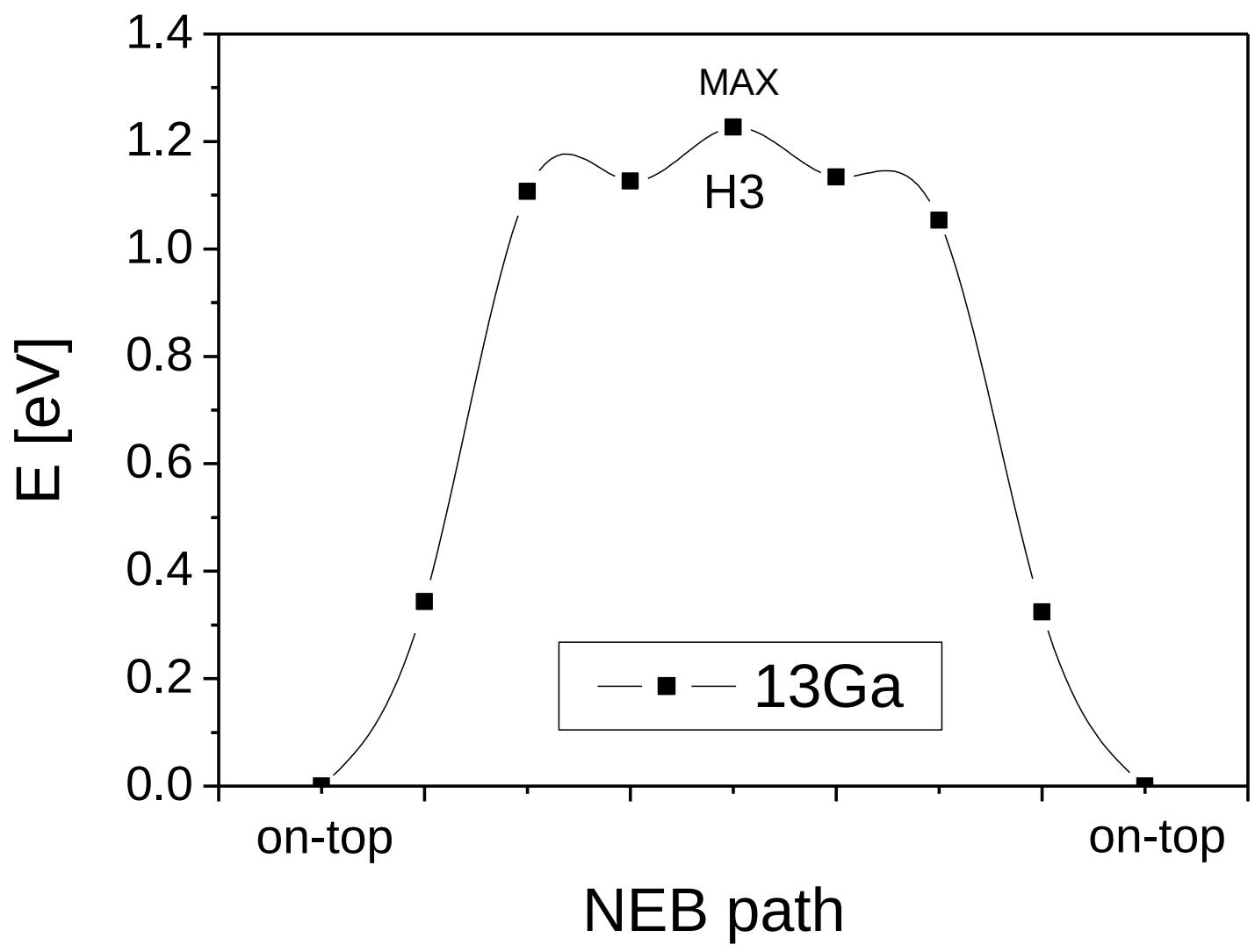

Fig. 40. System energy change during motion of the nitrogen adatom along the jump path over GaN(0001) surface fully covered by Ga adatoms. The number of Ga atoms refer (13) to $(2\sqrt{3}\ \times 2\sqrt{3})$ slab representing GaN(0001) surface, thus it is excess of Ga single layer coverage $\theta_{Ga} = (13/12)\ ML$. Reproduced from Fig. 12 Ref. 14 [14].

From these data, it follows that the energy barrier for full Ga coverage was $\Delta E_{bar} = 1.23\ eV$. Thus it is higher than to $\Delta E_{bar} = 0.92\ eV$ for $(1/6)ML$ Ga fractional coverage. Therefore, the existence of a diffusion channel for Ga coverage was not confirmed. It has to be added that the low energy barrier is related to relatively low energy of the broken bond state of N adatoms. Therefore, the entire picture is related to quantum statistics and quantum bonding, which affect the properties of the N adatom.

## IX. Summary

This review summarizes the *ab initio* investigations of the charge influence of charge on the properties of semiconductor surfaces due to quantum effects in the recent years period. These results are summarized by presenting the state of the art before these investigations took place, the results reported in this review, and the state of the art after these investigations were accomplished. The results reported results may be divided into the methodological and the physical contributions.

In summary, the state of the art before the research presented in this review was completed may be presented as follows (10):

1. The surface dipole charge leads to the emergence of potential coupling of the slab copies via PBC, which can be counterbalanced by the introduction of the compensating opposite dipole in the vacuum between the slab copies.
2. Separation of the real and termination surfaces by several atomic layers leads to asymptotic behavior in which these surfaces are quantum independent, such that the system properties obtained are independent of the number of layers in the slab for a relatively limited slab thickness.
3. A surface state band diagram can be obtained by the direct projection of all states in the slab on the plane of the momentum parallel to the surface.
4. The adsorption energy of the adsorbate was obtained by direct calculation of the energy difference between the total energy values of the far distance position of the adsorbate and the adsorbate attached to the surface.
5. Adsorption energy is the inherent value of adsorbate and surface bonding, which can be affected by adsorbate-adsorbate interactions only. In the absence of these interactions, the adsorption energy can be described by a single value of the energy gain in the process.
6. Reconstruction is related to the minimization of the surface energy. The reconstruction period is an intrinsic property of the surface and is not related to the charge balance.
7. ECR analysis can be successfully used to determine the charge distribution and Fermi level at a surface.
8. Gallium nitride and other wurtzite nitride semiconductors are bonded by the overlap of $sp^3$ hybridized orbitals, leading to the emergence of valence and conduction bands separated by the bandgap.

9. Thermalization of the adsorbate is relatively minor issue in adsorption in which the dissipation of the excess kinetic energy proceeds via motion of “hot” atoms along the surface.
10. The diffusion energy barrier is a single value, independent of the Fermi level and quantum statistical properties of the bonding.

The results obtained in the reported period are summarized as follows (16).

1. The existence of a dipole in the surface slab was confirmed, thus justifying the incorporation of the opposite, compensating the dipole to the slab models as a physically sound approach. The problem that may arise is the precision of the calculated dipoles. The alternate method, known as Laplace correction, could zero the electric field in the space between slab copies by adding the solution of the Laplace equation to the electric potential, which may possibly lead to better precision.
2. A system consisting of two presumably isolated subsystems, such as the slab and the species at a far distance, may possibly have two different Fermi energy levels. Standard *ab initio* procedures use a single Fermi energy and accordingly distribute the electrons among the quantum states using their energies equally for both subsystems. This may lead to the charging of the slab and the far distance species (adsorbate) by the opposite charge. In this case, the energy of the system composed of two parts was different from the energy calculated separately for both parts, which affected the obtained adsorption energy values.
3. The asymptotic far-distance behavior of the wavefunction of any finite system is exponential with no cutoff value. The separation of the real and termination surfaces leads to a mere decrease in the overlap magnitude of their wavefunctions, and not to complete isolation. Therefore, in the asymptotic regime, only the wavefunction overlap can be reduced. Thus, the separation of the two surfaces of the slab (real and terminated) is always incomplete.
4. The electric potential and semiconductor bands are controlled by the charge distribution in the slab. The profile has a position dependent slope in the case of Fermi level pinning and the presence of a net charge on the surface states. In the presence of net bulk charge within the slab, the potential profile has an additional parabolic component. Thus, the best method of analysis is to use the real-space electric potential and band distribution. The nonzero electric-field slab solution reflects a real semiconductor surface in which a subsurface charge layer exists.
5. In most cases, the Fermi level is pinned at the real surface of the slab. Therefore, the potential profile has a slope and, accordingly, the bands. Thus, the effective bandgap obtained by the direct projection of the slab states is smaller than that calculated for the bulk material.

Therefore, the projection-type representation of quantum states at the surface is burdened by field-related errors.

6. Long-range variation in the potential may be obtained by appropriate averaging procedures. The resulting spatial variation is in agreement with the spatial variation of the bands.

7. Reconstruction is correctly recognized as a minimization energy process. In addition, the reconstructed surface patterns and periodicity stem from the charge balance within the surface states, which can be determined using the ECR.

8. ECR analysis may be extended to the charge of the adsorbate and occupation of the new and old states. Thus, ECR may be extended to EECR.

9. Adsorption leads to the emergence and disappearance of new quantum states that are occupied by the electrons of the slab and adsorbate. This may lead to the total occupation of some states and a drastic change  the adsorption energy value at some coverages. At this point, the Fermi level was not pinned by the surface states and changed freely. This is followed by drastic changes in the equilibrium pressures under specific conditions. Accordingly, crystal growth is conducted under conditions at which the Fermi level is not pinned.

10. Entropy contribution may play important role in the thermodynamic properties of the adsorption.

11. Gallium nitride bonding occurs via the creation of two separate sub-bands: the upper due to the overlap of the $sp^3$ hybridized gallium states and nitrogen $p$ states, and the lower due to the overlap of gallium $d$ states and nitrogen $s$ states. The wurtzite tetrahedral symmetry of the lattice is stable because of the resonant states of nitrogen bonding, where four nitrogen states are bonded to gallium neighbors according to the resonant states created from three nitrogen p orbitals and occupied with fractional probability.

12. Thermalization of the adsorbate occurs predominantly via tunneling of electrons from the adsorbate to the solid interior induced by the external dipole electric field. After the electron loss, the positively charged adsorbate is decelerated by this field, losing excess kinetic energy; thus, smooth landing is possible.

13. The diffusion energy barrier may be affected by the change in the energy of the bonding states, which may be close to or even cross the Fermi level. This could drastically affect the energy barrier.

14. The diffusion energy barrier may be affected by the quantum statistics influence on the initial states, either via the resonant state or due to the Fermi energy change.

The present state of the art is different from that of the previous state. The main differences are (11):

1. The Coulomb interaction of the separate subsystems, both the slab copies and the slab, and a distant adsorbate are described using the electric potential distribution, showing its possible consequences for surface modelling.

2. The asymptotic exponential behavior of the wavefunction is used to explain its consequences for the slab models.

3. The existence of a subsurface charge dipole layer and its role in the *ab initio* representation of the wavefunctions are demonstrated, with an emphasis on the electric potential profile and the charge distribution close to the surface.

4. The electric field at the surface is usually present when the Fermi level is pinned at the surface. In addition, it could be present in the slab simulations even if the Fermi level is not pinned at the termination surface. Thus, the projection of band states across the slab may lead to an artificially small bandgap.

5. The charge role in the clean and adsorbate-covered surfaces is associated with the direct application of EECR in the explanation of the structure and periodicity of the reconstructed surfaces.

6. The role of charge balance in adsorption: The discovery of the adsorption energy jump at the coverage corresponding to the Fermi level is free.

7. The jump-like change the adsorption energy plays an important role in the thermodynamics of the adsorbate. The fractions of the equilibrium states for the Fermi level pinned and free are reduced and dominant in the pressure coordinate, respectively. The opposite effect is observed for surface diagrams in coverage coordinates: the fraction of surface states with the Fermi level pinned and free is extended and contracted, respectively.

8. The above connection has a direct consequence on the growth that occurs predominantly for the Fermi level free.

9. The role of the external charge dipole layer in the thermalization of the adsorbate via electron tunneling was formulated.

10. The hypothesis of resonant bonding of nitrogen is important, both in bulk nitrides and in some configurations of nitrogen as an adsorbate.

11. Quantum effects play an important role in bonding and statistics in the determination of the surface diffusion barrier, both in the initial state and in the activated energy complex.

In summary, the reported developments have changed the scientific landscape of semiconductor surfaces to a large extent. It must be added that  the progress is not complete in any sense. The present state of the art in the field requires research horizontally directed towards extending the scope of semiconductor systems, surfaces and adsorbates. The finished examples constitute a small fraction of all cases that require further investigations. Fortunately, the formulation of fundamentals was advanced, which provides a good starting point for the application of a massive calculation approach based on artificial intelligence (AI) algorithms.

The second research direction should be in- depth, that is, to investigate completely new features of semiconductor surface systems. These include the surfaces with different structures and compositions. This includes vicinal surfaces, that is, surface structures with parallel step sequences. From a more distant perspective, this will incorporate kink structures, that is, the local jumps on the steps. In these cases, the number of assumptions in the presently used model must be considerably modified. These require a creative approach that cannot be solved using AI-based methods. In fact, investigations of these steps have already started [210,211]. At this time, they used the old approach, which has a limited chance of success. Therefore, a considerable extension of these models is necessary in the future.

**Acknowledgement**

The paper was critically read by Victor M. Bermudez, for which the authors would express their gratitude, particularly for the critical comments and corrections to the text.

The calculations reported in this paper were performed using the computing facilities of the Interdisciplinary Centre for Mathematical and Computational Modelling of Warsaw University (ICM UW) under several grants over the reported period.